\documentclass[apj]{emulateapj}
\slugcomment{ApJ, Accepted March 17 th 2016}
\usepackage[dvipsnames,usenames]{color}
\pdfoutput=1
\usepackage{graphicx,amsmath,natbib, hyperref,amsfonts}
\usepackage{appendix}
\usepackage{longtable}
\usepackage{graphicx,natbib}

\shorttitle{KLIP Forward Modeling}
\shortauthors{Pueyo}

\begin{document}

\title{Detection and Characterization of Exoplanets\\ using Projections on Karhunen-Lo\`eve Eigenimages: Forward Modeling }

\slugcomment{ApJ, Accepted March 17 th 2016}

\author{Laurent Pueyo}
\affil{Space Telescope Science Institute, 3700 San Martin Drive, Baltimore MD 21218, USA}
\email{email: pueyo@stsci.edu}

\begin{abstract}

A new class of high-contrast image analysis algorithms that empirically fit and subtract systematic noise has lead to recent discoveries of faint exoplanet /substellar companions and scattered light images of circumstellar disks. These methods are extremely efficient at enhancing the detectability of faint astrophysical signal, but they do generally create systematic biases in their observed properties. This paper provides a general solution for this outstanding problem. We present the analytical derivation of a linear expansion that captures the impact of astrophysical over-subtraction and/or self-subtraction these image analysis techniques. We examine the general case for which the reference images of the astrophysical scene move azimuthally and/or radially across the field of view as a result of the observation strategy.  Our new method is  based on perturbing the covariance matrix underlying any least-squares speckles problem, and propagating this perturbation through the data analysis algorithm. Most of the work in this paper is presented in the Principal Component Analysis framework, but it can be easily generalized to methods relying on linear combination of images (instead of eigenmodes). Based on this linear expansion, obtained in the most general case, we then demonstrate practical applications of this new algorithm. We first consider the case of the spectral extraction of faint point sources in IFS data  and illustrate, using public Gemini Planet Imager commissioning data, that our novel perturbation-based Karhunen-Lo\`eve Image Processing Forward Modeling (KLIP-FM) can indeed alleviate algorithmic biases. We then apply KLIP-FM to the problem associated with the detection of  point sources. We show how it decreases the rate of false negatives (e.g missed planets) while keeping the rate of false positives unchanged when compared to classical least-squares fitting methods. This can potentially have important consequences on the design of  follow-up strategies of ongoing direct imaging surveys.
\end{abstract}


\maketitle 
\section{Introduction}
\label{sect:intro}

Progress in the domain of high-contrast image analysis has spearheaded recent discoveries of faint exoplanets /substellar companions,   
and resulted in spectacular scattered light images of cirsumstellar disks 
around nearby stars. This progress has been mostly driven by a new class of direct imaging data analysis algorithms \citep{lmd07,2012MNRAS.427..948A,2012ApJ...755L..28S} that empirically fit and subtract systematic noise in coronagraph data  (also called speckle noise). Speckles stems from light diffracted by the optics in the telescope and the instrument. They are a major nuisance when seeking to detect faint circumstellar point or extended sources, due to their characteristic temporal  and spatial scales (respectively of the order of the exposure time and of the size of the image of a point source). Modern coronagraph data analysis methods  calibrate this noise by using local estimates of the speckles' correlation between the science exposures and a library of noise realizations. This speckle fitting has been so far carried out in the least-squares sense. The collection of reference images is sometimes obtained using observations of calibration stars that act as true references (Reference Differential Imaging;  hereafter RDI). However, in most ground-based cases, the library of noise realizations is assembled using exposures of the source of interest in configurations for which the observer knows a priori that the location of the faint astrophysical source moves in the frame attached to the speckles. In these cases, each image can both be treated as a science frame and also included in the reference stack corresponding to other exposures and/or wavelengths in the sequence. Observation strategies enabling this feature include azimuthal motion of the astrophysical signal with respect to the speckles (Angular Differential Imaging ADI; \cite{mld06},  radial motion (with Integral Field Spectrograph observations, Spectroscopic Spectral Differential Imaging SSDI; \cite{sf02}, or they are based on the intrinsic properties of the hypothetical sources surveyed for (Polarization Differential Imaging, PDI, or presence of sharp spectral feature for Spectral Differential Imaging, SDI  \cite{bcl04}.\\

Once a library of noise realizations, or Point Spread Functions (PSF), has been assembled according to one or more of these strategies, least-squares fitting algorithms can be finely tuned. This is achieved in a variety of ways including optimizing how the field of view is partitioned before speckle fitting (e.g adapting the analysis to how locally one thinks the speckles are correlated), varying the selection criteria that select the ``best'' noise realizations from the ensemble of references and regularizing the inverse problem. While implementations and choice of algorithmic parameters vary amongst authors, the consensus emerging in the community is that these methods are extremely efficient at enhancing the detectability of faint astrophysical signals, but do generally create systematic biases in their observed properties (namely: photometry, spectra, astrometry of point sources and morphology, surface brightness of circumstellar disks). Now that large surveys based on Extreme Adaptive Optics Coronagraph instruments are hitting their full stride \citep{bfd08,hoz11,2014PNAS..11112661M}, these biases are  becoming one of the chief problems in high-contrast image analysis.\\

During the past few years several authors have proposed algorithmic modifications in order to mitigate such biases \citep{mmv10,pcv12,2014ApJ...794..161F,2014SPIE.9148E..0UM,2014arXiv1409.6388P}. Forward Modeling  in the context of exoplanet imaging was first proposed by \citet{mmv10} and \citet{lbc10}. It aims at jointly estimating the instrument response and the astrophysical signal. To do so, negative synthetic sources are injected in the raw data across the entire observing sequence. This new data set, with both positive astrophysical and negative synthetic signals, is then propagated through the reduction algorithm. Jointly minimizing the residuals in such processed images (by exploring the range of possible astrophysical properties for the synthetic negative sources) retrieves in principle the unbiased observables of the astrophysical signal. \citet{2012ApJ...755L..28S} suggested that carrying out least-squares speckle subtraction using  Karhunen-Loeve Image Processing (KLIP, or Principal Component Analysis, PCA)  provides a simple and computationally efficient framework to carry out astrophysical inference in a way that is equivalent to injecting a negative synthetic source in the raw data. However, that paper did not fully describe how to implement this Forward Modeling with KLIP (hereafter, KLIP-FM) in the most general case. \citet{2014arXiv1409.6388P} revisited this problem and described how to apply  KLIP-FM in the context of RDI, when the library of reference images is built using calibrator stars (with no astrophysical signal in the library). That paper then discussed how to modify the ADI/SDI problem so it mimics the RDI configuration, and thus in principle reduces biases on astrophysical estimates. That technique was used in \citet{2013ApJ...779..153H,2013ApJ...768...24O} and \citet{2015ApJ...798L..43C}.  In parallel, \citet{2013ApJ...764..183B} and \citet{2014ApJ...780...25E}, for point sources and disks , respectively, discussed how the presence of astrophysical signal in PSF libraries obtained using ADI can be accounted for as a small perturbation of the least-squares coefficients. They then showed how these small perturbations could be included in a Forward Modeling framework to self-calibrate biases on astrophysical observables a posteriori. \\  

{\color{black} In the present manuscript we generalize this class of perturbation analysis to all type of observations. Our main objective is to describe the principles underlying KLIP-FM in the most general case (e.g without the strong hypothesis previously discussed in the literature). The novelty of our method relies on an analytical expansion for the Principal Components, when astrophysical signal is present in the reference images. Because of their high technicality, we leave both the proof of this analytical expansion and the algorithmic details regarding its implementation out of main body of the paper. Instead \S 2 provides a high-level description of our main result and places it into the context of previously published work. We then demonstrate the advantages of our approach by applying  it to two key exoplanet imaging applications: spectral characterization with an Integral Field Spectrograph  (\S 3) and point source detection  (\S 4). We limit the scope of this paper to these  two practical examples. In \S 5 we conclude by listing other science cases for which our method could be potentially beneficial. The technical background underlying our results is then discussed in depth in the Appendices:
\begin{itemize}
\item Appendix A provides the most general formalism for an ADI + SSDI observing sequence and lays out the formal foundations for our work. 
\item Appendix B summarizes the notations Appendix A in a table format. In order to facilitate numerical implementation, it provides the dimensions of the various matrices discussed in this paper. 
\item Appendix C introduces Forward Modeling in the most general case, and then discusses the specific configuration of RDI. This was already presented in \citet{2014arXiv1409.6388P}, but serves here to set up the stage for Appendix F.  
\item Appendix D describes Forward Modeling for astrometry and photometry of point sources using the linear algebra notations introduced in Appendix A and C. It also set up the stage for the spectral estimation algorithm described in Appendix F. 
\item Appendix E contains the proof of our main result. It heavily relies on the notations introduced in Appendix A and summarized in Appendix B.
\item Appendix F describes how to take advantage of the result in Appendix E to carry out Forward Modeling for the estimation of point source' s spectra using IFS data. 
\end{itemize} 
}


\section{Generalized Forward Modeling}

\subsection{Over-subtraction and Self-subtraction}
\begin{figure*}[t]
\begin{center}
\includegraphics[width=1\textwidth]{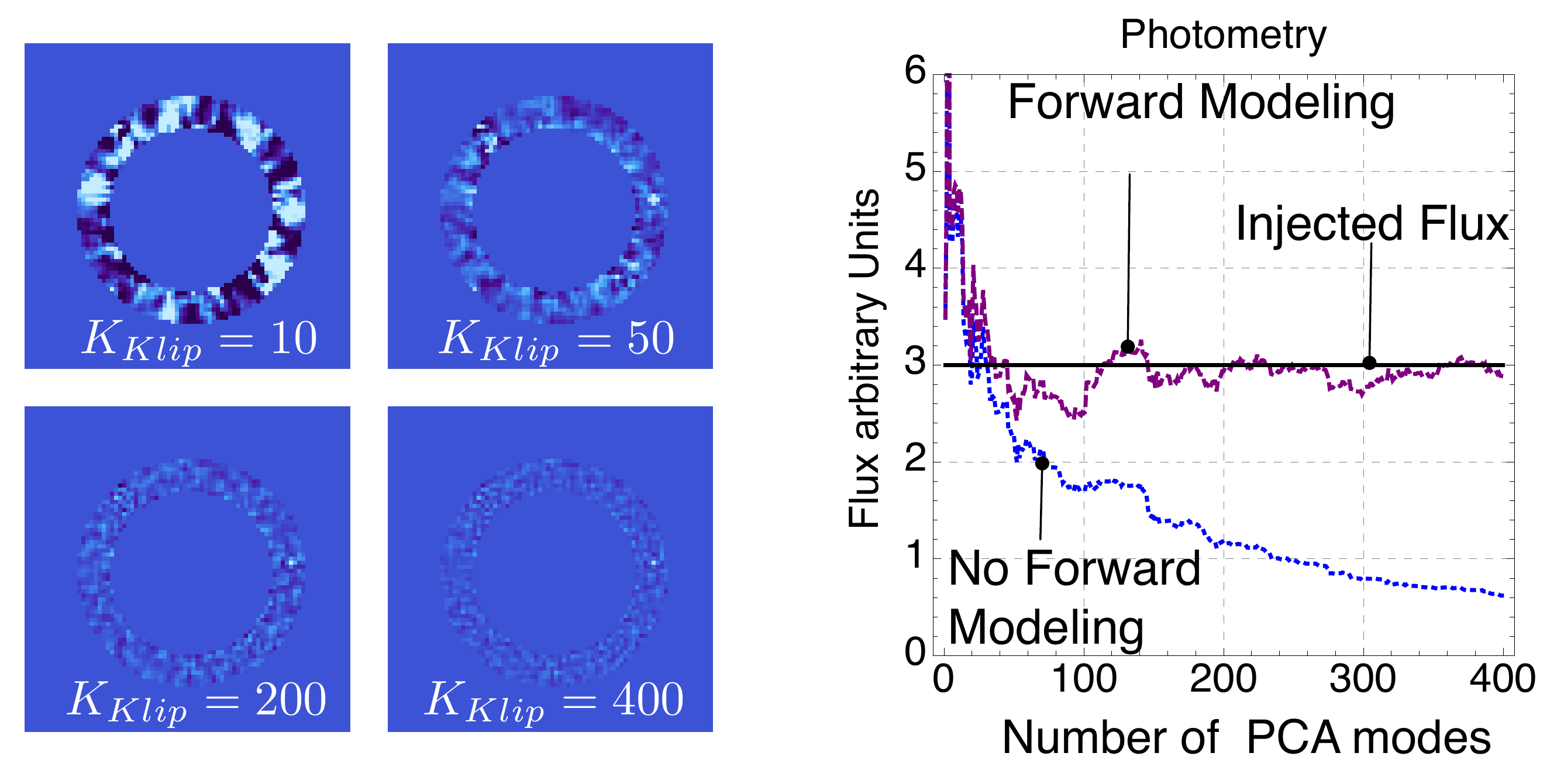} 
\caption{{\bf  Forward Modeling for a point source with RDI. Based on injecting a synthetic point source of known brightness in HST-NICMOS data}. {\em Left,} reduced images obtained for four values of $K_{Klip}$: the detectability of the point source changes with this parameter. When $K_{Klip}$ is too small the point source is not detected. It becomes apparent for larger values $K_{Klip} = 50$, albeit with some residual spatially correlated speckle noise. When $K_{Klip} =200-400$, spatially correlated residual speckles disappear but the point source has been significantly over-subtracted. {\em Right,} estimated flux as a function $K_{Klip}$ with and without Forward Modeling. Without Forward Modeling the estimated flux decreases as the over-subtraction becomes more prominent. With Forward Modeling the injected photometry is retrieved and stable when $K_{Klip}$  is large enough -  e.g when the residual speckle noise is well behaved.
}
\label{fig:ForwardRDI}
\end{center}
\end{figure*}

{\color{black} We start with the notations discussed in \citet{2012ApJ...755L..28S}, and assume the case of a target image $T(\mathbf{x})$ (where $\mathbf{x}$ is the spatial dimension) along with a set of reference images  $R_k(\mathbf{x})$. The details of how  $T(\mathbf{x})$ and  $R_k(\mathbf{x})$ are chosen among some generic coronagraph sequence are not discussed here. We refer the reader to Appendix A for a thorough presentation of the parameters associated with building a target/reference library in the most general case. An orthonormal basis $Z_k(\mathbf{x})$ is then obtained based on the eigenvectors of the references' covariance matrix. The associated eigenvalues $\Lambda_k$ are ranked in decreasing order. They quantify how prevalent each mode $Z_k(\mathbf{x})$ is  in the reference stack. When $\Lambda_k \gg 1$ the mode is present in most of the reference images, and conversely when $\Lambda_k \ll 1$ it is absent from most references. Again, linear algebra details are given in Appendix A. When astrophysical signal $A(\mathbf{x})$ is present in the target -- e.g., $T(\mathbf{x}) = I_{\psi}(\mathbf{x}) + A(\mathbf{x})$, with $I_{\psi}(\mathbf{x}) $ standing for the speckle noise realization in the target image to remain consistent with \cite{2012ApJ...755L..28S} -- the resulting processed image $P(\mathbf{x})$ is given by the sum of two terms $P(\mathbf{x}) = P_{spe}(\mathbf{x}) +P_{sig}(\mathbf{x}) $:
\begin{itemize} 
\item The residual speckles that have not been fully captured by the PCA:
\begin{equation}
P_{spe}(\mathbf{x})  = I_{\psi}(\mathbf{x}) - \sum_{k =1}^{K_{Klip}} <I_{\psi}(\mathbf{x}),Z_k(\mathbf{x})>_{\mathcal{S}} Z_k(\mathbf{x})
\end{equation}
where $K_{Klip}$ corresponds to the number of Principal Components over which the target image is projected and $< \bullet,\bullet >_{\mathcal{S}}$ stands for the $L_2$ inner product on the portion of the field of view over which the speckle fitting is carried out (also called the $\mathcal{S}$ zone). 
\item The astrophysical signal, corrupted by the  KLIP algorithm:
\begin{equation}
P_{sig}(\mathbf{x})  = A(\mathbf{x}) - \sum_{k =1}^{K_{Klip}} <A(\mathbf{x}),Z_k(\mathbf{x})>_{\mathcal{S}} Z_k(\mathbf{x})
\end{equation}
\end{itemize} 
This latter term is the source of the biases that we seek to calibrate with Forward Modeling.\\

When the $Z_k(\mathbf{x})$ {\em do not depend  on the astrophysical signal}, then the corruption of $A(\mathbf{x})$ is {\em a linear process}. It can be interpreted as confusion: namely the algorithm fits astrophysical signal with speckle noise. This occurs for instance in RDI. In this configuration, the $R_k(\mathbf{x})$ were built using images of other stars and thus do not contain the astrophysical signal of interest. We call this phenomenon {\em over-subtraction}. On the other hand, when the references, and thus the  $Z_k(\mathbf{x})$, {\em do depend on the astrophysical signal}, the corruption of $A(\mathbf{x})$  is {\em a nonlinear process}. Because the astrophysical signal in the reference images $R_{k}(\mathbf{x})$  is added to the speckle noise, its impact on the covariance matrix, which scales as the square of the references, is quadratic. As a consequence, the Principal Components also depend quadratically on the astrophysical signal. We call this phenomenon {\em self-subtraction}. In the context of the Locally Optimized Combination of Images algorithm--LOCI, \citet{lmd07}--this effect can be interpreted as the subtraction of the astrophysical object with itself as it rotates across the field of view during an ADI sequence. In this case, we write $Z_k(\mathbf{x}) = Z_k^{\mathcal{A}}(\mathbf{x})$ to denote the dependence of the Principal Components on the astrophysical signal.  

\subsection{Forward modeling complications due to Self-subtraction}
When a true astrophysical source is present in the data, a detection algorithm is first used to discriminate the $P_{sig}(\mathbf{x})$ and $P_{spe}(\mathbf{x})$ components. If  $P_{sig}(\mathbf{x})$ is corrupted by the speckle fitting algorithm, then Forward Modeling is used in an attempt to estimate the detected faint source's underlying astrophysical properties. This is often done by injecting a synthetic negative astrophysical source in the data $\widehat{A}(\mathbf{x})$ and carrying out a joint minimization over both properties of this negative source and the speckle noise, as discussed in \S~1. This minimization can be formally written as:  
\begin{equation}
\min_{\widehat{ \mathcal{A}}}  \left| \left| P(\mathbf{x}) - \widehat{A(\mathbf{x})} + \sum_{k =1}^{K_{Klip}} <\widehat{A}(\mathbf{x}),Z_k^{\widehat{\mathcal{A}}}(\mathbf{x})>_{\mathcal{S}} Z_k^{\widehat{\mathcal{A}}}(\mathbf{x})\right| \right|^2  
\label{Eq:BasicForwardModelling0}
\end{equation}
where $\min_{\widehat{ \mathcal{A}}}$ stands for the minimization over the observable properties of the negative synthetic signal, and $Z_k^{\widehat{\mathcal{A}}}(\mathbf{x})$ for the Principal Components resulting from injecting this negative source in the observing sequence. Note that even if the discussions in this section rely on the example in  Eq.~\ref{Eq:BasicForwardModelling0}, which uses the formalism of \citet{2012ApJ...755L..28S}, they are applicable to any least-squares speckle fitting algorithm. The last sub-section of Appendix E discusses this more general  framework. Direct inspection of  Eq.~\ref{Eq:BasicForwardModelling0} shows that Forward Modeling is a nonlinear optimization, in which the speckle subtraction (the determination of the Principal Components in our example),  is nested within an outer nonlinear loop. As a consequence, KLIP has to be carried out every time the cost function in Eq.~\ref{Eq:BasicForwardModelling0} is evaluated. One hopes that this two steps  process breaks degeneracies, and extensive tests using low-dimensional configurations by a variety of authors have shown this to be true in most cases (see \citet{mmv10} or \citet{2015ApJ...815..108M}). However, this approach presents two main limitations. First it becomes quickly untractable numerically when the number of astrophysical observables is large ($\sim 30$ in the case of IFS data). Second, and more importantly, there is no guarantee that the optimization in Eq.~\ref{Eq:BasicForwardModelling0} will converge to its global minimum, for which $\widehat{A(\mathbf{x})} =A(\mathbf{x})$. Indeed, because $Z_k^{\widehat{\mathcal{A}}}(\mathbf{x})$ is a nonlinear function of the negative synthetic  signal, there is no guarantee that Eq.~\ref{Eq:BasicForwardModelling0} is strictly convex with respect to the astrophysical properties captured in  $\widehat{A(\mathbf{x})}$. In other words, there is no mathematical certainty that the Forward Modeling cost function contours are always similar to the convex parabolas shown in \citet{2015ApJ...815..108M}. Under such pathological cases (which are more prone to occur in high-dimensional IFS data), the minimization can easily stall in local minima, thus yielding biased observables.  This is an important and fundamental drawback stemming from {\em self-subtraction}. Up until now it could only be addressed using sophisticated nonlinear optimizers.\\

\subsection{Forward modeling for Over-subtraction}

In the case of {\em over-subtraction} the Principal Components do not depend on the astrophysical signal. As a consequence, the propagation of  $\widehat{A(\mathbf{x})}$ through the algorithm is linear. Using the example of the KLIP algorithm, this can also be written as $\mathcal{KLIP} [ T(\mathbf{x})  - \widehat{A(\mathbf{x})}] = P(\mathbf{x}) - \mathcal{KLIP} [\widehat{A(\mathbf{x})}]$. In that case, the Forward Modeling cost function is truly quadratic and convex. Unbiased astrophysical observables can be readily retrieved by direct application of Eq.~\ref{Eq:BasicForwardModelling0}. Appendix C and D describe how this can be implemented in practice, and how in the particular case of point sources with RDI there exist numerical algorithms more tractable than a brute force minimization of Eq.~\ref{Eq:BasicForwardModelling0}.\\  

Figure~\ref{fig:ForwardRDI} illustrates how in this configuration KLIP-FM yields unbiased photometry. This result was obtained using Hubble Space Telescope-NICMOS data and the KLIP algorithm when injecting a synthetic point source of known flux. The left panel shows the reduced images for four values of $K_{Klip}$ (the number of Principal Components used for the data analysis) and illustrates how the detectability of the point source changes with this parameter. When $K_{Klip}$ is too small, the point source is not detected. It only becomes apparent for larger values $K_{Klip} = 50$ albeit with some residual spatially correlated speckle noise in the image (e.g $P_{spe}(\mathbf{x}) \neq 0$). This noise obviously contaminates the astrophysical observables. When $K_{Klip} =200-400$ the residual noise disappears ($P_{spe}(\mathbf{x}) \sim  0$) but the point source has been significantly over-subtracted. The right panel of Figure~\ref{fig:ForwardRDI} illustrates over-subtraction increasing with $K_{Klip}$ when Forward Modeling is not used. {\color{black} In this case, only the numerator of Eq.~\ref{Eq::RefFreeForwardModellingPhoto} is taken into account. This corresponds to a matched filter or to cross-correlating of the reduced image of a point source, which captures the corrugations due to the data analysis algorithm, with the uncorrugated instrument PSF. Without Forward Modeling and for large $K_{Klip}$, the photometric estimate is wrong by a factor of three. However, when using Forward Modeling (e.g., Eq.~\ref{Eq::RefFreeForwardModellingPhoto}), we find that the injected photometry is retrieved at the $\sim 5 \%$ level for $K_{Klip}$ that is large enough. This corresponds to the regime for which the residual speckle noise is sufficiently well behaved.} {\color{black} Unfortunately, most modern high-contrast instruments often privilege strategies combining ADI+SSDI. In this case, the reference images do contain astrophysical signals and $\mathcal{KLIP} [ T(\mathbf{x}) - \widehat{A(\mathbf{x})}] \neq P(\mathbf{x}) - \mathcal{KLIP} [\widehat{A(\mathbf{x})}]$; as a consequence the method outlined in Appendix C, which relies on only considering {\em over-subtraction}, is most often not applicable.}

\subsection{Propagation of astrophysical signal through KLIP}

{\color{black}  In this paper we introduce an analytical expansion that quantifies the propagation of the astrophysical signal through KLIP, even the presence of {\em self-subtraction}. Moreover, we show that when the astrophysical signal is small, this expansion only depends on $A(\mathbf{x})$ in a linear fashion. The proof of this result is described in Appendix E, along with the linear algebra formalism necessary to implement it in computer calculations. We do not provide these technical details here, and we only focus on the implications of this expansion. Moreover, instead of discussing the most general framework of Appendix E, we here discuss the example of a faint point source detected in IFS data (as in Appendix F) using the KLIP algorithm. The spectrum of this point source is $f$. If the source is faint enough with respect to the speckles, then the Principal Components associated with any reference stack picked within a most general ADI+SSDI observing sequence, can simply be written as:
\begin{equation}
 Y_k(\mathbf{x})  =  Z_k^{\mathcal{A}}(\mathbf{x}) = Z_k(\mathbf{x}) + f^T \mathbf{\Delta Z_k}^{\lambda}  (\mathbf{x}) 
 \label{Eq:PerturbePCASimple0}
 \end{equation} 
where $\mathbf{\Delta Z_k}^{\lambda}  (\mathbf{x}) $ is a matrix that only depends on the $Z_k(\mathbf{x}) , \Lambda_k$ eigenpair and on the instrument PSF: it does not depend on the point source's spectrum. In other words, in our example of a point source seen through an IFS, our analytical expansion captures the propagation of astrophysical signal through a least-squares speckles fitting algorithm (KLIP here) in a linear fashion: $\mathcal{KLIP} [ I_{\psi}(\mathbf{x}) +A(\mathbf{x})] =  \mathcal{KLIP} [ I_{\psi}(\mathbf{x})] + f^T \Delta\mathcal{KLIP} [I_{\psi}(\mathbf{x}),PSF(\mathbf{x})]$. The actual expression of this expansion is given in Appendix E: Eqs.~\ref{Eq:PerturbePCASimple} and \ref{Eq:LinearKLPerturbations}. While this result is here presented in the context of PCA-based algorithms, it can also be applied to algorithms that rely on linear combinations of images (e.g  LOCI). This in virtue of the direct equivalence between LOCI and KLIP discussed by  \citet{2015ApJ...800..100S}. The three main terms in this expansion of $ Z_k^{\mathcal{A}}(\mathbf{x}) $ have already been discussed in the literature in the context of LOCI. We describe them qualitatively here:
\begin{itemize}
 \item the unperturbed Principal Components $Z_k(\mathbf{x})$ that capture the correlations of the instrument PSF. These are normalized such that $||Z_k(\mathbf{x})|| = 1$ and are responsible for {\em over-subtraction}.
 \item the perturbation to the Principal Components that captures the {\em direct self-subtraction} associated with the presence of an astrophysical source  at various parallactic angles and wavelengths in the observing sequence. If $\epsilon$ is the brightness of the astrophysical source, then this term scales as $\epsilon/ \sqrt{\Lambda_k}$. In the case of LOCI, this term can be modeled by multiplying  images of the astrophysical source at various parallactic angles and wavelengths by their corresponding LOCI coefficients. This is the term that \cite{2014ApJ...780...25E} correct in the case of disk imaging with ADI.
\item the perturbation to the Principal Components that captures the {\em indirect self-subtraction} associated with correlations between the astrophysical signal and the speckles. This term scales as $\epsilon / \Lambda_k$. In the case of LOCI+ADI this term can be quantified by conducting the perturbation analysis of the LOCI coefficients introduced by \cite{2013ApJ...764..183B}.  
\end{itemize}
Because the unperturbed eigenvalues are ordered by decreasing magnitude we can readily identify three regimes of astrophysical biases:
\begin{itemize}
\item when $K_{Klip}$ is small, {\em over-subtraction} dominates the biases. Provided that the astrophysical source can be detected, the solution described in \S 2.3 can be applied. 
 \item when $K_{Klip}$ has an intermediate value, {\em direct self-subtraction} dominates the biases. Provided that the astrophysical source can be detected, methods based on linear combinations of images (e.g LOCI), along with the method described in \cite{2014ApJ...780...25E} are best suited. 
 \item  when $K_{Klip}$ is large, which might be the only recourse for very faint astrophysical sources,  {\em indirect self-subtraction} dominates the biases. In this configuration one can take advantage of our expansion to predict the influence of a synthetic negative source of spectrum $\widehat{f}$:
 \begin{equation}
  Z_k^{\widehat{\mathcal{A}}}(\mathbf{x}) = Y_k(\mathbf{x}) + \widehat{f}^T \mathbf{\Delta Y_k}^{\lambda} (\mathbf{x}) 
 \end{equation} 
In other words, here we have applied our analytical expansion to propagate to the synthetic negative source through the data analysis algorithm: $\mathcal{KLIP} [T(\mathbf{x}) -\widehat{A(\mathbf{x})}] =  \mathcal{KLIP} [ T(\mathbf{x})] - \widehat{f}^T \Delta\mathcal{KLIP} [T(\mathbf{x}),PSF(\mathbf{x})]$. Substituting this expression for $  Z_k^{\widehat{\mathcal{A}}}(\mathbf{x}) $ into Eq.~\ref{Eq:BasicForwardModelling0} yields a quadratic Forward Modeling cost function. This ensures that the Forward Modeling optimization will converge toward the global minimum (e.g no pathological biases). Moreover, because each evaluation of Eq.~\ref{Eq:BasicForwardModelling0} is calculated only via a simple matrix multiplication (see Appendix F for details), Forward Modeling  becomes numerically tractable even with highly dimensional astrophysical observables (such as IFS data). 
 \end{itemize}
 }
 
This latter case is of course the most interesting one, for which previously published methods fail. We will highlight this configuration when presenting practical applications of Eq.~\ref{Eq:PerturbePCASimple0} in \S~3 and \S~4.
 
}

\begin{figure*}[t!]
\includegraphics[width=1\textwidth]{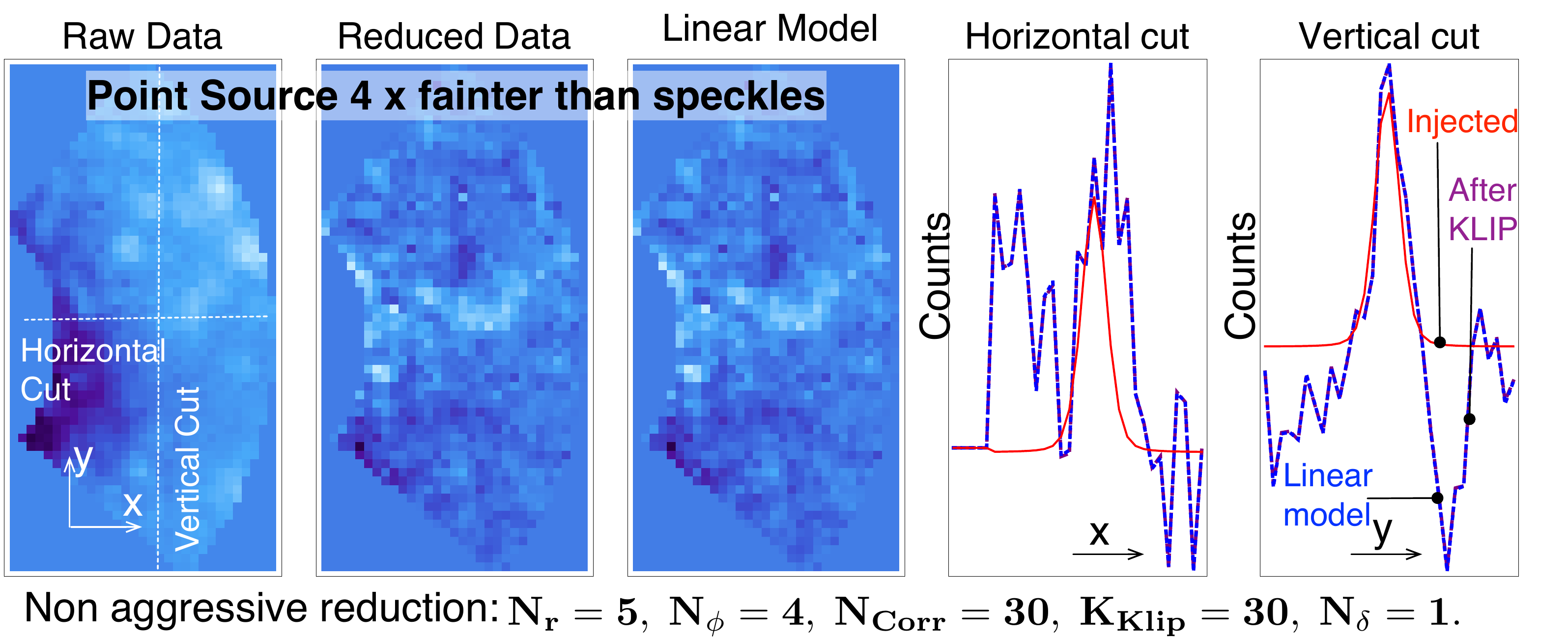}
\includegraphics[width=1\textwidth]{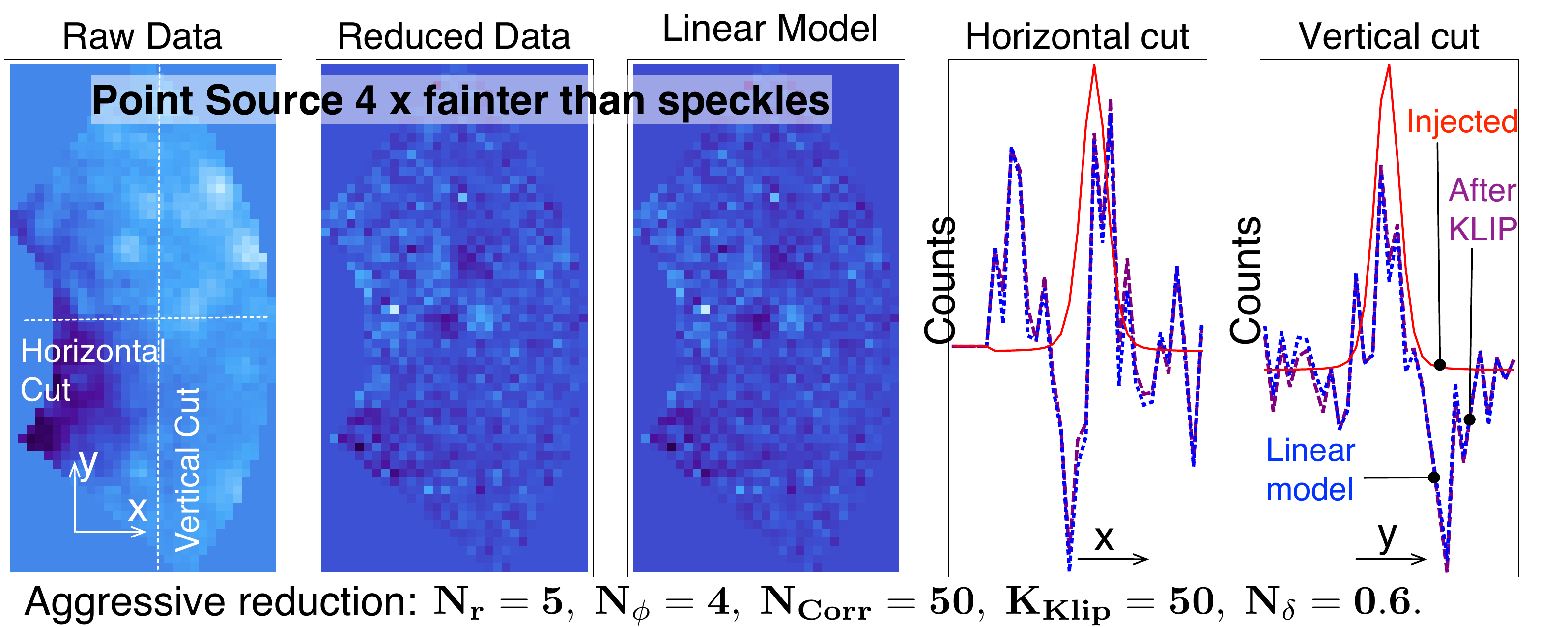}
\caption{{\bf When the point sources are fainter than the local speckles Eq.~\ref{Eq:PerturbePCASimple0} always holds.} Two leftmost panels show the raw and KLIP processed images in the spectral channel centered around $1.26  \; \mu$m of the GPI J-band filter.  The middle panel shows a model of the KLIP processed image, predicted by Eq.~\ref{Eq:PerturbePCASimple0}based on the Principal Components calculated in the absence of  the synthetic source in the reference library and the model of the point source moving azimuthally/radially across the PSF library. The two rightmost panels then show the horizontal and vertical cross section of both the actual and model-based KLIP images at the location of the injected synthetic source. When the reduction is not aggressive enough (top panel), the linear model fares very well but the injected point source is barely seen by eye and cannot be distinguished from local speckles. On the other hand, it is detected when using the same geometry but more aggressive settings (bottom panel). The cross sections illustrate how the PSF morphology is very much altered by KLIP, thus resulting into biases of astrophysical estimates when inference is carried out on these reduced images. However, in this case the results from numerical KLIP and the linear model are in very good agreement, even with an aggressively selected PSF library ($N \delta =  0.6$ PSF FWHM).}
\label{fig:PerturbationDoesWorkFintNoDetec}
\end{figure*}

\subsection{Validity of this expansion}
\label{Sec:validity}
\begin{figure}[h]
\includegraphics[width=0.5\textwidth]{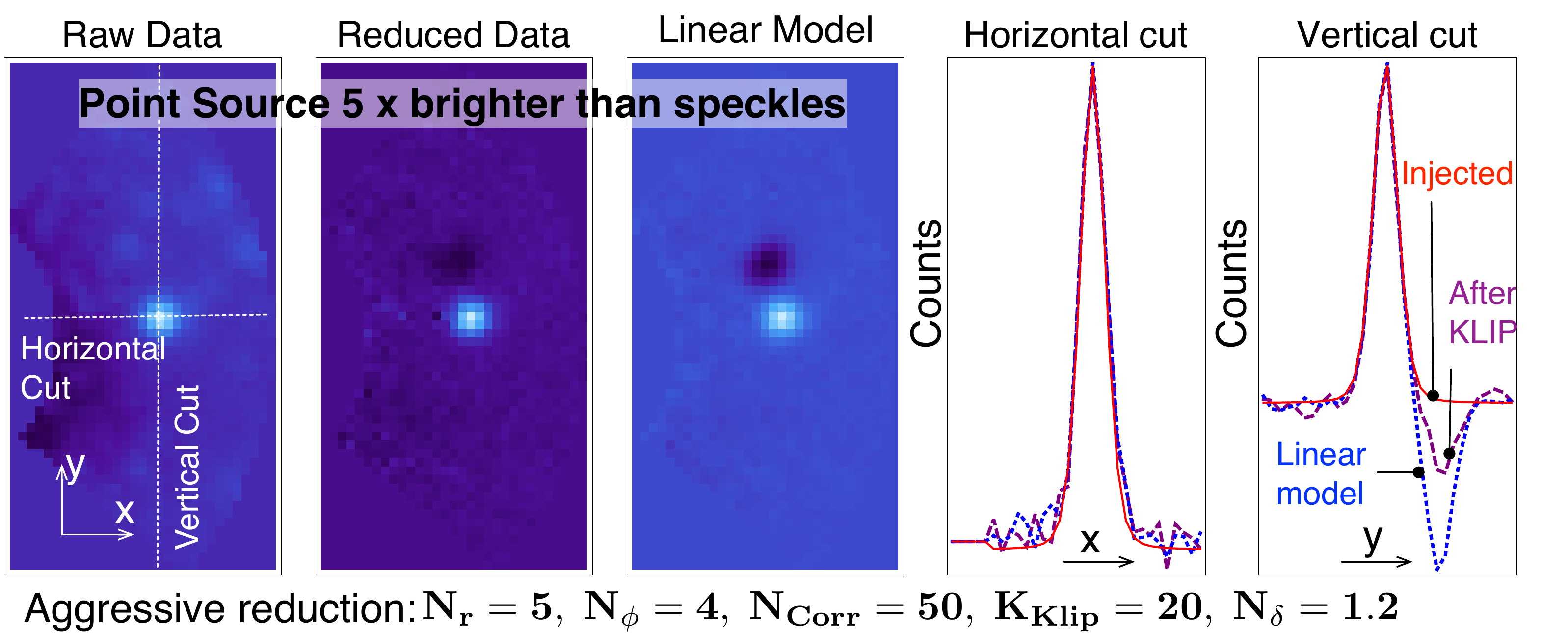}
\includegraphics[width=0.5\textwidth]{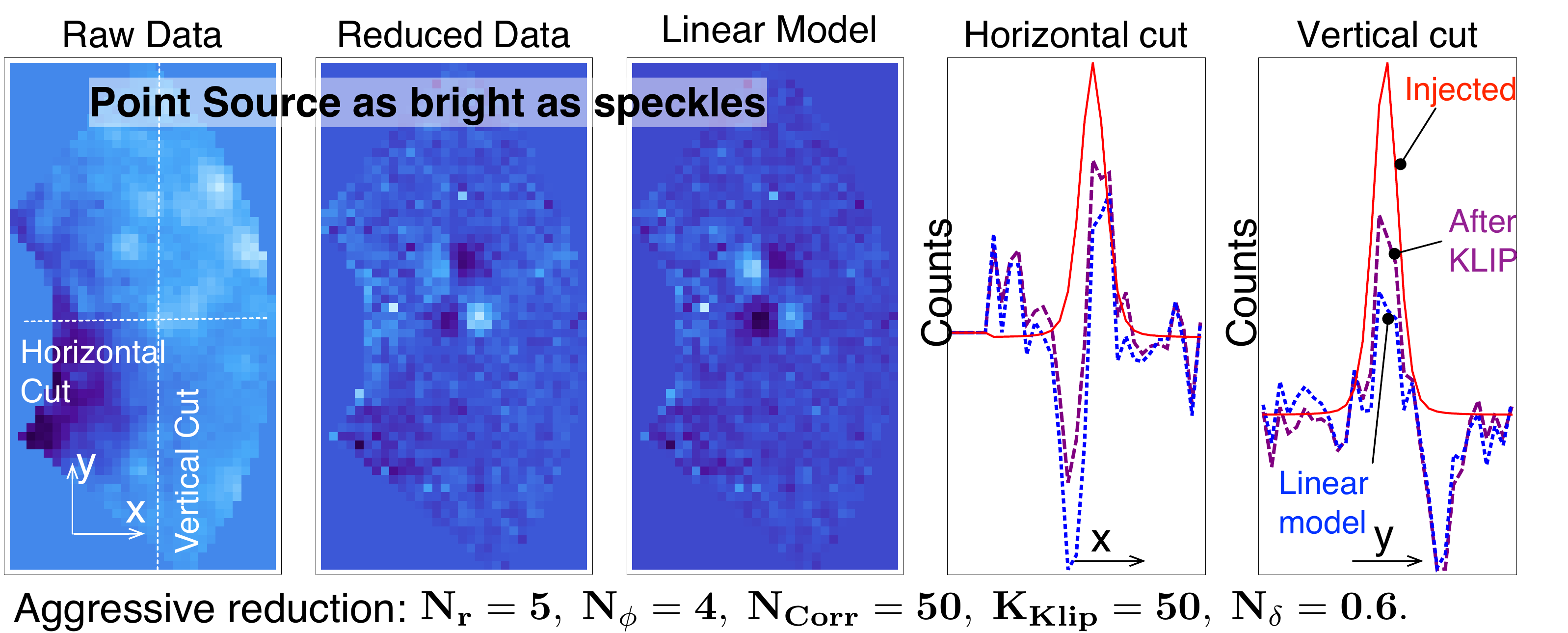}
\caption{{\bf Example configuration for which Eq.~\ref{Eq:PerturbePCASimple0} does not hold for bright point sources}. The two leftmost panels show the raw and KLIP processed images in the spectral channel centered around $1.26  \; \mu$m of the GPI J-band filter.  The middle panel shows a model of the KLIP processed image, predicted by Eq.~\ref{Eq:LinearKLPerturbations} based on the Principal Components calculated in the absence of  the synthetic source in the reference library and the model of the point source moving azimuthally/radially across the PSF library.  The two rightmost panels then show the horizontal and vertical cross section of both the actual and model-based KLIP images at the location of the injected synthetic source. While the images look like good matches, the cross sections do no perfectly overlap: the linear approximation does not hold in the case of point sources brighter (top panel) or as bright  (bottom panel) as the local speckles and with relatively aggressive KLIP parameters.}
\label{fig:PerturbationDoesNOTWork}
\end{figure}

\begin{figure}[t]
\includegraphics[width=0.5\textwidth]{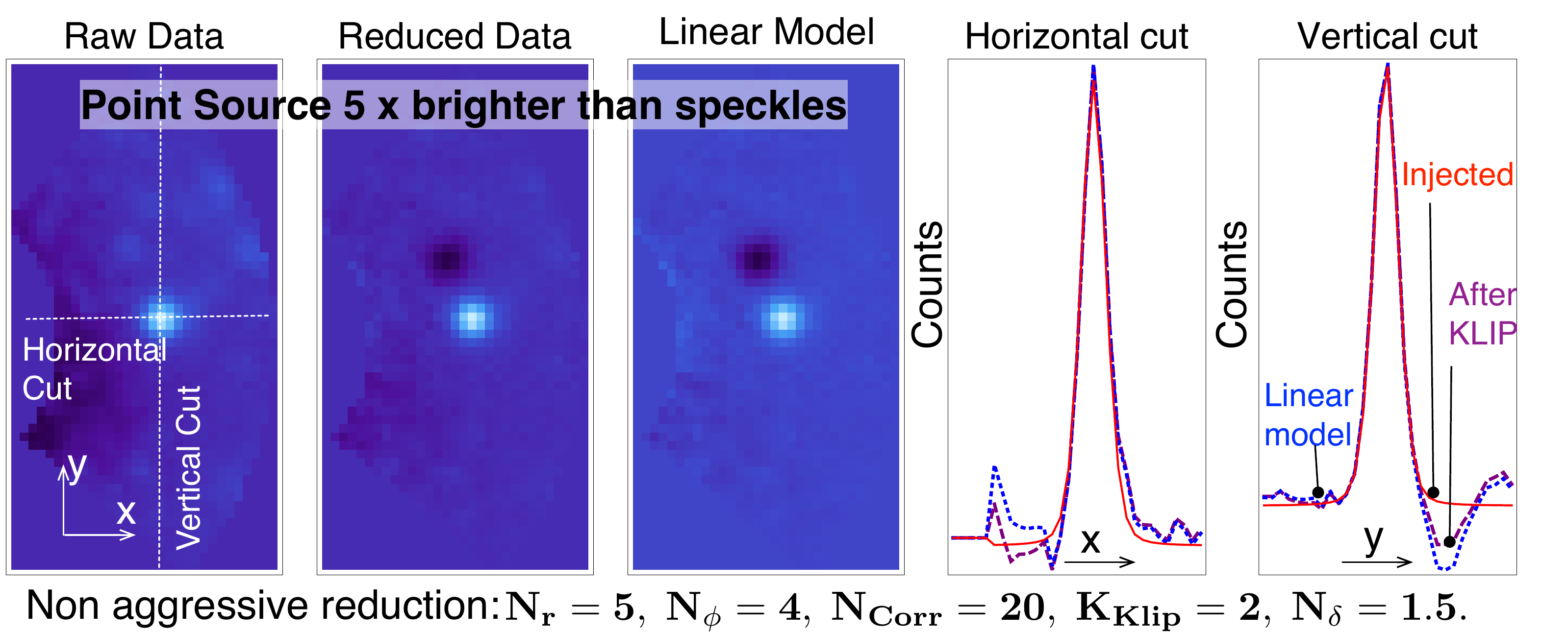}
\includegraphics[width=0.5\textwidth]{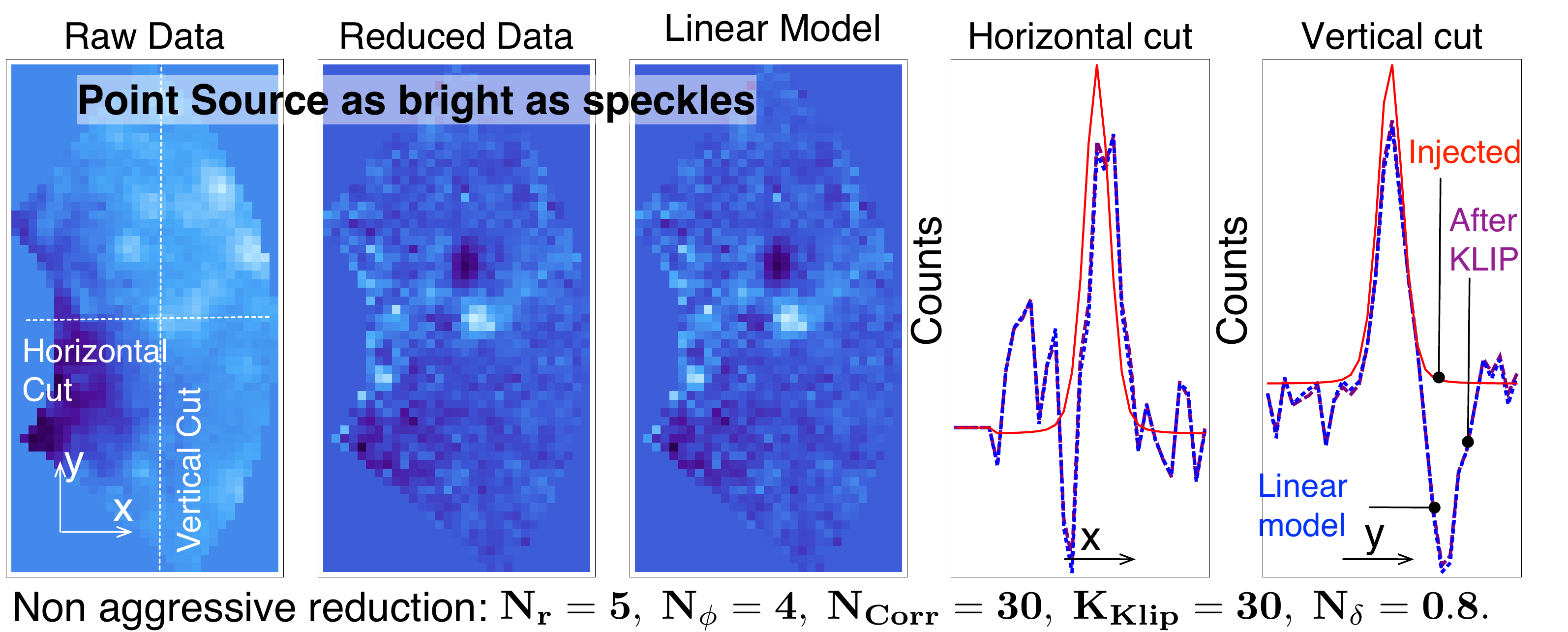}
\caption{{\bf Example of configurations for which Eq.~\ref{Eq:PerturbePCASimple0} holds for bright point sources.} Two leftmost panels show the raw and KLIP processed images in the spectral channel centered around $1.26  \; \mu$m of the GPI J band filter.  The middle panel shows a model of the KLIP processed image, predicted by Eq.~\ref{Eq:LinearKLPerturbations} based on the Principal Components calculated in the absence of  the synthetic source in the reference library and the model of the point source moving azimuthally/radially across the PSF library. The two rightmost panels then show the horizontal and vertical cross section of both the actual and model based KLIP images, at the location of the injected synthetic source. On these cross-sections, both the data and mode overlap. When compared to Figure~\ref{fig:PerturbationDoesNOTWork} this example illustrates that simply changing the KLIP parameters to less aggressive settings yields better agreement between the actual reduced KLIP image and the linear model, both in the brighter than and as bright as speckles configurations.}
\label{fig:PerturbationDoesWork}
\end{figure}

Before delving into practical examples, we first study the validity of our linear approximation. The mathematical rationale associated with this aspect is described Appendix E.  We in particular direct the reader toward Eq.~\ref{Eq:MainApprox} which can be used priori to decide  whether or not Eq.~\ref{Eq:PerturbePCASimple0} is valid. We illustrate the various regimes of this approximation using public Gemini Planet Imager J-band data on Beta Pictoris, obtained in December 2013 as part of GPI commissioning activities. Details about the observations and scientific implications are presented in \citet{2014A&A...567L...9B}. In this paper we do not discuss the exoplanet in this system and instead we inject synthetic planets at other locations in the GPI field of view. We  chose this data for our numerical examples because GPI raw J-band data is dominated by speckles (when compared to the results reported in \citet{2014ApJ...794L..15I,2015ApJ...798L...3C}) and because Beta Pictoris is the brightest star with GPI public commissioning data in this filter. Figures \ref{fig:PerturbationDoesNOTWork} to \ref{fig:PerturbationDoesWorkFintNoDetec} illustrate how the linear model described by Eq.~\ref{Eq:PerturbePCASimple0} fares when compared with propagating numerically (without any approximation) a synthetic source through the KLIP algorithm. We carried out this test using target images from a single GPI exposure, without limiting the ensemble of potential references (e.g. the target image is chosen for a given $t_0$ but the references are picked among all $t_1 ... t_{N_{exp}}$). There is no loss of generality associated with using a single exposure to illustrate the validity of Eq.~\ref{Eq:PerturbePCASimple0}, because it can easily be generalized by derotation and summation and over all exposures of an ADI sequence.

We start with Figure~\ref{fig:PerturbationDoesWorkFintNoDetec}, which addresses the case of a point source that is too faint to be detected in raw IFS data. The top panel compares numerical KLIP data with the linear model for non-aggressive parameters. {\color{black} A detailed description of algorithm parameters is given in Appendix A. For the sake of our discussions here (and for the remainder of the paper) the main consideration to remember is that ``aggressive'' corresponds to parameters that are tuned to reduce speckles very efficiently, and thus reveal the faintest underlying point sources.} In the case of the top row of Figure~\ref{fig:PerturbationDoesWorkFintNoDetec}, while the model fares very well, the injected point source (at the same location as in the bottom panel) is barely seen by eye and cannot be distinguished from local speckles. On the other hand, when using the same geometry but more aggressive settings, the faint point source is  detected. The cross sections in the  bottom panel of Figure~\ref{fig:PerturbationDoesWorkFintNoDetec} illustrate how the PSF morphology of the injected source is very much altered by KLIP. This results in biases on the astrophysical estimates when the inference is carried out on these reduced images and in the absence of Forward Modeling. However, the results from numerical KLIP and the linear model are in very good agreement, even with an aggressively selected PSF library, the cross sections corresponding to the reduced data and the linear model are completely indistinguishable. This demonstrates that the analytical expansion in Eq.~\ref{Eq:PerturbePCASimple0} does indeed capture with high fidelity the degradation of the astrophysical signal due to the speckle noise fitting algorithm. As a consequence one can in principle predict this degradation prior to any measurements and use our analytical model for unbiased astrophysical inference.\\

On the other hand, Figure~\ref{fig:PerturbationDoesNOTWork} illustrates two cases for which the linear approximation in Eq.~\ref{Eq:PerturbePCASimple0} does not hold. Indeed, as predicted in Eq.~\ref{Eq:MainApprox} the linear approximation is not valid for point sources brighter (top panel of Figure~\ref{fig:PerturbationDoesNOTWork}) or as bright as (bottom panel) as the local speckles when using relatively aggressive KLIP parameters. Fortunately, Figure~\ref{fig:PerturbationDoesWork} shows that simply changing the KLIP parameters to less aggressive settings yields better agreement between the actual reduced KLIP image and the linear model in both configurations (point source brighter and as bright as speckles). In this case, the cross sections corresponding to the reduced data and the linear model are much closer one to another on Figure~\ref{fig:PerturbationDoesWork} than on Figure~\ref{fig:PerturbationDoesNOTWork} (albeit not matching perfectly). These cases are somewhat of limited interest because they operate in configurations for which the point source can be detected in raw IFS data (either in a single slice or in an IFS cubes where it would stand immobile when compared to the speckles). For those brightnesses, aggressive KLIP might not be needed. This illustrates the limitations of the perturbation method presented in \S 2.4. It also emphasizes how the applicability of our analytical result can be extended to ``bright'' objects provided that the least-squares PSF subtraction parameters are chosen to be non-aggressive. \\

\section{Example 1: IFS spectroscopy of point sources}

 \subsection{Forward modeling with astrophysical signal in the PSF library. }

\begin{figure*}[t]
\includegraphics[width=1\textwidth]{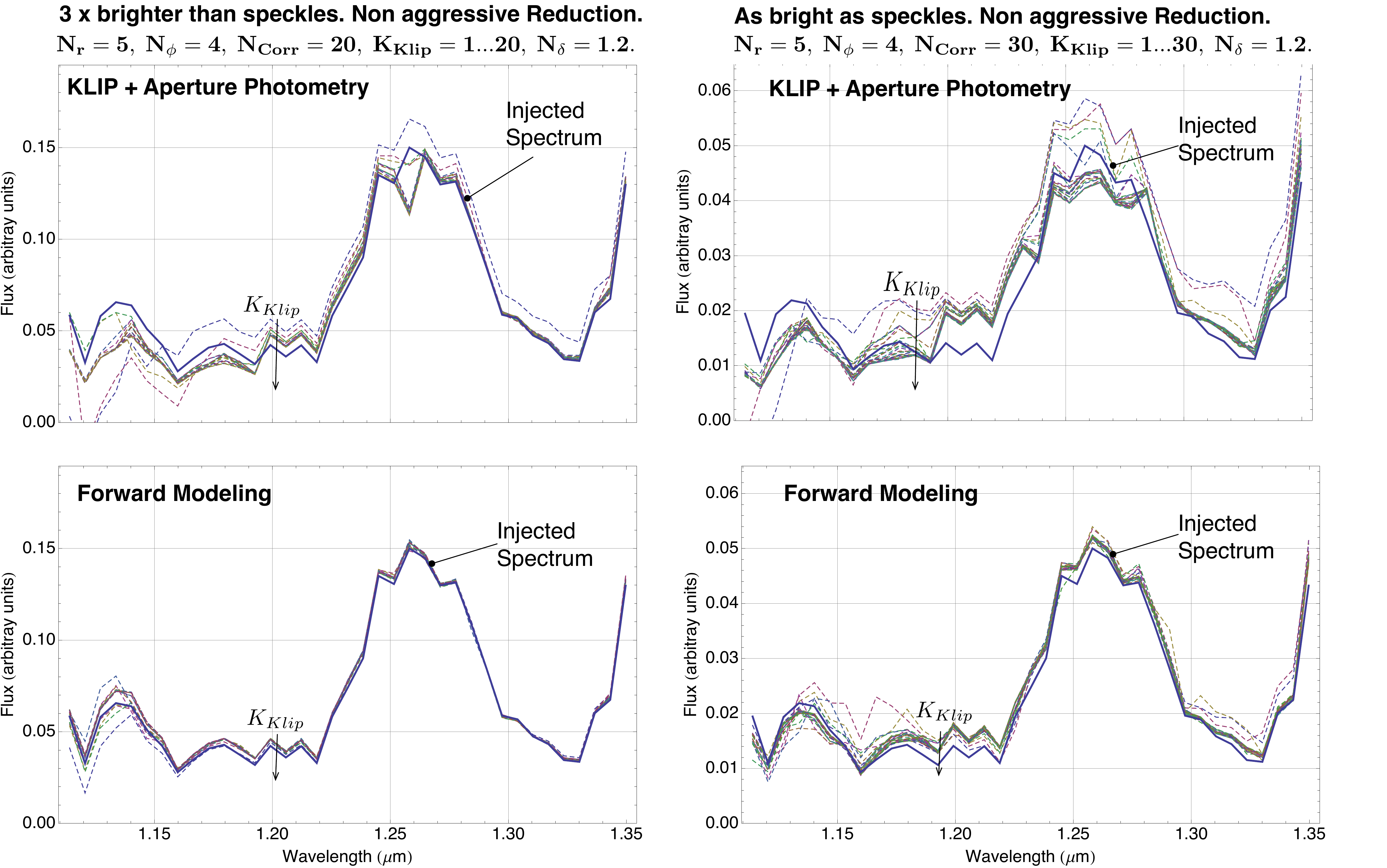} 
\caption{{\bf Comparison between KLIP+aperture photometry and KLIP-FM with a source featuring sharp spectral features and with flux, brighter than and as bright, as the speckles.  Non-Aggressive reductions.} {\em Top Left,} KLIP+aperture photometry with a source brighter than speckles. {\em Top Right,} KLIP+aperture photometry with a source as bright as speckles. {\em Bottom Left,} KLIP-FM  with a source brighter than speckles. {\em Bottom Right,} KLIP-FM  with a source as bright as speckles. The solid tick line represents the injected spectrum and each thin dashed line is an estimated spectrum corresponding to $K_{Klip} = 1 .... N_{Corr}$. {\color{black} The downward arrow indicates the variations of the estimated spectrum as a function of $K_{klip}$. As predicted by our analytical expansion, self-subtraction scales at $1/\Lambda_k$ and gets more and more severe in the absence of Forward Modeling as $K_{klip}$ increases.} With KLIP-FM this sensitivity is greatly reduced and the estimated spectrum is almost identical to the ground truth. However, in cases for which the point source is at least as bright as the speckles, Forward Modeling might not be absolutely necessary since aperture photometry after KLIP with non-aggressive reductions still yields an estimate of the spectrum within $\sim 10 \%$ of the injected signal.}
\label{fig:BrightAndMiddleNonAgressive}
\end{figure*}

Up until this point, our discussion, and in particular the analytical perturbation of Principal Components discussed in \S.~2 (and appendix E), was general and could be applied to extended objets.  We now consider a more specific example, in which we show that Eq.~\ref{Eq:PerturbePCASimple0} can be used to estimate the spectrum of faint point sources in IFS data. Because of the high dimensionality ($N_{\lambda}$) of the astrophysical observables potentially affected by self-subtraction, this problem is often considered as one of the most challenging in coronagraph data analysis. The injection of negative synthetics point sources and/or the direct minimization of Eq.~\ref{Eq:BasicForwardModelling0} can be made tractable in the cases of RDI or ADI \citep{mmv10,2014A&A...569A..29M}. However, the presence of astrophysical signal at {\em other wavelengths} in the reference library (e.g when using SSDI) renders the spectral estimation problem very degenerate. These degeneracies, along with the large number of unknown astrophysical quantities, are an important obstacle to the spectral characterization of the fainter substellar companions discovered using modern high-contrast instruments (see \cite{mmv10,pcv12} for examples).\\ 

 {\color{black} The linear expansion in  Eq.~\ref{Eq:PerturbePCASimple0} can alleviate this problem entirely.  Indeed, the negative synthetic source can be propagated through the algorithm a priori, and thus inference can occur without having to compute multiple times the costly matrix inversion associated with KLIP. Moreover, under the assumption that  Eq.~\ref{Eq:PerturbePCASimple0} holds (see discussion in \S~2.5 and in Appendix E), it does capture in a linear fashion the actual degradation of the astrophysical signal due to over- and self-subtraction. When neglecting the higher order terms in $\widehat{f}$ in $ <\widehat{A}(\mathbf{x}),Z_k^{\widehat{\mathcal{A}}}(\mathbf{x})>_{\mathcal{S}} Z_k^{\widehat{\mathcal{A}}}(\mathbf{x})$  (e.g for $\widehat{f} \sim f$ is small), the Forward Modeling cost function in Eq.~\ref{Eq:BasicForwardModelling0} becomes quadratic. This ensures that there are no pathological cases for which KLIP-FM converges to a local minimum. In Appendix F we describe in greater detail how to carry out astrophysical inference by injecting Eq.~\ref{Eq:PerturbePCASimple0} into Eq.~\ref{Eq:BasicForwardModelling0}. There, we show that the spectral extraction consists of (1) running the KLIP algorithm, (2) building a wavelength and pixel dependent model of the point source propagated through the KLIP algorithm, (3) based on this model, building a $N_{\lambda} \times  N_{\lambda}$ matrix whose diagonal terms capture over-subtraction while off-diagonal terms capture self-subtraction, (4) invert this matrix to retrieve the spectrum of the detected point source. We insist here that while this method is mathematically equivalent to injecting a negative synthetic source into the data, we do not carry out the multidimensional minimization in Eq.~\ref{Eq:BasicForwardModelling0} ``as is'.' Instead we rely on the formalism of Appendix E and F to built a linear model of the corruption of the astrophysical signal and then invert this model. Our algorithm thus does not feature a series of iterations to find the global minimum of Eq.~\ref{Eq:BasicForwardModelling0} . Of course this method is not perfectly suited to quantify the stochastic uncertainties associated with the extracted spectrum. However this ``frequentist'' approach has the benefit of being computationally cheap: in this first paper, we limit our examples to this simple inversion algorithm. As discussed in \S 3.3, Forward Modeling can also be carried out in the Bayesian sense: in this case, Eq.~\ref{Eq:BasicForwardModelling0} (along with adequate weighting to capture the properties of the residual noise) becomes a likelihood function in which the contribution of the negative synthetic source is accounted for according to  Eq.~\ref{Eq:PerturbePCASimple0}.}

  \begin{figure*}[t]
\includegraphics[width=1\textwidth]{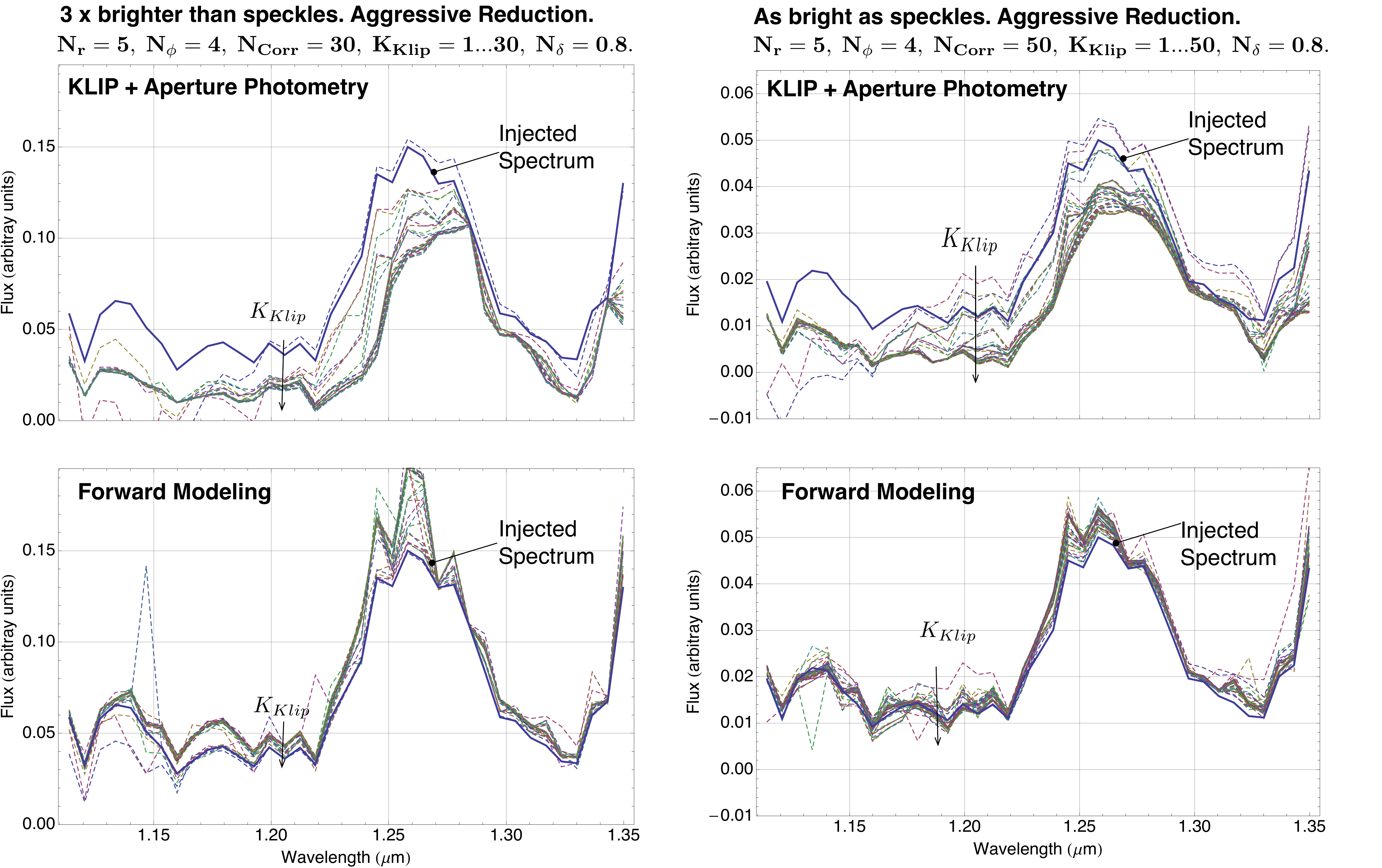} 
\caption{{\bf Comparison between KLIP+aperture photometry and KLIP-FM with a source featuring sharp spectral features and with flux brighter than and as bright, as the speckles.  Aggressive reductions.} {\em Top Left,} KLIP+aperture photometry with source brighter than speckles. {\em Top Right,} KLIP+aperture photometry with a source as bright as speckles. {\em Bottom Left,} KLIP-FM  with a source brighter than speckles. {\em Bottom Right,} KLIP-FM  with a source as bright as speckles. The solid thick line represents the injected spectrum and each thin dashed line is an estimated spectrum corresponding to $K_{Klip} = 1 .... N_{Corr}$. This figure serves as a cautionary tale in cases for which the point source can be identified in the raw data.{\color{black} The downward arrow indicates the variations of the estimated spectrum as a function of $K_{klip}$. As predicted by our analytical expansion, self-subtraction scales at $1/\Lambda_k$ and gets more and more severe in the absence of Forward Modeling as $K_{klip}$ increases. With KLIP-FM this sensitivity is greatly reduced, however some significant biases remain.} This was predicted in Figure~\ref{fig:PerturbationDoesNOTWork} where we found that in such configurations the linear model based KLIP deviates from numerical KLIP. This issue can be simply solved by using less aggressive parameters, see Figure~\ref{fig:BrightAndMiddleNonAgressive}}
\label{fig:BrightAndMiddleAgressive}
\end{figure*}
\subsection{Results with synthetic point sources}

\begin{figure*}[t]
\includegraphics[width=1\textwidth]{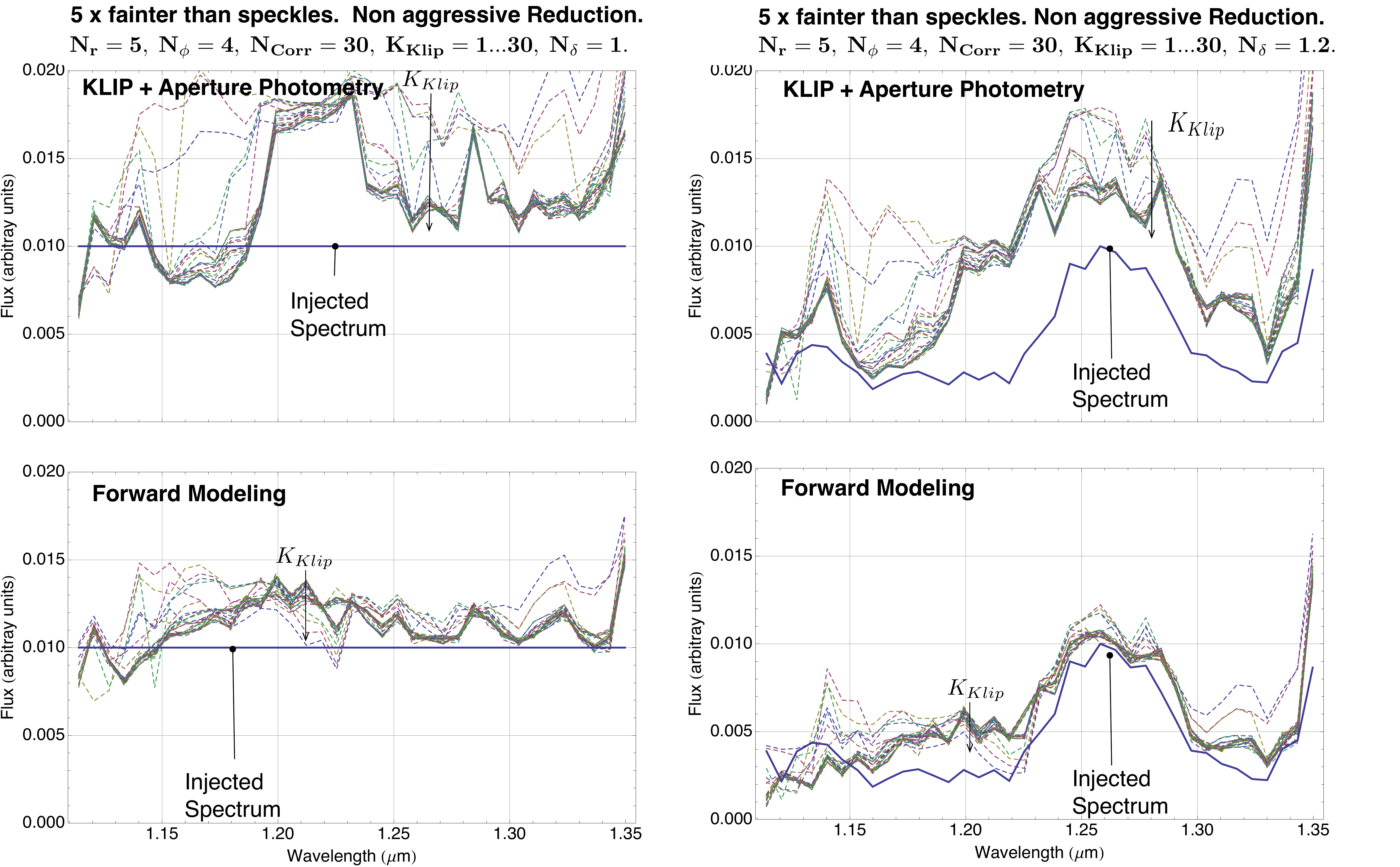} 
\caption{{\bf Comparison between KLIP+aperture photometry and KLIP-FM with a sources fainter than the speckles and both flat spectra and peaky spectra. Non-aggressive reductions.} {\em Top Left,} KLIP+aperture photometry, flat spectrum. {\em Top Right,} KLIP+aperture photometry, peaky spectrum. {\em Bottom Left,} KLIP-FM, flat spectrum. {\em Bottom Right,} KLIP-FM, peaky spectrum. {\color{black} \color{black} The downward arrow indicates the variations of the estimated spectrum as a function of $K_{klip}$. As predicted by our analytical expansion, self-subtraction scales at $1/\Lambda_k$ and gets more and more severe in the absence of Forward Modeling as $K_{klip}$ increases. With KLIP-FM this sensitivity is greatly reduced.} In the case of a point source much fainter than the speckles, significant biases remain because KLIP-FM still operates under the assumption that the residual speckle noise is well behaved, which is clearly not applicable to this non-aggressive configuration (correlated noise is still present in the reduced images). 
}
\label{fig:FaintNonAgressive}
\end{figure*}

In order to test the KLIP-FM IFS spectral estimation algorithm described in Appendix F, we inject point sources of known spectra in the GPI J-band Beta Pictoris public data set, and illustrate how this method can alleviate SSDI algorithmic biases. To do so, we study the three regimes illustrated in Figures \ref{fig:PerturbationDoesNOTWork} to \ref{fig:PerturbationDoesWorkFintNoDetec}, using two types of underlying spectra for the synthetic sources: a flat spectrum (in units of contrast: same point source to star flux ratio as a function of wavelength) and a sharp triangular spectrum. These two cases can be considered, respectively, as the most and least challenging in terms of high-contrast IFS spectral estimation. Indeed, the former is particularly difficult because the presence of the point source at other wavelengths in the reference images yields a ``local derivative'' of the spectrum after KLIP or LOCI, see \cite{doi:10.1117/12.2055245}. As a consequence, retrieving a flat spectrum might be difficult using the simple inversion described above. On the other hand, a sharp triangular spectrum, with a significant fraction of the bandpass for which the point source's flux is small, might be intuitively more amenable to this type of analysis. In this section we again  illustrate the performances of KLIP-FM in various configurations by using only one data cube for the target image.\\

 \begin{figure*}[t]
\includegraphics[width=1\textwidth]{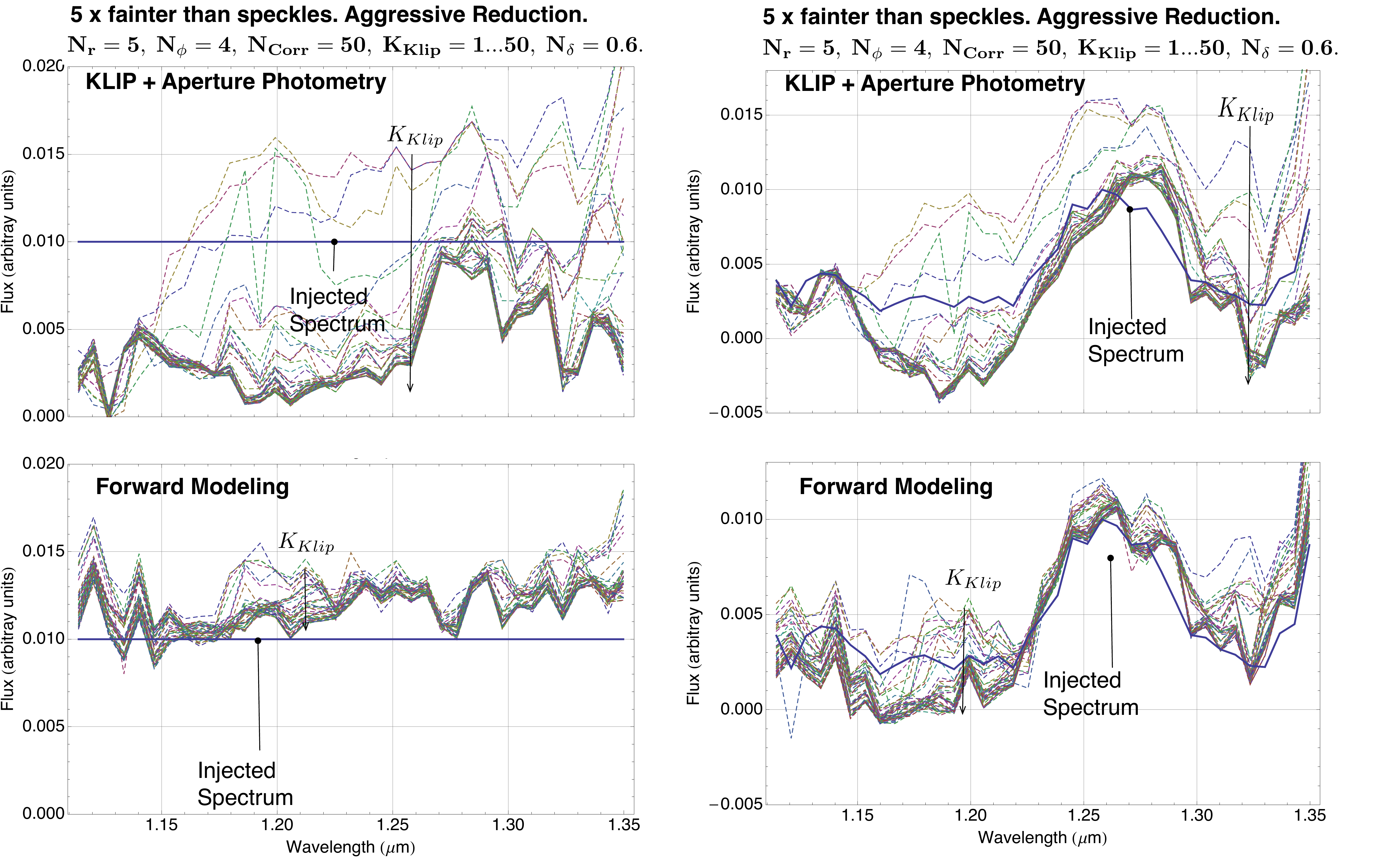} 
\caption{{\bf Comparison between KLIP+aperture photometry and KLIP-FM with a source fainter than the speckles and both flat spectra and peaky spectra. Aggressive reductions.} {\em Top Left,} KLIP+aperture photometry, flat spectrum. {\em Top Right,} KLIP+aperture photometry, peaky spectrum. {\em Bottom Left,} KLIP-FM, flat spectrum. {\em Bottom Right,} KLIP-FM, peaky spectrum.  {\color{black} \color{black} The downward arrow indicates the variations of the estimated spectrum as a function of $K_{klip}$. As predicted by our analytical expansion, self-subtraction scales at $1/\Lambda_k$ and gets more and more severe in the absence of Forward Modeling as $K_{klip}$ increases.  With KLIP-FM this sensitivity is greatly reduced.}In this configuration, the bright channels of the ``T Dwarf like'' KLIP-FM estimated spectrum are almost as close to the injected ground truth than in the case of Figure~\ref{fig:BrightAndMiddleNonAgressive}. This is quite remarkable considering that for this latter figure the flux of the injected companion was $15$ times brighter than in the faint regime presented here. Results using a flat spectrum show somewhat of a lesser fidelity when compared to the ground truth: in particular the both aggressive and non-aggressive setting seem to yield similar biases. It is important to note that even in this worse-case scenario of a flat spectrum point source only detected at the $\sim 1\sigma$ level (see images in the bottom panel of Figure~\ref{fig:PerturbationDoesWorkFintNoDetec}) KLIP-FM does correct for the $K_{klip}$ sensitive self-subtraction and yields residual biases at the $20-30 \%$ level that are well below the expected statistical uncertainties for such a low-significance detection. }
\label{fig:FaintAgressive}
\end{figure*}

Figure~\ref{fig:BrightAndMiddleNonAgressive} illustrates how the estimated spectrum varies as a function of the number of Principal Components ($K_{Klip}$) when the injected source has a sharp spectral feature whose maximum flux is brighter (left column) and as bright (right column) as the speckles. The top row shows the estimated spectrum when using aperture photometry and the bottom row when using KLIP-FM.  The solid thick line represents the injected spectrum and each thin dashed line is an estimated spectrum that corresponds to $K_{Klip} = 1 .... N_{Corr}$ (where $N_{Corr}$ is the number of images in the PSF library). It was generated using non-aggressive parameters close to the ones in Figure~\ref{fig:PerturbationDoesWork}. {\color{black} Because indirect self-subtraction, which scales as $1/\Lambda_k$, worsens when increasing the number of KLIP modes, one expects that the estimated spectrum after KLIP in the absence of Forward Modeling to vary as $K_{Klip}$ increases. This is exactly the behavior that the top two panels of Figure~\ref{fig:BrightAndMiddleNonAgressive} (and subsequent figures) exhibit. On the other hand, self-subtraction is accounted for with Forward Modeling and thus the estimated spectrum should in principle not depend on $K_{Klip}$ (provided that the residual speckle noise is small enough and that the linear approximation is valid). Again, this is exactly what happens on the bottom panels of Figure~\ref{fig:BrightAndMiddleNonAgressive}: KLIP-FM reduces the sensitivity of the estimated spectrum to $K_{Klip}$ and brings the estimated spectrum closer to the injected one}. However, in cases for which the point source is at least as bright as the speckles, Forward Modeling is not absolutely necessary because aperture photometry after KLIP with non-aggressive reductions still yields an estimate of the spectrum within $\sim 10 \%$ of the injected signal (top panels). This is because over-subtraction does not depend on $\Lambda_k$, and direct self subtraction only scales as $1/\sqrt{\Lambda_k}$. In this context KLIP-FM is a tool to reduce uncertainties. This is of course only true when using algorithm parameters chosen so that Eq.~\ref{Eq:PerturbePCASimple0} holds. Indeed, when the point source is relatively bright and the speckle fitting is aggressive,  Eq.~\ref{Eq:PerturbePCASimple0}  does not hold and biases still remain after KLIP-FM when comparing the estimated and injected spectra. This is illustrated on Figure~\ref{fig:BrightAndMiddleAgressive}: in spite of a somewhat reduced sensitivity to $K_{Klip}$ a significant offset remains between injected and extracted spectra. \\

Such considerations do not apply to point sources fainter than the speckles, for which Eq.~\ref{Eq:PerturbePCASimple0} is always valid. We discuss results obtained in this configuration using both flat (left columns) and peaky (right columns) spectra along with non-aggressive (Figure~\ref{fig:FaintNonAgressive}) and aggressive (Figure~\ref{fig:FaintAgressive}) KLIP settings. In the case of non-aggressive subtractions, the post-KLIP aperture photometry spectra are significantly biased by residual speckle noise. Because KLIP-FM still operates under the assumption that the residual speckle noise is well behaved (e.g $P_{spe}(\mathbf{x}) \sim 0$), which is clearly not applicable to this case, Forward Modeling does still yield biases in both the flat and peaky spectra configurations (Figure~\ref{fig:FaintNonAgressive}). This ought to be expected for such algorithm settings that do not seek to reach the absolute best speckle least-squares fitting, yielding the type of correlated residuals illustrated in the top panel of Figure~\ref{fig:PerturbationDoesWorkFintNoDetec}. On the contrary, Figure~\ref{fig:FaintAgressive}, which was obtained using aggressive settings this time, shows that the bright portion of a peaky spectrum becomes very close to the injected spectrum when using KLIP-FM. As a matter of fact, the spectral fidelity in this case is almost as good as Figure~\ref{fig:BrightAndMiddleNonAgressive}, even though the injected point source is $15$ times fainter. As predicted, results using a flat spectrum exhibit somewhat lesser fidelity: in particular both the aggressive and non-aggressive settings seem to yield similar biases. However, these biases are much smaller in the case of KLIP-FM than without using Forward Modeling. Even in this worse-case scenario of a point source with a flat spectrum, only detectable at the $\sim 1\sigma$ level (see images in the bottom panel of Figure~\ref{fig:PerturbationDoesWorkFintNoDetec}),  residual biases are at the $20-30 \%$ level, which is well below the statistical uncertainties that ought to be expected for the low-significance detections we simulated here. 
 \subsection{Residual biases and statistical uncertainties.}
%
%

\begin{figure}[h]
\includegraphics[width=0.5\textwidth]{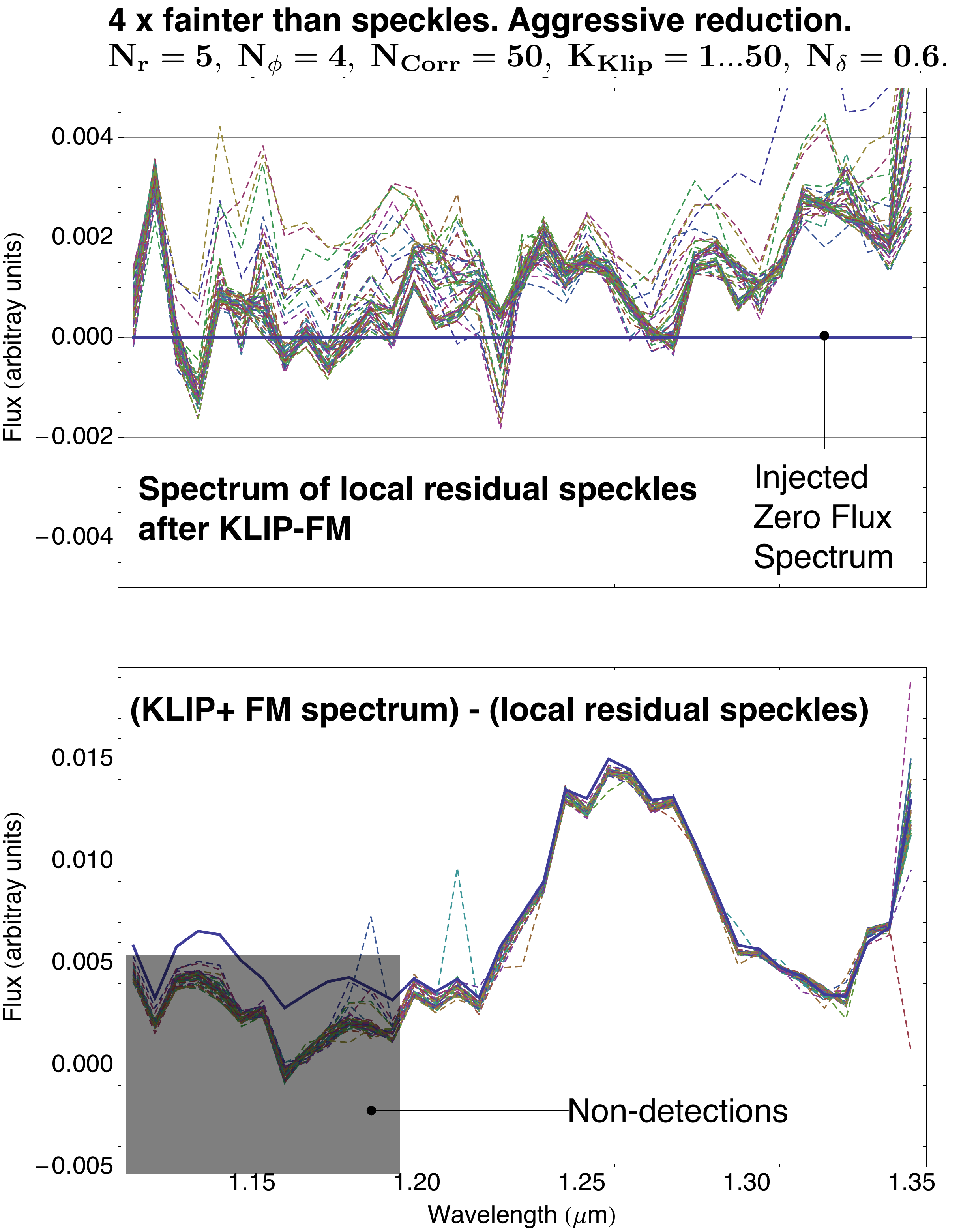} 
\caption{{\bf Investigating residual biases after KLIP-FM.} {\em Top,} null test where we carried out the same exact analysis as in Figure~\ref{fig:FaintAgressive} except that the flux of the synthetic point source was set to zero. {\em Bottom,} when subtracting this ``residual speckle noise flux'' to the KLIP-FM spectral estimate we find indeed that we can obtaine a bias-free extracted spectrum in all the spectral channels for which there is a statistical significant detection 
}
\label{fig:ExtraBias}
\end{figure}

{ Figures~\ref{fig:BrightAndMiddleNonAgressive} to \ref{fig:FaintAgressive} clearly demonstrate how KLIP-FM reduces the systematic biases associated with spectral extraction of faint point sources in IFS coronagraph data. However deviations from the injected spectrum are noticeable in the case of fainter point sources. This is in spite of the fact that the final estimated spectrum is a much weaker function of $K_{Klip}$, which we use here as a proxy to establish the ability of KLIP-FM to correct for over- and self-subtrcation. We investigated this feature by carrying out the same analysis as in Figures~\ref{fig:BrightAndMiddleNonAgressive} to \ref{fig:FaintAgressive} except that in a first step we set the flux of the injected point source to zero to quantify the residual speckles floor. For illustration, the estimated spectrum obtained under this null hypothesis is given in the top panel of Figure~\ref{fig:ExtraBias}.  Subtracting this ``residual speckle noise flux'' to the KLIP-FM spectral estimate yields the bottom panel of Figure~\ref{fig:ExtraBias}. We indeed obtain a bias-free estimated spectrum in the bright channels of the spectrum. We confirm by eye inspection that the bluer end of the spectrum corresponds to non-detections, which explains the remaining offset. We find a similar outcome when repeating this test for all the configurations shown on Figures~\ref{fig:BrightAndMiddleNonAgressive} to \ref{fig:FaintAgressive}. The test on Figure~\ref{fig:ExtraBias} illustrates that with KLIP-FM, spectral estimation is not limited by over- and self-subtraction. By and large the post post-KLIP-FM biases stem from the residual speckles in the reduced images. 

Of course in practice this null test cannot be carried out and the estimated spectrum, along with its associated uncertainties, will be affected by poorly subtracted speckles. When using a full ADI sequence this will be alleviated by co-adding cubes over time, in virtue of the central limit theorem, as described in \citet{2008ApJ...673..647M}. In that respect, our non-ADI single cube test is somewhat of a pessimistic configuration. If the observing strategy does not include ADI, the brightness of the residual speckles can also be minimized by adjusting KLIP parameters (within the range for which the linear approximation remains valid, as discussed in \S 2.5). Regardless of these adjustments, the estimation of confidence intervals associated with the now unbiased KLIP-FM extracted spectrum is of critical importance for astrophysical inference. Most often, these confidence intervals are calculated by injecting and extracting synthetic point sources at various positions in the coronagraph field of view. Errors bars associated with this process include contribution of both the possible algorithmic biases and of the residual post-processed speckles. Using KLIP-FM makes the former term negligible. Correlations between  spectral channels (associated with the latter term) can also be estimated based on residual noise statistics at other locations than the one of the detected point source. All of these approaches yield realistic confidence intervals under the assumption that the noise properties are spatially uniform across the field of view (or at least  over all azimuths at a given angular separation). When using a ground-based Adaptive Optics system, this assumption is not always true due to signatures of the wind direction in the coronagraph PSF.  Our Forward Modeling approach alleviates this assumption since it enables the estimation of confidence intervals based the contribution of the residual speckles {\em at the location of the point source}. This can be achieved by estimating spatial and spectral co-variances at positions where astrophysical signal is absent and introducing them into Eq.~\ref{Eq:BasicForwardModelling0} so it becomes a true likelihood function in the Baysian sense (see \citet{gb2016} for details on how the spectral correlation can be included). This represents significant progress when compared with the present state, in particular for data sets that feature significant residuals associated with atmospheric wind. However a full end to end demonstration of this approach is beyond the scope of this paper and we leave out this analysis to an upcoming publication (Wang et al., in preparation).} 

\section{Example 2: Detectability of faint point sources in IFS data}
\subsection{Detection threshold and completeness}
We now tackle the actual detection problem, which is the decision process that chooses whether or not to trigger a detection alarm and take action accordingly (in the case of a first epoch this action consists of carrying out confirmation observations). This problem has been extensively discussed by \citet{Caucci:07,2008ApJ...673..647M,2009JOSAA..26.1326M,2013A&A...551A.138Y,2014ApJ...792...97M,2015arXiv150203092W,2015A&A...582A..89C,2016arXiv160208381G}. Here we revisit these results in the context of KLIP-FM. A detection algorithm can be seen as an observer \footnote{Here ``observer'' refers to an image analysis algorithm such as the Hotelling Observer described in \citet{Caucci:07}, not an individual collecting astronomical data with a telescope} that estimates the probability that flux in a given set of pixels originates from an astrophysical signal rather than scattered starlight (speckles) or other sources of noise. Because the characteristic scales of speckle noise mimic the presence of a planet for most classes of observers, a preliminary routine aimed at calibrating this noise is necessary. In this paper we discuss algorithms based on least-squares PSF fitting  for this denoising step. After this has been carried out, statistical inference regarding the presence of a certain class of point sources, or lack thereof, then occurs using the chosen observer. If a given combination of pixels (as defined by the observer) is above a given threshold (often chosen to minimize the false positive rate), then an alarm is triggered and follow-up observations are carried out. On the other hand, if no alarm is triggered, the range of astrophysical objects that are absent from the data (e.g completeness) is then quantified in order to inform the statistical distribution of such object across the ensemble of stars observed  (see work by \cite{2012A&A...544A...9V,2013ApJ...776....4N,2014ApJ...794..159B} for recent exoplanet surveys). In this section we describe how, for a given set of chosen algorithm parameters, KLIP-FM can keep the false positive rate similar to the one obtained without Forward Modeling while significantly increasing completeness. 
%
 \begin{figure}[h]
\includegraphics[width=0.45\textwidth]{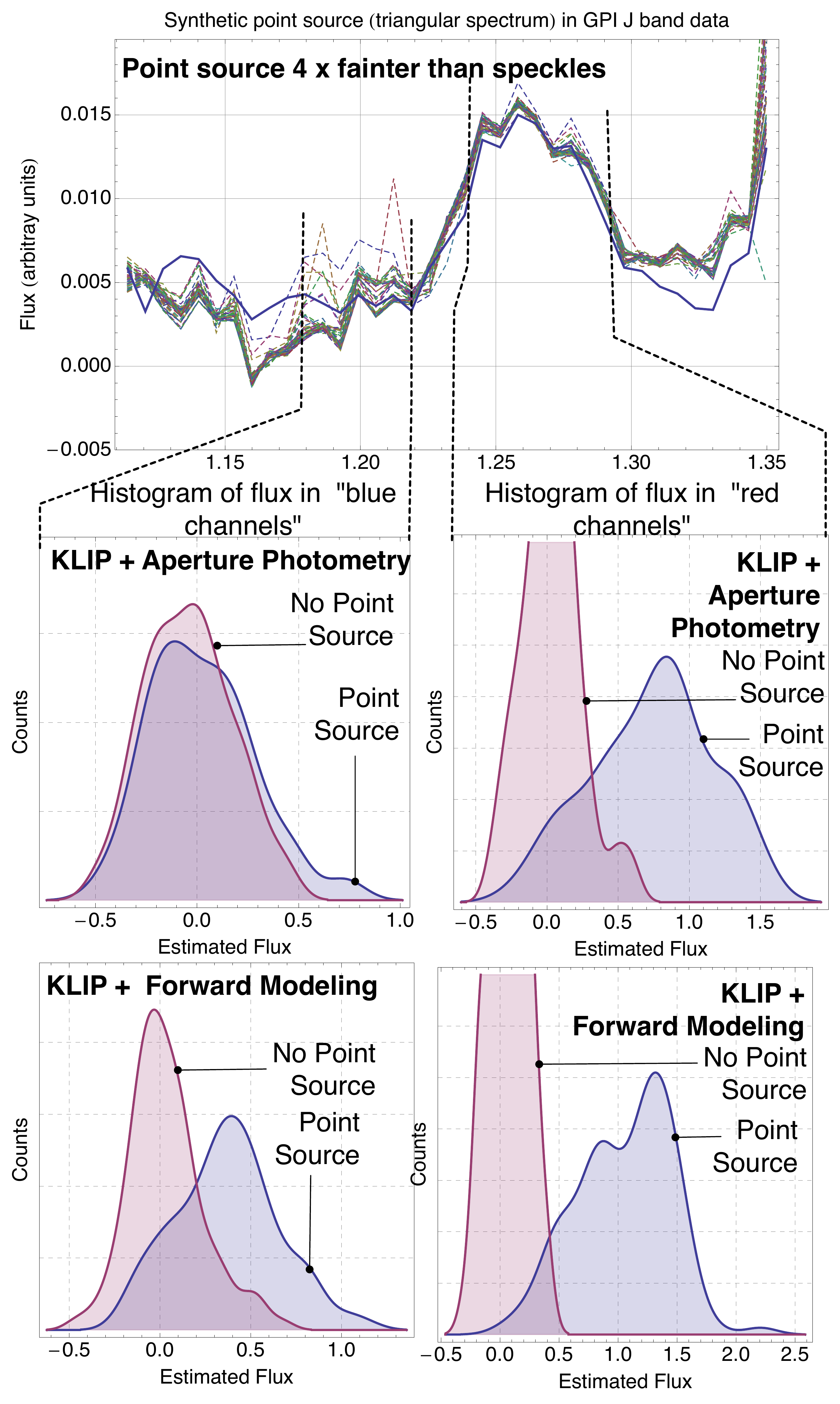} 
\caption{{\bf Histograms of observed counts the case of synthetic companions fainter than the speckles.} This figure was generated using  $N_{synthetic} = 50$ injection of true positives, at separation of $0.4''$ and random azimuthal positions. {\em Left:} the observer measures the maximum of the counts in an aperture the size of the FWHM of a PSF in the ``blue'' channels, at the base of the spectral feature, in both the KLIP+ApPhot and the KLIP-FM cases. {\em Right:} the observer measures the maximum of the counts in an aperture the size of the FWHM of a PSF in the ``red'' channels, at the peak of the spectrum, in both the KLIP+ApPhot and the KLIP-FM cases. In both cases we find that using Forward Modeling in conjunction with KLIP fares better that simply using aperture photometry after KLIP. KLIP-FM shifts to the right the ``point source present'' histogram while only slightly changing the tail of the histogram associated with no signal. This reduces the area of the ``confusion zone'', where the two histograms overlap.}
\label{fig:ProbablitiesFaint}
\end{figure}


\subsection{Maximizing true positives while minimizing false negatives}
\begin{figure}[h]
\includegraphics[width=0.45\textwidth]{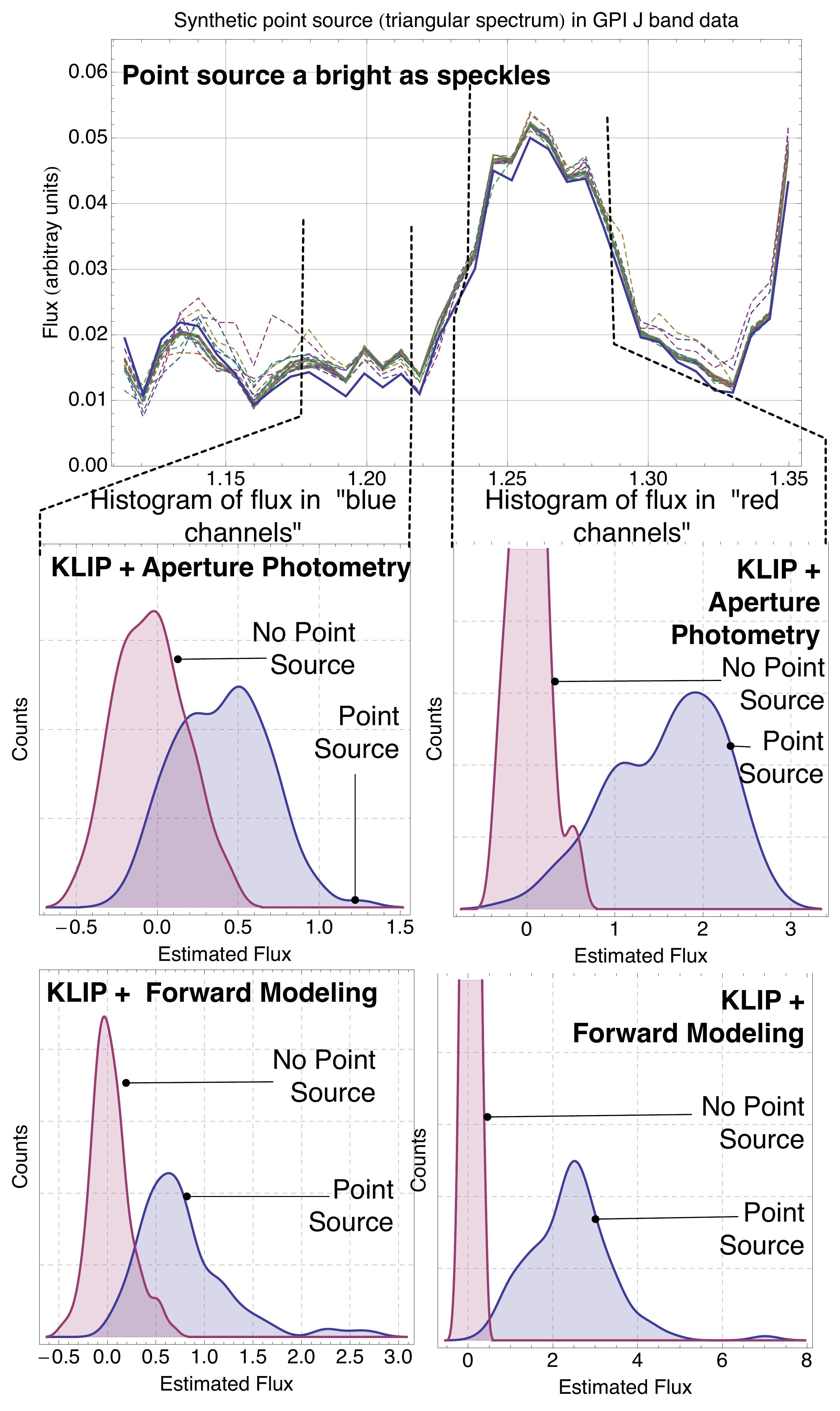} 
\caption{{\bf Histograms of observed counts in the case of synthetic companions as bright as the speckles.} This figure was generated using  $N_{synthetic} = 50$ injections of true positives, at separation of $0.4''$ and random azimuthal positions. {\em Left:} the observer measures the maximum of the counts in an aperture the size of the FWHM of a PSF in the ``blue'' channels at the base of the spectral feature in both the KLIP+ApPhot and the KLIP-FM cases. {\em Right:} the observer measures the maximum of the counts in an aperture the size of the FWHM of a PSF in the ``red'' channels at the peak of the spectrum in both the KLIP+ApPhot and the KLIP-FM cases. In both cases we find that using Forward Modeling in conjunction with KLIP fares better than simply using aperture photometry. KLIP-FM shifts to the right the histogram of counts associated with point source signal injected in the data while only slightly changing the tail of the histogram associated with no signal. This reduces the area of the ``confusion zone,'' where the two histograms overlap.}
\label{fig:ProbablitiesBright}
\end{figure}

Our goal is to quantify the efficiency of KLIP-FM when applied to the detection problem. To do so we compare two types of observers:
\begin{itemize}
\item  De-noising with KLIP and then aperture photometry (denoted KLIP+ApPhot). 
 \item De-noising with KLIP and then inversion of  Eq.~\ref{Eq:SimpleMat} (KLIP-FM).
\end{itemize} 
We depart from the common practice in the high-contrast imaging community that consists of comparing the local signal-to-noise ratio (S/N) of images obtained using various algorithms. Instead we follow the prescription described in \citet{Caucci:07} and recently revisited by \citet{2016arXiv160208381G}. Using such metrics is now becoming common practice in high-contrast imaging. Note that this approach was also pointed out by \citet{2015arXiv150203092W}, who demonstrated that the rigorous way to assess the efficiency of  various joint denoising and detection methods is to study them under the paradigm of minimizing the False Positive Fraction (FPF) while maximizing the True Positive Fraction (TPF). For the sake of brevity we do not recall the formal definition of these quantities, and refer the reader to the excellent presentations in \cite{2014ApJ...792...97M,2015arXiv150203092W}. In practice the FPF and TPF can be estimated using numerical simulations as follows :
\begin{itemize}
\item We choose a hypothesis for the underlying population of astrophysical signal we want to test: this includes the brightness of the point sources, their separation from the star, and their underlying spectrum. 
\item We generate a series of $2 \times N_{synthetic}$ data sets:  $ N_{synthetic}$ of them have a signal as prescribed in the previous step, the other $ N_{synthetic}$ do not have signal (e.g their brightness has been set to zero). This latter data set serves as a null hypothesis test in the assessment of the data analysis algorithm.   
\item We propagate these data sets through each denoising+observer algorithm whose performance we want to assess. 
\item Based on these simulations, we build the empirical Probability Density Function (PDF) of the scalar metrics given by each observer under both the signal present and absent hypotheses. This yields the histograms shown in Figures~\ref{fig:ProbablitiesFaint} and ~\ref{fig:ProbablitiesBright}. 
\item The FPF captures the probability that, for a given threshold, the observer will classify an event as an astrophysical detection while it is actually stemming from noise realizations. It is thus calculated as the area under the curve of the ``no point source'' histogram, from the threshold to $+\infty$.
\item The TPF measures completeness (i.e.,the probability that astrophysical signal will be classified as such and not as noise). It is thus calculated as the area under the curve of the ``point source'' histogram, from the threshold to $+\infty$.
\item We then move the threshold  from the left to the right of each histogram and compute the FPF and TPF at each threshold value. This yields the Receiver Operating Characteristic (ROC), which is parametric curve describing $TPF = roc ( FPF) $. This ROC can then be used to compare denoising+observers methods. 
\end{itemize}

\begin{figure}[h]
\includegraphics[width=0.5\textwidth]{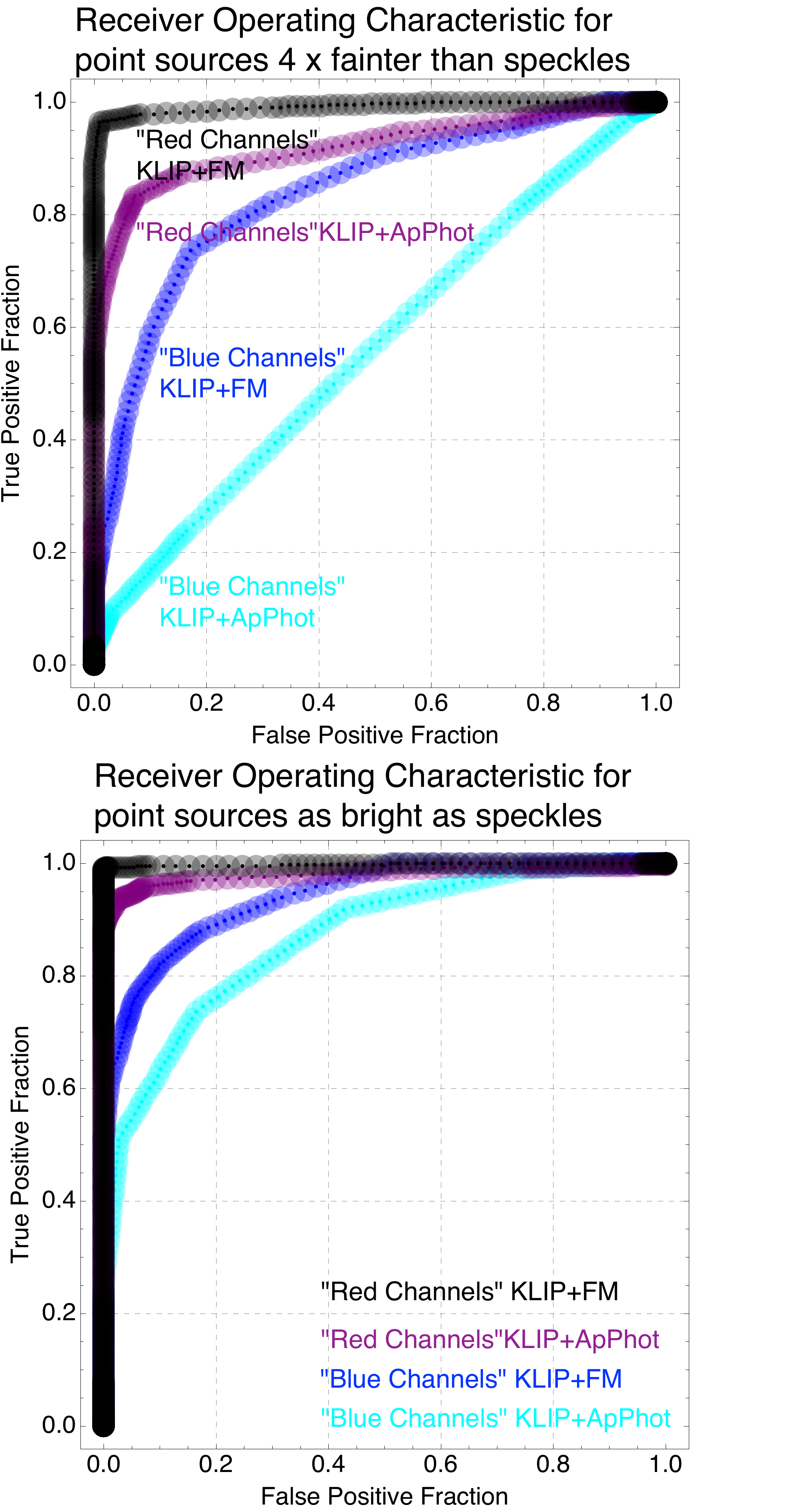} 
\caption{{\bf Receiver Operating Characteristics (ROC) obtained for synthetic companions located at $0.4''$ separation: fainter than the speckles (top) and as bright as the speckles (bottom).}  The area integrated under the ROC curve (AUC) is the figure of merit that quantifies the performances of a given denoising+observer combination. In all cases we find that using using Forward Modeling in conjunction with KLIP increases this figure of merit. For instance, in the case of the ``blue'' channels of companions that are fainter than the speckles (top) then $FPF = TPF = 50 \%$ for all possible values of thresholds without Forward Modeling. On the other hand, KLIP-FM results in the existence of an ``optimal threshold'' (at the elbow of the ROC) for which $FPF  = 20 \%, TPF  = 70 \%$. { The thickness of the lines indicated the uncertainties of the ROC due to the coarse resolution of our numerical experiment ($N_{synthetic} = 50$). However this does not change our conclusion that, by and large,  KLIP-FM does increase the TPF  for a given FPF when compared to KLIP without Forward Modeling.}
}
\label{fig:ROCs}
\end{figure}

 As described extensively in the Imaging Science literature (\citet{Caucci:07} and references therein), having the ROC follow a straight line between the $(0,0)$ and $(1,1)$ implies that TPF and FPF are always equal for all values of threshold: the observer is no better than a coin toss. On the other hand, having the ROC follow a perfect elbow from $(0,0)$ to $(0,1)$ to $(1,1)$ implies that there exists an optimal value for the threshold for which the ``no point source" and ``point source" histograms do not overlap at all, thus enabling the recovery of all possible astrophysical signal without any false positive. In this case the observer is ideal. In this framework the area integrated under the ROC curve (AUC) is the figure of merit that quantifies the performances of a given denoising+observer combination. We ran a series of numerical tests to compare the performance of KLIP+Aperture Photometry and KLIP-FM under this metric. Figures \ref{fig:ProbablitiesFaint} and \ref{fig:ProbablitiesBright} show the histograms of observed counts in the wavelengths around the base and the maximum of a sharp triangular spectrum for synthetic companions fainter and as bright as the speckles. Similar extractions were also conducted without injecting companions and the null hypothesis histograms as also reported in Figures \ref{fig:ProbablitiesFaint} and \ref{fig:ProbablitiesBright}. These figures were generated using  $N_{synthetic} = 50$ injections of true positives at separation of $0.4''$ and at random azimuthal positions. Based on these, we then calculate the ROC corresponding to each configuration, as shown in Figure~\ref{fig:ROCs}. In all cases, we find that using Forward Modeling in conjunction with KLIP fares better than simply using aperture photometry after this algorithm. A closer look at Figures \ref{fig:ProbablitiesFaint} and \ref{fig:ProbablitiesBright}  illustrates how KLIP-FM does shift to the right the histogram of counts when an astrophysical signal is present, while only slightly changing the tail of the histogram associated with and absent signal. This reduces the area of the ``confusion zone',' where the two histograms overlap, and increases the area under the ROC. This is particularly striking in the left panel of Figure~\ref{fig:ProbablitiesFaint} for which both histograms without Forward Modeling almost completely overlap and result in a straight ROC between  $(0,0)$ and $(1,1)$ (e.g. coin toss). KLIP-FM, under similar conditions, does yield an ROC that can operate at $70 \%$ completeness only with $20 \%$ FPF. Note that these tests were carried out for aggressive least square subtraction settings and that the difference between Aperture Photometry and KLIP-FM is less striking when using less aggressive KLIP configurations. \\
 
It is important to remember that here we do not  discuss a new algorithm to remove speckles more efficiently (such as the one presented in \citet{2016arXiv160208381G}); as a matter of fact the actual images underlying the two methods compared in this section are {\em identical}. The only difference between these methods resides in analyzing these images using a Forward Model for over- and self-subtraction. Inverting this model then yields a retrieved signal for true astrophysical sources that is now less impacted by flux losses. This reduces the ``confusion zone'' illustrated in Figures \ref{fig:ProbablitiesFaint} and \ref{fig:ProbablitiesBright}. Here comparisons were limited to KLIP with and without Forward Modeling in the case of an Aperture Photometry observer (e.g the fitting zone $\mathcal{F}$ in KLIP-FM was chosen to be equal to the aperture in KLIP+ApPhot). Future investigations are needed to assess the gain when using Forward Modeling with more sophisticated observers, such as the ones presented in \citet{0004-637X-646-2-1260,Caucci:07}.  Our work was limited to speckle fitting in the least-squares sense, applying perturbation methods to the more sophisticated costs functions such as presented \citet{2016arXiv160208381G} would also be of great interest.

\subsection{Toward point-wise KLIP-FM}
\begin{figure*}[t]
\begin{center}
\includegraphics[width=1\textwidth]{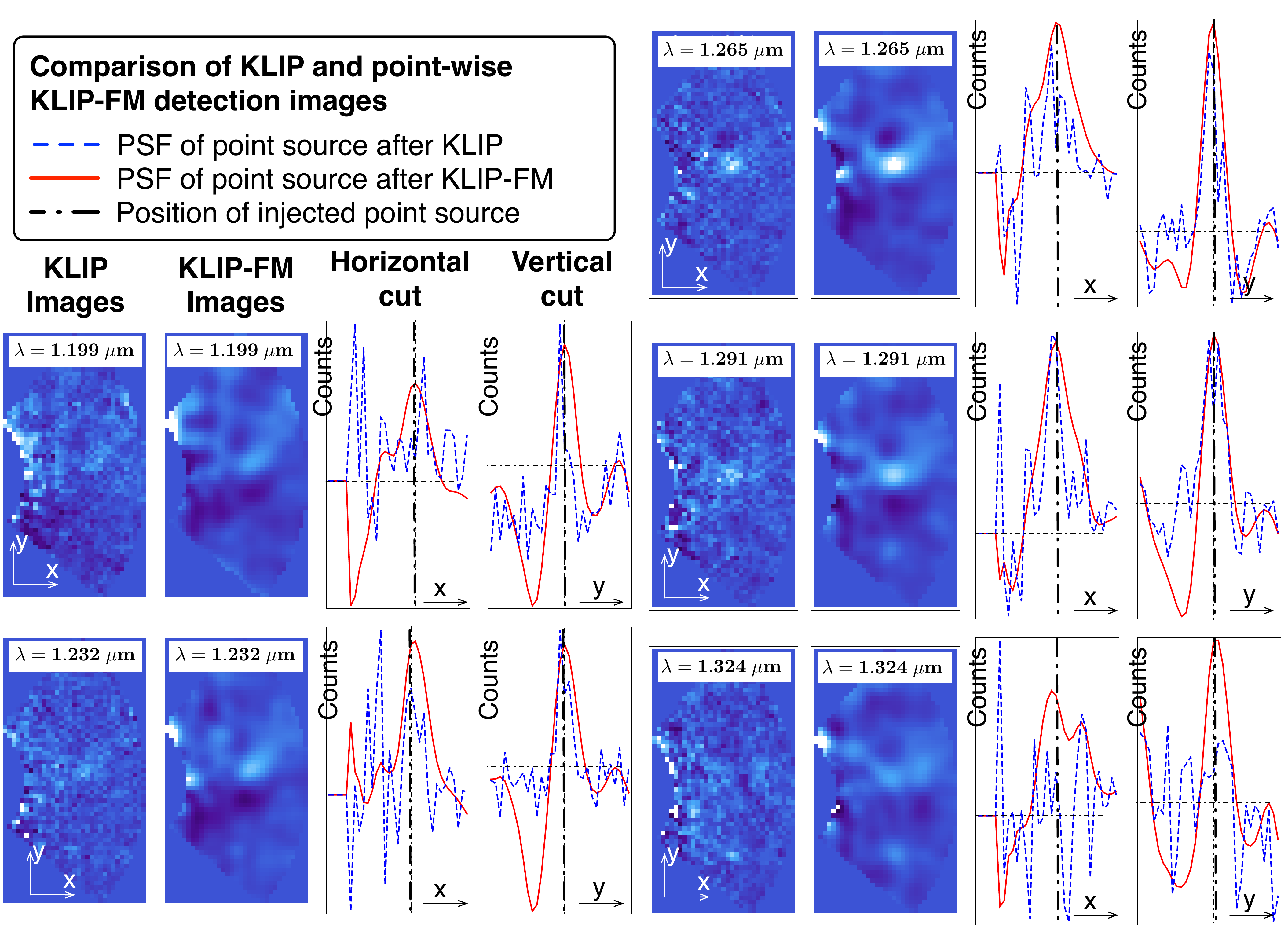} 
\caption{{\bf Comparison of images obtained with KLIP  and a point-wise implementation of KLIP-FM. A point source five times fainter than the local speckles has been injected in the data.} We only used a portion of the GPI field of view along with one datacube . The algorithm parameters are set to aggressive ($N_{\delta} = 0.6$, $N_{Corr} = 30$, $K_{Klip} = 30$).   In the low SNR channels, denoted as  $\lambda = 1.199, 1.232,1.291,1.324 \; \mu$m, the KLIP image features hints of faint flux at the location of the injected point source but the single wavelength detection is much more convincing with the point-wise implementation of KLIP-FM. } 
\label{fig:FInalComparisonKLIPFMImage}
\end{center}
\end{figure*}

Finally we illustrate how this procedure can also be used in a more systematic manner for ``planet search.'' Instead of building a model for hypothetical point sources scattered across the field of view, we present here a point-wise implementation of KLIP-FM that inverts Eq.~\ref{Eq:SimpleMat} {\em at each point the the astrophysical scene, one at a time}. Figure~\ref{fig:FInalComparisonKLIPFMImage} was generated using this method over a portion of the GPI field of view, with a synthetic companion that is five times fainter than the local speckles, aggressive PSF subtraction parameters, and we only used one GPI data-cube. The leftmost column of each panel Figure~\ref{fig:FInalComparisonKLIPFMImage} shows the images from the KLIP algorithm, and the next column shows the flux across wavelengths obtained with point-wise KLIP-FM. The two right columns show the horizontal and vertical cross section of both images at the location of the injected point source. This figure highlights the four advantages of point-wise KLIP-FM:
\begin{itemize}
\item  In the low flux channels, denoted as  $\lambda = 1.199, 1.232,1.291$, and $1.324 \mu$m, the KLIP image features hints of faint flux at the location of the injected point source, but the single wavelength detection is much more convincing in the left column with point-wise KLIP-FM. This is in part due to the convolution by the instrument PSF, which would occur regardless of Forward Modeling. However this is also due to the fact that inverting Eq.~\ref{Eq:SimpleMat} does shift to the right the signal present histogram (as shown in Figure.~\ref{fig:ProbablitiesFaint}), thus helping to discriminate faint astrophysical signals from residual speckle noise.
\item In all channels the centroid of the signal after KLIP is very much affected by the over/self-subtraction. On the other hand, with point-wise KLIP-FM, the maximum of the retrieved signal is at the location of the injected source in the channels for which the residual speckle noise is uncorrelated. This is highly beneficial when using detection metrics that rely on the stability of the point source location as a function of wavelength. It also has significant advantages for astrometry. 
\item In all channels the point-wise KLIP-FM counts correspond to the unbiased spectrum of the point source (see \S~4). 
\item For all locations corresponding to non-detections, over/self-subtraction have also been corrected; such images can be readily used to derive detection limits. 
\end{itemize}
In spite of all these advantages, point-wise KLIP-FM, as implemented to generate  Figure~\ref{fig:FInalComparisonKLIPFMImage}, presents  one major drawback: it is painstakingly slow and memory hungry (generating Figure~\ref{fig:FInalComparisonKLIPFMImage},  for only one GPI data cube and a small fraction of the field of view, required a day of computations on a standard macbook pro laptop). Carrying out the forward model construction and inversion ``as is'',  over the entire field of view of a coronagraph imager and over an entire observing sequence is thus prohibitively expensive computationally. However, simplifications exist: in particular the exceedingly large intermediate matrices of Appendix E and F need not to be evaluated in all cases.  One can show that temporary variables of lower dimensionality can be used instead. The linear algebra details associated with these simplifications is beyond the scope of this paper and will be presented in a latter communication (Ruffio et al., in preparation). Here we simply use our  computationally inefficient implementation point-wise KLIP-FM to illustrate on Figure~\ref{fig:FInalComparisonKLIPFMImage} that this algorithm has the potential to become a very powerful tool for exoplanet detection via direct imaging.
%

\section{Conclusion and perspectives}
In this paper we introduced a linear expansion that captures the impact of over/self-subtraction in high-contrast imaging data. This is done in the most the general case for which the reference images of the astrophysical scene move azimuthally and/or radially across the field of view (ADI and/or SSDI). This method is based on perturbing the covariance matrix underlying any least-squares speckles problem and propagating this perturbation through the data analysis algorithm. Most of the work in this paper has been presented in the PCA framework, but it can be easily generalized to methods relying on linear combinations of images (instead of eigenmodes). Based on this linear expansion, we then demonstrated how this new algorithm could be used in practice. We first considered the case of the spectral extraction of faint point sources in IFS data (under the ADI+SSDI observation strategy) and illustrated, using public Gemini Planet Imager commissioning data, that our novel perturbation-based Forward Modeling can indeed alleviate algorithmic biases. We then applied KLIP-FM to the detection of  point sources and showed how it decreases the rate of false negatives while keeping the rate of false positives unchanged, when compared to classical KLIP.\\

Beyond these two examples, our analytical result is broadly applicable to a wide range of high-contrast science:
\begin{itemize}
\item {\em Planet detection:} should the point-wise KLIP-FM described in \S 4.3 be improved upon so it can be implemented in an efficient manner, it will facilitate the detection of the faintest end of the point sources buried in the residual speckles of ongoing surveys. Note however that this gain will only occur if astronomers relax their standards to trigger follow-up observations. Indeed, for very faint planets if the threshold is solely based on a FPF $<10^{-3}$ then the TPF  will be close to $0 \%$ regardless of whether or not Forward Modeling is used. This is illustrated in Figure~\ref{fig:ROCs}, for the faintest example considered in this paper. However, if a FPF of $20 \%$ is tolerated, then Forward Modeling will bring the completeness, or TPF, up $70 \%$. Without Forward Modeling, increasing the allowable FPF to $20 \%$ only brings completeness up to $20 \%$, which is still no better than a coin toss. This demonstrates loosening detection threshold might be benefitial, now that we are equipped with a tool that can greatly increase completeness at only a modest cost in observing efficiency (in our example, one in five follow-up observations is triggered based on a bright speckle instead of a true astrophysical point source). Given the paucity of currently directly imaged exoplanets, we argue that the experimental design of ongoing surveys should consider such an option. 
\item {\em Planet detection:} in this manuscript we only considered observers that integrate flux over an aperture (with and without Forward Modeling). More sophisticated observers such as the ones presented in \citet{0004-637X-646-2-1260,Caucci:07} could be used. Moreover, the linear model developed here could also be used in a Bayesian framework. It was recently shown that a simple model of dual-image ADI subtraction in such a framework was an effective method for the detection of faint sources \citep{2015A&A...582A..89C}. Because of the analytical derivation presented herein, similar work can now be conducted using more sophisticated PCA-based denoising algorithms. 
 \item {\em Detection limits in IFS data:} In the case of non-detections, completeness is estimated for each hypothesis regarding the potential astrophysical signal that was not observed. In the case of broadband imaging with RDI or ADI, this ensemble of astrophysical hypothesis is of  low dimensionality because the observables are simply separation and integrated brightness. In the case of IFS observations, this dimensionality significantly increases. Moreover, when using SDI or SSDI, the ROC, and thus the completeness, varies as a function of the hypothesized underlying spectrum. This significantly  complicates the population statistics for this type of observation, and is one of the outstanding problems for the statistical analysis of ongoing large high-contrast surveys with IFS \citep{bfd08,hob08,2014PNAS..11112661M}. In Appendix F we briefly describe how KLIP-FM could be used to address this issue, but leave out numerical examples for future work. 
\item {\em Astrometry:} point-wise Forward Modeling can be carried out at the subpixel level around a first guess for the location of a detected point source. This in principle should yield high precision astrometry, even when using ADI and/or SSDI. We will demonstrate the performances of this method in a future paper (Wang et al. 2016, in preparation). 
\item {\em Retrieval of physical properties of planets}:  as discussed in \S 3, because of its simplicity Eq.~\ref{Eq:ForwardModellingSpectra} is amenable to be used within a Bayesian framework  to evaluate correlated uncertainties associated with the spectrum of a faint point source. One could also directly fit the physical model `` in the data'' by using Eq.~\ref{Eq:ForwardModellingSpectra} as the cost function in retrieval codes, such as the one recently presented in \citet{2015ApJ...807..183L}.
\item {\em Disk imagery and characterization:} the biggest practical difference between high-contrast disk imagery and point source detection resides in the choice of optimal least-squares reduction parameters. In this paper we presented a Forward Modeling that is parameter independent, and as a consequence all our discussions are in principle applicable to disk imaging and characterization. Note however that in practice Forward Modeling with disks is complicated by the fact that Eq~\ref{Eq:convol} cannot be simplified to by using a simple PSF as the astrophysical model: every hypothetical disk morphology instead has to be explored. 
\end{itemize} 
The amount of work required to robustly devise these algorithmic improvements and thoroughly test them  goes beyond the scope of what can be achieved by a single individual. It is our hope that the community will conduct the potentially important investigations in data analysis development outlined here in a collaborative manner and include promising advances in publicly available tools, such as \citet{2015ascl.soft06001W}. 
\section*{Acknowledgements}
We thank the two anonymous referees for their feedback and extremely important suggestions which made this paper significantly more readable. This manuscript also benefited from critical inputs from the entire Gemini Planet Imager team. In particular Dimtry Savransky who carefully went over the linear algebra presented in the Appendices; Abhiji Rajan, Kim Ward-Duong and Marshall Perrin who provided insightful suggestions on the writing and presentation; Christian Marois and Kate Morzinski who provided important suggestions regarding the context of the paper. The vast majority of the results in \S~4 stemmed from initial discussions during the Exoplanet Imaging Workshop whose findings are presented in \citet{2012SPIE.8447E..22L}. Finally this paper would have never been written without Jason Wang and Jean Baptiste Ruffio who were able to confirm the findings presented herein using a separate implementation of KLIP-FM and made their code available to the community \citep{2015ascl.soft06001W}.

\clearpage

\appendix

{\bf Reminder of the structure of the Appendices:}
\begin{itemize}
\item Appendix A provides the most general formalism for an ADI + SSDI observing sequence and lays out the formal foundations for our work. 
\item Appendix B summarizes the notations Appendix A in a table format. In order to facilitate numerical implementation, it provides the dimensions of the various matrices discussed in this paper. 
\item Appendix C introduces the formalism underlying Forward Modeling in the most general case, and then discusses the specific configuration of RDI. This was already presented in \citet{2014arXiv1409.6388P}, but serves to set up the stage for Appendix F.  
\item Appendix D describes how to carry out Forward Modeling for the astrometry and photometry of point sources for RDI within the framework of the linear algebra notations introduced in Appendix A and C.
\item Appendix E contains the proof of our novel analytical expansion. It heavily relies on the notations introduced in Appendix A. It contains the key innovation of the present manuscript.
\item Appendix F describes how to take advantage of the result in Appendix E to carry out Forward Modeling in order to estimate the spectrum of point sources in IFS data. 
\end{itemize}


\clearpage

\section{Appendix A: Narrative explanation of the various notations}
\begin{figure}[h]
\begin{center}
\includegraphics[width=0.7\textwidth]{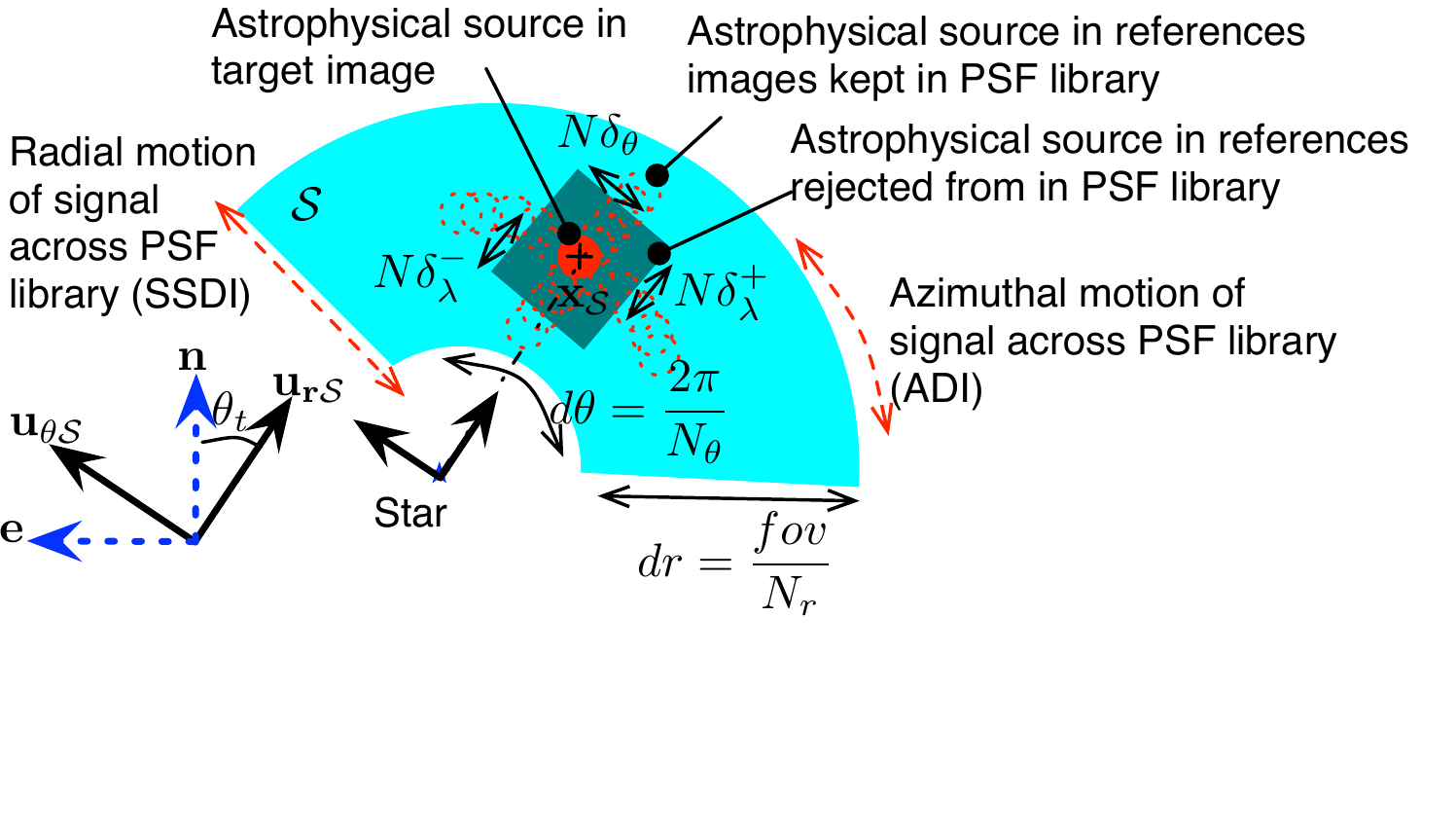} 
\caption{{\bf  Parametrization of the least-squares PSF subtraction algorithm for the most general case of ADI+SSDI discussed in this paper.} Even if this choice of zone geometry and reference PSF selection criteria is not be applicable to all high-contrast science cases, the perturbation analysis presented herein is general and can be ported to other applications. Note that here we simplified the geometry by choosing $\mathbf{x_{\mathcal{S}}} = \mathbf{x_{A}}$.
}
\label{fig:CartoonParameters}
\end{center}
\end{figure}

In this appendix we provide a detailed description of our mathematical notations regarding the most general case of a high-contrast observing sequence and a generic implementation of  the KLIP algorithm  (e.g we do put Table \ref{TabNotations} into words). Note that all the material in this appendix has already been discussed in the literature, but is revisited here in order to provide a rigorous framework for the latter introduction of the KLIP-FM algorithm. An illustration of the algorithm parameters  discussed here is given in Fig.~\ref{fig:CartoonParameters}

\subsection{Observing sequence}

We consider the general case of an observing sequence with an Integral Field Spectrograph (IFS) and in the presence of field rotation (ADI). An image, $I_{\lambda,t}(\mathbf{x})$ at the wavelength $\lambda$ and at time $t$ (parallactic angle $\theta_{t}$), within the observing sequence, can be written as:
\begin{equation}
I_{\lambda,t}(\mathbf{x}) = S_{\psi_{\lambda,t}}(\frac{\mathbf{x}}{\lambda}) +\epsilon a_{\lambda} A_{\lambda}(\mathit{R}_{\theta_t}[\mathbf{x}])
\label{Eq:GenrealFrame}
\end{equation}
where:
\begin{itemize}
\item $  S_{\psi_{\lambda,t}}$ is the focal plane intensity associated with speckles (integrated over the narrow bandpass around $\lambda$  at time $t$) that results from a random realization $\psi_{\lambda,t}$ of the telescope + instrument wavefront. 
\item $\mathbf{x}$ are the 2D coordinates across the field of view. 
\item $\epsilon$ is equal to zero if there is no astrophysical signal. If an object is present, it is equal to the integrated photometry of the astrophysical source over the entire bandpass of the IFS. 
\item $a = [ a_1.. a_p ...  a_{N_\lambda}]$ is the normalized spectrum of the object. Namely, if the actual spectrum of the object at the resolution of the IFS  is $f = [f_1... f_p ... f_{N_\lambda}]$ then $ \epsilon = \sum_{p = 1}^{N_{\lambda}}  f_p $ and $  a_p =  f_p / \epsilon$. 
\item  $\epsilon a_{\lambda} A_{\lambda}(\mathbf{x})$ is the image at $\lambda$ of the astrophysical source, at the spatial resolution of the instrument, rotated north up. 
\item $\mathit{R}_{\theta_t}$ corresponds to the 2D rotation matrix --with respect to the stellar location-- associated with the azimuthal motion of the astrophysical source across the ADI observing sequence. {\color{black} $\theta_t$ corresponds to the parallactic angle (direction of north in the images) which varies across an ADI sequence}. 
\item {\color{black} throughout the paper $\mathbf{x}$  corresponds to the 2D coordinates across the field of view. In the linear algebra formalism discussed in the appendices, and for practical implementations, these two dimensions can be collapsed onto one. For instance if $\mathbf{x}$ describes all the possible pixel coordinates across the field of view (of size $N_{fov} \times N_{fov}$), then x is a $1 \times N_{fov}^2 \time 2 $ array.} 

\end{itemize}
PCA-based reduction algorithms use a well-chosen library of images to build an empirical model of the speckle noise realization associated with each target image within the observing sequence. Each empirical model is then subtracted from its corresponding target image in order to increase the S/N of potential astrophysical sources. Without a loss of generality, we choose here the target image at wavelength $\lambda_{0}$ at at the exposure starting at $t_0$ (parallactic angle $\theta_{0}$).  
\begin{equation}
T(\mathbf{x}) = I_{\lambda_0,t_0}(\mathbf{x}) = S_{\psi_{\lambda_0,t_0}}(\frac{\mathbf{x}}{\lambda_{0}}) +\epsilon a_{\lambda_{0}} A_{\lambda_{0}}(\mathit{R}_{\theta_{0}}[\mathbf{x}]). 
\end{equation}
The corresponding reference library is then assembled by choosing among all other possible images with $(\lambda,t) \neq (\lambda_0,t_0)$. This captures the most general configuration discussed in this paper. Of course there exists observing scenarios for which it greatly simplifies:
\begin{itemize}
\item when using RDI (a PSF library built using images of other sources) and under the assumption that the library has been built to be ``signal free'' (see for instance \cite{2014SPIE.9143E..57C}), then the astrophysical signal is only present in the target image $T = S_{\psi_0}(\mathbf{x}) + \epsilon A(\mathbf{x})$.  In this case $\epsilon = 0$ for all images in the reference in the PSF library and thus $R_{k} =  S_{\psi_k}(\mathbf{x})$. These are the shorthanded notations described in \cite{2012ApJ...755L..28S} (here the state of the telescope+instrument $\psi$ does not depend on time  and wavelength, it is simply indexed over the ensemble of reference stars). 
\item when using Angular Differential Imaging (ADI) with non-IFS data  (or when using IFS data that excludes images at other wavelengths from the PSF library), we can drop the wavelength dependence for both the spatial scaling of the speckle noise and the brightness of the potential astronomical objects. Then Eq.~\ref{Eq:GenrealFrame} reduces to:
\begin{equation}
I_{t}(\mathbf{x})  = S_{\psi_{t}}(\mathbf{x}) +\epsilon a  A(\mathit{R}_{\theta_t}[\mathbf{x}]).
\end{equation}
where $a$ is now a scalar instead of a vector. 
\end{itemize}

\subsection{Reference PSF Library}

\subsubsection{Spatial Rescaling and image plane motion of a point source}

The first step in least-squares speckles fitting is to build for each $T(\mathbf{x}) = I_{\lambda_0,t_0}(\mathbf{x}) $ its corresponding ensemble of reference PSFs --$R_{\lambda,t}(\mathbf{x})$-- by choosing among all other possible images with $(\lambda,t) \neq (\lambda_0,t_0)$. In the most general case (e.g when using SSDI and ADI) all IFS slices are first spatially rescaled to $\lambda_{0}$, so that the characteristic scale of the speckle noise in the references matches the noise in the target image. Thus, the reference images are:
%
\begin{equation}
R_{\lambda,t}(\mathbf{x})= I_{\lambda,t}(\mathbf{x} \frac{\lambda}{\lambda_0}) = S_{\psi_{\lambda,t}}(\frac{\mathbf{x}}{\lambda_0}) +\epsilon a_{\lambda} A_{\lambda}(\mathit{R}_{\theta_t}[\mathbf{x} \frac{\lambda}{\lambda_0}])
\label{Eq:ImagePlanet} 
\end{equation}
%
%
In the case of  ADI only, this rescaling is not necessary and the wavelength dependence can be dropped. The flux normalized astrophysical scene seen by the instrument can be written as the convolution of the sky --$Sky(\mathbf{x})$-- by the instrument PSF:
\begin{equation}
A_{\lambda}(\mathbf{x}) = \int  Sky_{\lambda}(\mathbf{u}) PSF_{\lambda,\mathbf{x}}(\mathbf{u}-\mathbf{x}) d \mathbf{u},
\label{Eq:convol}
\end{equation}
where the wavelength and field dependence of the coronagraphic PSF are captured in the subscripts of $PSF_{\lambda,\mathbf{x}}$. Using these notations and neglecting PSF field dependence, the motion of the astrophysical signal at given field point $\mathbf{x_\mathcal{S}}$ associated with wavelength scaling and field rotation -- $A_{\lambda}(\mathit{R}_{\theta_t}[\mathbf{x_\mathcal{S}} \frac{\lambda}{\lambda_0}])$-- is captured by $PSF_{\lambda}(\mathbf{u}- \mathit{R}_{\theta_t}[\mathbf{x_\mathcal{S}} \frac{\lambda}{\lambda_0}])$, with: 
\begin{eqnarray}
\mathit{R}_{\theta_t}[\mathbf{x_\mathcal{S}} \frac{\lambda}{\lambda_0}] &=& \mathbf{x_\mathcal{S}} +  \delta^{(\lambda_{0},t,\lambda)} \mathbf{x_\mathcal{S}}\\
 \delta^{(\lambda_{0},t,\lambda)} \mathbf{x_\mathcal{S}} &=& ||\mathbf{x_\mathcal{S}}||  \frac{\lambda}{\lambda_{0}}( \cos(\theta_t) \mathbf{n} + \sin(\theta_t)  \mathbf{e})
\end{eqnarray}
%
where $\mathbf{n} ,\mathbf{e}$ are the unit vectors pointing north and east. $\delta^{(\lambda_{0},t,\lambda)} \mathbf{x_\mathcal{S}}$ is a 2D vector that relates the position in the field of view of a hypothetical point source in the science image of interest --at $(t_0,\lambda_0)$-- to its position in each one of the spatially the rescaled reference images (at $(t_0,\lambda_0) \neq (t,\lambda)$).

This motion of the astrophysical scene with respect to the speckle noise across the instrument field of view is key to building PSF libraries for which the signal in the reference PSFs is not located at $\mathbf{x_\mathcal{S}}$ (thus enabling local empirical fitting of the speckles only, with ``minimal contamination from the signal''). 

\subsubsection{Reference selection criteria}

\label{Ref:SelCrit}
More formally, we build such a collection of reference images by ensuring that $\mathbf{ \delta^{(\lambda_{0},t,\lambda)} x_\mathcal{S}}$ is large enough so that there is no (or little) astrophysical signal at $\mathbf{x_\mathcal{S}}$ in the PSF library. We write these references as $\mathcal{R} = \{ R_{k}(\mathbf{x}), k = 1...N_{\mathcal{R}} \}$. This library is constructed over subsections of the image, or subtraction zones $\mathcal{S}$, centered on $\mathbf{x_\mathcal{S}}$. Here we parameterize these zone in polar coordinates: radial extent $dr$ (or $N_{r}$ annuli across the field of view) or azimuthal extent $d\phi$ (or $N_{\phi}$ sectors per annulus). Note that in this paper we do not follow the method described in \cite{lmd07}, which splits  the geometry of the problem between optimization ($\mathcal{O}$) and subtraction zones ($\mathcal{S}$)  and we solely focus on the case for which $\mathcal{O} = \mathcal{S}$. However, in principle, the KLIP-FM formalism is also applicable when $\mathcal{S}$ is a subregion  of $\mathcal{O}$. We write $\mathbf{u_{r \mathcal{S}}}$  and $\mathbf{u_{\theta \mathcal{S}}}$ as the radial and tangential unit vectors in the direction of $\mathbf{x_{S}}$. Whether or not an image is included in this library is then decided using a combination of the following criteria:
 \begin{itemize}
\item $k$ such that $ ( \delta^{(\lambda_{0},t_k,\lambda)} \mathbf{x_\mathcal{S}}-  \delta^{(\lambda_{0},t_0,\lambda)}\mathbf{ x_\mathcal{S}}).\mathbf{u_{\theta \mathcal{S}}} > N \delta_{\theta} *FWHM( PSF_{\lambda_{0}}) $ to account for the minimal motion of a source due to field rotation. $FWHM( PSF_{\lambda_{0}})$ is the Full Width at Half Maximum of the instruments's PSF at wavelength $\lambda_0$. 
\item $k$ such that $ \delta^{(\lambda_{0},t,\lambda_k)} \mathbf{ x_\mathcal{S}}.\mathbf{u_{r \mathcal{S}}} > N \delta_{\lambda}^{+} *FWHM( A_{\lambda_{0}})$ to account for the minimal outward motion of a source due to speckle chromaticity.  
 \item $k$ such that $ \delta^{(\lambda_{0},t,\lambda_k)} \mathbf{ x_\mathcal{S}}.\mathbf{u_{r \mathcal{S}}} < N \delta_{\lambda}^{+} *FWHM( A_{\lambda_{0}})$ to account for the minimal inward motion of a source due to speckle chromaticity. Very often $N \delta_{\lambda}^{-}  =  N \delta_{\lambda}^{+}$. However, as explained in \cite{2014SPIE.9148E..0UM}, it can be very beneficial to use different values when seeking to detect faint companions with sharp spectral features. In this case, $N \delta_{\lambda}^{-}, N \delta_{\lambda}^{+}$ can be chosen based on the hypothetical underlying sharp spectral feature of the hypothetical astrophysical signal. 
\item $k$ such that the reference belongs to the $N_{Corr}$ images with the largest correlation with the target. Note that we adopt the following notation for correlations for the remainder of the paper: $<R_{k},T>_{\mathcal{S}} = \int_{\mathcal{S}}R_{k}(\mathbf{x}) T(\mathbf{x}) d \mathbf{x}$. 
\end{itemize}
Note that in the case of point sources, when $ Sky_{\lambda}(\mathbf{u}) = a_{\lambda} P_{source} (\mathbf{u} - \mathbf{u_{source}})$, with $ P_{source}(0,0) = 1$ and zero otherwise, then the first three selection criteria above directly relate to the flux contamination across wavelengths and rotation angle. For all the examples in this manuscript, we simplify the reference selection by using $N_{\delta} =N \delta_{\lambda}^{-} =  N \delta_{\lambda}^{+} = N \delta_{\theta}$.  We also introduce the following shorthand notations:
\begin{itemize}
\item $\mathbf{E_{obs}}(\mathbf{x})$ is the overall ensemble of PSFs in the observing sequence. {\color{black} When folding the two-dimensional spatial variable $\mathbf{x}$ into a line vector it can be seen as a matrix with $N_{Exp} \times N_{\lambda}$ lines and $N_{Pix}$ columns ($N_{Exp}$ is the number of exposures in the observing sequence and $N_{Pix}$ the number of pixels in the $\mathcal{S}$ zones). Note that $\mathbf{x}$ corresponds to 2D coordinates that are folded into one dimension for practical reasons. That is, if the $\mathcal{S}$ zone was the entire field of view (of dimension $N_{fov} \times N_{fov}$ pixels), then one row entry of $\mathbf{E_{obs}}(\mathbf{x})$ would be of dimension $1 \times N_{fov}^2$. The same applies to $R(\mathbf{x})$.}
\item $\mathbf{R_{\mathcal{S},\lambda_0,t_0}}^{(N \delta_{\theta}, N \delta_{\lambda}^{+}, N \delta_{\lambda}^{-}, N_{Corr},N_{r},N_{\phi})}(\mathbf{x})$ is the ensemble of reference PSFs chosen to analyze a target image at $(\lambda_0,t_0)$. When folding $\mathbf{x}$ into a line vector, it becomes a matrix with $N_{\mathcal{R}}$ lines and $N_{Pix}$ columns ($N_{\mathcal{R}}$ is the number of frames selected in the PSF library). For clarity, we drop the dependence on algorithms parameters and write this matrix as  $\mathbf{R}(\mathbf{x})$.
\item $\mathbf{Sel_{\mathcal{S},\lambda_0,t_0}}^{(N \delta_{\theta}, N \delta_{\lambda}^{+}, N \delta_{\lambda}^{-}, N_{Corr},N_{r},N_{\phi})}$ is an $N_{\mathcal{R}}$ by $N_{Exp} \times N_{\lambda}$ selection matrix whose entries are defined by $\mathbf{S}[i,j] = 1$ if $\mathbf{E_{obs}}[j] (\mathbf{x})$ is the $i$ th entry in the reference library and $0$ otherwise. Or, more succinctly:
\begin{equation}
\mathbf{R}(\mathbf{x}) = \mathbf{Sel} \; \mathbf{E_{obs}}(\mathbf{x})
\end{equation}
where again, we write $\mathbf{Sel}$ dropping the dependence on $\mathcal{S},\lambda_0,t_0$ and $(N \delta_{\theta}, N \delta_{\lambda}^{+}, N \delta_{\lambda}^{-}, N_{Corr},N_{r},N_{\phi})$. 
 \end{itemize}

\subsection{Principal component analysis}
\label{Ref:PCABasic}
Once the reference library corresponding to a given target image has been assembled, the PCA  is carried out as follows:
\begin{enumerate}
\item Zero mean $T(\mathbf{x})$ and $R_{k}(\mathbf{x})$ over $\mathcal{S}$. 
\item Calculate the Karhunen-Lo\`eve transform of  $\mathbf{R}(\mathbf{x})$:
\begin{equation}
 Z_k(\mathbf{x}) = \frac{1}{\sqrt{\Lambda_k}}\sum_{m=1}^{N_{\mathcal{R}}} v_k[m]  R_m(\mathbf{x}) 
 \label{Eq:KLdefinition}
 \end{equation}
where the vectors $V_k = \left[ v_{k}[1] ... v_{k}[N_{\mathcal{R}}] \right]$ are the eigenvectors of the $N_{\mathcal{R}}$-dimensional covariance matrix of the reference library $C_{RR} = <\mathbf{R}(\mathbf{x}), \mathbf{R}(\mathbf{x}) >_{\mathcal{S}} =  \mathbf{R}(\mathbf{x}) \; \mathbf{R}(\mathbf{x})^{T}$, and correspond to its eigenvalues $\{\Lambda_k \}_{k=1...N_{\mathcal{R}}}$.
\item Choose a cutoff, $K_{Klip}$, for the number of modes the target image will be projected on. 
\item Project the target image on the Principal Components and subtract this projected speckle noise model from the target image:
 \begin{equation}
 \mathcal{KLIP}_{\mathcal{R}} [T(\mathbf{x})] =  P(\mathbf{x})   = T(\mathbf{x})  - \sum_{k = 1}^{K_{Klip}} <T(\mathbf{x}),Z_k(\mathbf{x})>_{\mathcal{S}}  Z_k(\mathbf{x})
  \label{Eq:ProjectionsDefinition}
  \end{equation}
\end{enumerate}
This algorithm was outlined in \cite{2012MNRAS.427..948A} and \cite{2012ApJ...755L..28S}. \citet{2012ApJ...755L..28S} suggested that such a formalism could serve as a foundation to calibrate potential systematic errors on the astrophysical observables due to the reduction algorithms, however they did not delve into the details of such a procedure. The present manuscript addresses this outstanding point. 

\clearpage

\section{Appendix A: Table summarizing our mathematical notations.}

Here we summarize the various notations introduced in the  discussions in the appendices (along with the dimensions of each variable). It is our hope that this summary will help the reader through the more technical arguments of this manuscript, along with helping interested parties to  implement the KLIP-FM algorithm. 

\begin{center}
\begin{longtable}{|l|l|l|p{6cm}|}
\caption[Notations]{Description of the mathematical notations in this Appendix} \label{TabNotations}\\
\hline \multicolumn{1}{|c|}{\textbf{Symbol}} &
\multicolumn{1}{c|}{\textbf{Expression}} &
\multicolumn{1}{c|}{\textbf{Dimensions}}&
\multicolumn{1}{c|}{\textbf{Comments}} \\\hline
\endfirsthead

\multicolumn{4}{c}%
{{\bfseries \tablename\ \thetable{} -- continued from previous page}} \\
\hline \multicolumn{1}{|c|}{\textbf{Symbol}} &
\multicolumn{1}{c|}{\textbf{Expression}} &
\multicolumn{1}{c|}{\textbf{Dimensions}}&
\multicolumn{1}{c|}{\textbf{Comments}} \\ \hline 
\endhead

\hline \multicolumn{4}{|r|}{{Continued on next page}} \\ \hline
\endfoot

\hline \hline
\endlastfoot

\multicolumn{4}{c}{Coordinates and Algorithm Parameters}\\
\hline
$\mathbf{x}$ &  & $1 \times 2$ & Coordinates in focal plane\\
$\mathbf{x_{\mathcal{S}}}$ &  & $1 \times 2$ & Center of PCA subtraction zone\\
$ \delta^{(\lambda_{0},t,\lambda)} \mathbf{x_\mathcal{S}}$ &&  $1 \times 2$ & Radial and azimuthal motion of an hypothetical source located at $\mathbf{x_\mathcal{S}}$ over an ADI+SSDI observing sequence\\
$\theta_t$ &&  $1 \times 2$  & ADI field rotation corresponding to the exposure at time $t$\\
$dr  = \frac{1}{N_r}$ & & $1 \times 1$ & Radial extent of the local $\mathcal{S}$ zone of the field of view over which the speckles least-squares fitting occurs\\
$d \theta = \frac{1}{N_{\theta}}$ & & $1 \times 1$ & Azimuthal extent of the $\mathcal{S}$ zone\\
$ \delta x_q$ & & $1 \times 1$ & Azimuthal displacement of an astrophysical source at the $q$ th exposure\\
$N \delta_{\theta}$& & $1 \times 1$ & ADI exclusion criterion\\
$ N \delta_{\lambda}^{-}$& & $1 \times 1$ & SDI exclusion criterion (inwards) \\
$ N \delta_{\lambda}^{+}$& & $1 \times 1$ & SDI exclusion criterion (outwards)\\
$ N_{Corr}$& & $1 \times 1$ & Number of most correlated \\
& & & references kept in PSF library\\
$ N_{\mathcal{R}}$& & $1 \times 1$ & Number of references in PSF library, in this paper we use $N_{Corr} = N_{\mathcal{R}}$\\
$ N_{pix}$& & $1 \times 1$ & Number of pixels in the $\mathcal{S}$ zone\\
\hline
\multicolumn{4}{c}{Astrophysical Quantities}\\
\hline
$\mathbf{x_{A}}$ &  & $1 \times 2$ & Location of an astrophysical point source\\
$\mathbf{\widehat{x_A}}$ && $1 \times 2$ & Location of synthetic negative point source underlying Forward Modeling\\
$\mathbf{\tilde{x}_A}$ && $1 \times 2$ & Estimated location of an astrophysical point source\\
$\epsilon$ &  & $1 \times 1$ & Photometry of astrophysical source\\
$\widehat{\epsilon}$ &  & $1 \times 1$ & Photometry f synthetic negative point source underlying Forward Modeling\\
$\widetilde{\epsilon}$ &  & $1 \times 1$ & Estimated photometry  of astrophysical source\\ 
$f$ & $\left[\begin{array}{c} f_{1} \\ ... \\  f_{\lambda} \\... \\  f_{N_{\lambda}}\end{array}\right] $&  $N_{\lambda} \times 1$ & Spectrum of astrophysical point source\\
$a$ & $ f_{\lambda}/\epsilon$& $N_{\lambda} \times 1$ & Normalized spectrum of astrophysical source\\
$\widehat{f}$ & & $N_{\lambda} \times 1$ &  Spectrum of synthetic negative source underlying Forward Modeling\\
$\widehat{a}$ & & $N_{\lambda} \times 1$ & Normalized spectrum of synthetic negative source underlying Forward Modeling\\
$\widetilde{f}$ & & $N_{\lambda} \times 1$ &  Estimated spectrum of astrophysical source \\
$\widetilde{a}$ & & $N_{\lambda} \times 1$ & Estimated normalized spectrum of astrophysical source \\
$\bar{a}$ & $ \left[\begin{array}{c} a_{\lambda_{1}} \\ ... \\   a_{\lambda_k}  \\... \\  a_{\lambda_{N_{\mathcal{R}}}} \end{array}\right] $& $N_{\mathcal{R}} \times 1$ & Vector of normalized flux of the astrophysical source corresponding to the signal contained in each one of $N_{\mathcal{R}}$ reference images\\
$\mathbf{a}$ & $\left[\begin{array}{ccccc} a_{\lambda_{p(1)}}  & ... & 0 & ... & 0 \\... & ... &  0 & ... & ... \\0 & 0 & a_{\lambda_{p(k)}}  & 0 & 0 \\... & ... & 0 & ... & ... \\0 & ... & 0 & 0 & a_{\lambda_{p(N_{\mathcal{R}})}} \end{array}\right] $ &  $N_{\mathcal{R}} \times N_{\mathcal{R}} $ & Matrix in whose diagonal elements have been populated by each one of the $N_{\mathcal{R}}$ entries of $\bar{a}$\\
\hline
\multicolumn{4}{c}{Vectors and Matrices in Data Space }\\
\hline
$S_{\psi_{\lambda,t}}(\mathbf{x})$ &  & $1 \times N_{pix}$ & Scattered starlight (speckles) at wavelength $\lambda$ in the zone $\mathcal{S}$ corresponding to the state of the instrument at time $t$ \\
$\mathbf{S}(\mathbf{x})$ & $\left[\begin{array}{c} S_{\psi_{\lambda_1,t_1}}(\mathbf{x})\\ ... \\ S_{\psi_{\lambda_k,t_k}}(\mathbf{x}) \\... \\  S_{\psi_{\lambda_{N_{\mathcal{R}}},t_{\mathcal{R}}}}(\mathbf{x})\end{array}\right] $ &$N_{\mathcal{R}}\times N_{pix}$ & Matrix of the concatenated speckles realizations kept in the reference PSF library\\
$\mathbf{S}$ & & $N_{\mathcal{R}}\times N_{pix}$ & Short-handed notation for $\mathbf{S}(\mathbf{x})$ \\
$A_{\lambda}(\mathit{R}_{\theta_{t}}[\mathbf{x}])$ &  & $1 \times N_{pix}$ & Astrophysical image at wavelength $\lambda$  that has been rotated by $\theta_t$ due to ADI field rotation\\
$T(\mathbf{x})$ &  $S_{\psi_{\lambda_0,t_0}}(\frac{\mathbf{x}}{\lambda_{p_0}}) +\epsilon a_{\lambda_{0}} A_{\lambda_{0}}(\mathit{R}_{\theta_{0}}[\mathbf{x}])$ & $1\times N_{pix}$& Target image at $(\lambda_0,t_0)$\\
$R_{\lambda,t}(\mathbf{x})$ & $S_{\psi_{\lambda,t}}(\frac{\mathbf{x}}{\lambda_0}) +\epsilon a A_{\lambda}(\mathit{R}_{\theta_t}[\mathbf{x} \frac{\lambda}{\lambda_0}])$ & $1\times N_{pix}$ & k th image in the reference PSF library for the target image\\
$\mathbf{R}(\mathbf{x})$ & $\left[\begin{array}{c} R_{\lambda_1,t_1}(\mathbf{x})\\ ... \\ R_{\lambda_k,t_k}(\mathbf{x})\\... \\
R_{\lambda_{N_{\mathcal{R}}},t_{N_{\mathcal{R}}}}(\mathbf{x}) \end{array}\right]$ &$N_{\mathcal{R}} \times N_{pix}$ & Matrix of the concatenated references in the PSF library\\
$P_{\lambda,t}(\mathbf{x}) $& & $1 \times N_{pix}$& Processed image at wavelength $\lambda$ and time $t$\\
$\mathbf{R}$ & $\mathbf{S} + \epsilon \mathbf{a} A_{\delta}$& $N_{\mathcal{R}} \times N_{pix}$ & Short-handed notation for $R_{k}(\mathbf{x}) $\\
$\mathbf{A_{\delta}}(\mathbf{x}) $ & $ \left[\begin{array}{c}  A_{\lambda_1}(\mathit{R}_{\theta_{t_1}}[\mathbf{x} \frac{\lambda_1}{\lambda_0}])\\ ... \\ A_{\lambda_k}(\mathit{R}_{\theta_{t_k}}[\mathbf{x} \frac{\lambda_k}{\lambda_0}]) \\... \\  A_{\lambda_{N_{\mathcal{R}}}}(\mathit{R}_{\theta_{t_{N_{\mathcal{R}}}}}[\mathbf{x} \frac{\lambda_{N_{\mathcal{R}}}}{\lambda_0}])\end{array}\right] $ & $N_{\mathcal{R}}\times N_{pix}$ & Matrix of the concatenated $N_{\mathcal{R}}$ astrophysical images in the reference library, at wavelength $\lambda_k$  that have been rotated by $\theta_t$ due to ADI field rotation and rescaled to wavelength $\lambda_0$\\
$\mathbf{A_{\delta}}$ & & $N_{\mathcal{R}}\times N_{pix}$ & Short-handed notation for $\mathbf{A_{\lambda_{p_0},\delta}}(\mathbf{x}) $ \\
$Z_{k}(\mathbf{x}) $& & $1 \times N_{pix}$& KL modes of speckles in the $\mathcal{S}$ zone \\
$\mathbf{Z}(\mathbf{x}) $ & $\left[\begin{array}{c} Z_1(\mathbf{x})\\ ... \\Z_k(\mathbf{x}) \\... \\  Z_{N_{\mathcal{R}}}(\mathbf{x})\end{array}\right] $ &$N_{\mathcal{R}}\times N_{pix}$ & Matrix of the concatenated  KL modes associated with speckles\\\
$\Delta Z_{k}(\mathbf{x}) $& & $1 \times N_{pix}$& Perturbation of speckles' KL modes due to astrophysical signal in the reference library\\
$\mathbf{\Delta Z_k}^{\lambda}  (\mathbf{x})$ & $\left[\begin{array}{c} \Delta Z_k^{\lambda_1}  (\mathbf{x}) \\ ... \\\Delta Z_k^{\lambda_p}  (\mathbf{x}) \\... \\  \Delta Z_k^{\lambda_{N_{\lambda}}}  (\mathbf{x})\end{array}\right] $ & $N_{\lambda} \times N_{pix}$ &  Perturbation of the speckles' KL modes decomposed as a function of wavelength.\\ 
$Y_{k}(\mathbf{x}) $& $Z_{k}(\mathbf{x}) +\epsilon \Delta Z_{k}(\mathbf{x}) $ & $1 \times N_{pix}$& KL modes of the Instrument PSF  perturbed by astrophysical signal   = KL modes of the actual data\\
$\mathbf{Y}(\mathbf{x}) $ & $\left[\begin{array}{c} Y_1(\mathbf{x})\\ ... \\Y_k(\mathbf{x}) \\... \\  Y_{N_{\mathcal{R}}}(\mathbf{x})\end{array}\right] $ &$N_{\mathcal{R}}\times N_{pix}$ & Matrix of the concatenated  KL modes calculated based on the data (contains perturbation from astrophysical source)\\
$\widehat{\Delta Y_{k}}(\mathbf{x}) $& & $1 \times N_{pix}$& Perturbation of the $Y_{k}$'s due to a negative synthetic source \\
$ \widehat{\mathbf{\Delta Y_k}^{\lambda} } (\mathbf{x})$ & $\left[\begin{array}{c} \widehat{\Delta Y_k^{\lambda_1}}  (\mathbf{x}) \\ ... \\ \widehat{\Delta Z_k^{\lambda_p}}  (\mathbf{x}) \\... \\  \widehat{\Delta Z_k^{\lambda_{N_{\lambda}}} } (\mathbf{x})\end{array}\right] $ & $N_{\lambda} \times N_{pix}$ &  Perturbation of the speckles' KL modes decomposed as a function of wavelength.\\ 
$\mathbf{F}_{\lambda,t} (\mathbf{x})$& see text, too ugly & $N_{\lambda} \times N_{pix}$& Model of the astrophysical source propagated through the data analysis algorithm at $(\lambda,t)$, decomposed as a function of wavelength\\
\hline
\multicolumn{4}{c}{Eigenvalues, eigevectors, covariance matrices }\\
\hline
$\mathbf{C_{SS}}$ & $\mathbf{S}\mathbf{S}^{T}$ & $N_{\mathcal{R}} \times N_{\mathcal{R}}$ & Covariance of the speckles\\
$\Lambda_k$ & & $1 \time 1$ & $k$ th eigenvalue of $\mathbf{C_{SS}}$\\
$V_k$ &  $\left[\begin{array}{c}  v_k[1]\\ ... \\ v_k[m] \\... \\  v_k[N_{\mathcal{R}}]\end{array}\right]  $ & $N_{\mathcal{R}} \time 1$ & $k$ th eigenvector of $\mathbf{C_{SS}}$\\
$\mathbf{V_k}$  & $ \left[ \begin{array}{ccccc} v_k[1]  & ... & 0 & ... & 0 \\... & ... &  0 & ... & ... \\0 & 0 &  v_k[m] & 0 & 0 \\... & ... & 0 & ... & ... \\0 & ... & 0 & 0 &   v_k[N_{\mathcal{R}}] \end{array}\right] $  &  $ N_{\mathcal{R}} \times N_{\mathcal{R}} $ & Elements of the  $k$ th eigenvector of $\mathbf{C_{SS}}$ arranged on the diagonal of a square matrix\\
$\mathbf{V}$  & $\left[ \begin{array}{ccccc} V_1, ... , V_{k}, ... , V_{N_{\mathcal{R}}} \end{array} \right]$ &  $ N_{\mathcal{R}} \times N_{\mathcal{R}} $ & Eigenvector of $\mathbf{C_{SS}}$ concatenated to build a square matrix\\
$\mathbf{C_{RR}}$ & $\mathbf{R}\mathbf{R}^{T}$ & $N_{\mathcal{R}} \times N_{\mathcal{R}}$ & Covariace matrix of the reference library\\
$\mathbf{C_{A_{\delta} S}}$ & $\mathbf{a} \mathbf{A_{\delta}} \mathbf{S}^{T} +    \mathbf{S} \mathbf{A_{\delta}}^{T}   \mathbf{a}^{T}$ & $ N_{\mathcal{R}} \times N_{\mathcal{R}} $ & Cross term between speckles and  astrophysical signal \\
$\Gamma_k$ & & $1 \time 1$ & $k$ th eigenvalue of $\mathbf{C_{RR}}$ as defined by the perturbation analysis of this paper\\
$U_k$ &  $\left[\begin{array}{c}  u_k[1]\\ ... \\ u_k[m] \\... \\  u_k[N_{\mathcal{R}}]\end{array}\right]  $ & $N_{\mathcal{R}} \time 1$ & $k$ th eigenvector of $\mathbf{C_{RR}}$ as defined by the perturbation analysis of this paper\\
\hline
\multicolumn{4}{c}{Miscellaneous linear algebra} \\
\hline
$\mathbb{I}_{N_{\mathcal{R}}}$ & $ \left[ \begin{array}{ccccc}1  & ... & 0 & ... & 0 \\... & ... &  0 & ... & ... \\0 & 0 &  1 & 0 & 0 \\... & ... & 0 & ... & ... \\0 & ... & 0 & 0 &   1 \end{array}\right] $ & $ N_{\mathcal{R}} \times N_{\mathcal{R}} $ & Identity matrix of size $ N_{\mathcal{R}}$\\
$\mathbf{L}$ & $\left[ \begin{array}{ccc} \mathbb{I}_{N_{\lambda}}... \mathbb{I}_{N_{\lambda}}\end{array} \right]$ &  $\begin{array}{c} N_{\lambda} \times \\ (N_{\lambda} N_{Exp}) \end{array} $ & Rectangular matrix that relates each slice of each exposure to the its wavelength \\
 $ \mathbf{Sel}$ & $\mathbf{S}[i,j] = 1$ if $ E_{p(j),q(j)}$  & $ \begin{array}{c}  N_{\mathcal{R}}\times \\ (N_{Exp}  N_{\lambda})  \end{array} $ & Selection matrix that relates each slice of each exposure to its position in the reference library (note it depends on algorithm parameters  ( $\mathbf{S_{(p_0,q_0)}}^{(N \delta_{\theta}, N \delta_{\lambda}^{+}, N \delta_{\lambda}^{-}, N_{Corr})}$)\\
 $ \mathbf{Sel_{\lambda}}$ & $\mathbf{L}\;\mathbf{Sel}^{T}$ & $N_{\lambda} \times N_{\mathcal{R}}$ &  Selection matrix that relates each slice in the reference library to its wavelength \\
\end{longtable}
\end{center}

\clearpage

\section{Appendix C: Forward modeling in the case of RDI}

In this appendix we provide a detailed description of our mathematical notations and Forward Modeling implementation in the case of RDI, in configurations for which there is no astrophysical signal in the PSF library. Most of the material in this appendix has already been discussed in \citet{2014arXiv1409.6388P}, albeit in somewhat less detail. We revisit it here in order to provide a rigorous context for the latter introduction of the KLIP-FM algorithm in the general case of ADI and/or SSDI. 

\subsubsection{Basic principle of Forward Modeling}

Forward Modeling  in the context of exoplanet imaging was first proposed by \citet{mmv10} and aims at jointly estimating the instrument response and the astrophysical signal. To do so, negative synthetic sources are injected in the raw data across the entire observing sequence. This new data set, with both positive astrophysical and negative synthetic signals, is then propagated through the reduction algorithm. Jointly minimizing the residuals in such processed images (by exploring the range of possible astrophysical properties for the synthetic negative sources) retrieves in principle the properties of the astrophysical signal. We call $\widehat{ \mathcal{A}}$ the ensemble of estimated astrophysical observables  $\widehat{ \mathcal{A}} = \{ \widehat{\epsilon}, \widehat{\mathbf{x_A}},  \widehat{a_{\lambda_1}}, ..., \widehat{a_{\lambda_{N_{\lambda}}}}, \widehat{A_{\lambda_1}(\mathbf{x})}, ..., \widehat{A_{\lambda_{N_{\lambda}}}(\mathbf{x})} \}$. These are the quantities corresponding to the synthetic negative astrophysical signal injected in the data, while $ \mathcal{A}$ are the quantities corresponding to the actual signal. This notation covers the most general case  (i.e., resolved source, not centered on the star, and whose morphology changes with wavelength). Of course in practice one never faces such a challenge and the dimensionality of astrophysical estimates is much smaller. We can then write the Forward Modeling problem at a given wavelength $\lambda_0$ and time $t_0$ as the following minimization:
\begin{equation}
\min_{\widehat{ \mathcal{A}}}  || \mathcal{LSQ}_{\mathcal{R}(\mathcal{A},\widehat{ \mathcal{A}}) } \left[ T(\mathbf{x})  - \widehat{\epsilon} \widehat{a_{\lambda_{0}}} \widehat{A_{\lambda_{p_0}}} (\mathit{R}_{\theta_{0}}[\mathbf{x}] )\right] ||_{\mathcal{F}}^2  
\label{Eq:BasicForwardModelling}
\end{equation}
where $\mathcal{LSQ}_{\mathcal{R}(\mathcal{A},\widehat{ \mathcal{A}})}$ describes a most general and generic least-squares speckle fitting algorithm (using  reference images that depend both on the astrophysical observables, $\mathcal{A}$, and their synthetic negative counterparts, $\widehat{ \mathcal{A}}$ ).  $\mathcal{F}$ is a ``fit'' region of the field of view that does not necessarily correspond to the $\mathcal{S}$ zone. In the context of this paper, we assume that  $\mathcal{LSQ}_{\mathcal{R}(\mathcal{A},\widehat{ \mathcal{A}}) }$ corresponds to $\mathcal{KLIP}_{\mathcal{R}(\mathcal{A},\widehat{ \mathcal{A}}) }$ described in \S \ref{Ref:PCABasic}, with reference images chosen according to the rules described in \S \ref{Ref:SelCrit}. Note that in practice, images from a sequence of exposures (and/or wavelengths) are co-added before the signal is estimated, and as a consequence Eq.~\ref{Eq:BasicForwardModelling} is only representative of realistic cases up to one or two summations. However for the sake of clarity we will present our work without these summations and only discuss them when outlining practical implementations for spectral extraction in Appendix F. 

\subsubsection{RDI of a point source}

We review here applications of Eq.~\ref{Eq:BasicForwardModelling} to the case of RDI. In this configuration the signal can be {\em over-subtracted}, that is, the image of a point source can be fitted using some combination of instrument noise realizations (e.g it can be over-subtracted by a speckle at the same exact location in the reference library). In this case the reference library does not depend on the astrophysical ($\mathcal{A}$) or on the synthetic negative ($\widehat{ \mathcal{A}}$) signals. The least-squares speckles fitting algorithm can then be written as an operator on any arbitrary image $I(x)$:  $\mathcal{KLIP}_{\mathcal{R}(\mathcal{A},\widehat{ \mathcal{A}}) } = \mathcal{KLIP}_{\mathcal{R}}[I (x)] = I(x)   - \sum_{k = 1}^{K_{Klip}} <I(x) ,Z_k(\mathbf{x})>_{\mathcal{S}}  Z_k(\mathbf{x})$. Here the simplification $\mathcal{R}(\mathcal{A},\widehat{ \mathcal{A}}) = \mathcal{R}$ is key because it alleviates all issues associated with self-subtraction: only over-subtraction remains. Assuming that the morphology of the PSF  is known (e.g $\widehat{A}(\mathbf{x}) = A(\mathbf{x}) = PSF(\mathbf{x})$), and that is not field dependent, then there are only three unknowns: the photometry over the bandwidth of interest $\epsilon$ and the location of the point source (astrometry) in the scene $\widehat{\mathbf{x_A}}$. We call $(\tilde{\epsilon},\mathbf{\tilde{x}_A})$, the values of $(\widehat{\epsilon}, \widehat{\mathbf{x_A}})$ that actually minimize Eq.~\ref{Eq:BasicForwardModelling}. Under these assumptions the Forward Modeling problem, Eq.~\ref{Eq:BasicForwardModelling} reduces to the following minimization:
%
\begin{equation}
 (\tilde{\epsilon},\mathbf{\tilde{x}_A})  = \arg \min_{(\widehat{\mathbf{x_A}},\widehat{\epsilon})} || \mathcal{LSQ}_{\mathcal{R}}[T (x)]- \left( \widehat{\epsilon} A(\mathbf{x} - \widehat{\mathbf{x_A}})  - \sum_{k = 1}^{K_{Klip}} <\widehat{\epsilon} A(\mathbf{x} - \widehat{\mathbf{x_A}})  ,Z_k(\mathbf{x})>_{\mathcal{S}}  Z_k(\mathbf{x}) \right)||_{\mathcal{F}}^2
 \label{Eq:RefFreeForwardModelling}
\end{equation}
%
%
where Eq.~\ref{Eq:RefFreeForwardModelling} means that because  the PSF library (and thus the Karuhnen-Loeve modes) does not contain any astrophysical signal, estimating the astrophysical observables of a detected point source can occur in processed image space.  This implies that one does not need to reprocess the data for every evaluation of the Forward Modeling cost function (e.g every value of $(\tilde{\epsilon},\mathbf{\tilde{x}_A})$ that is hypothesized while iterating to find the global minimum of Eq.~\ref{Eq:RefFreeForwardModelling}). In particular,  the CPU intensive matrix inversion associated with the determination of the Principal Components is only carried out once. Moreover Eq.~\ref{Eq:RefFreeForwardModelling} also provides insights regarding the signal estimation algorithm. Indeed, the  quantity that is being minimized can be decomposed into three terms:
\begin{itemize}
\item A noise term that represents the remaining speckle noise that has not been captured by the PCA:
\begin{equation}
 P_{spe}(\mathbf{x}) = S_{\psi_{0}}(\mathbf{x})  - \sum_{k = 1}^{K_{Klip}} <S_{\psi_{0}}(\mathbf{x}),Z_k(\mathbf{x})>_{\mathcal{S}}  Z_k(\mathbf{x})
\end{equation}
Algorithms such as LOCI or KLIP are aimed at changing the PDF of this noise from spatially-correlated+ Rician in the raw data \citep{sa04}, to spatially-uncorrelated+Gaussian \citep{marois08} (with the smallest standard deviation). 
\item A term capturing  the difference between the astrophysical signal $ \epsilon  A(\mathbf{x} - \mathbf{x_A}) $ and the negative synthetic source, $\widehat{\epsilon}   A(\mathbf{x} - \widehat{\mathbf{x_A}})$,  both are propagated through the Principal Component Analysis:
\begin{equation}
\left( \epsilon  A(\mathbf{x} - \mathbf{x_A})   - \widehat{\epsilon}   A(\mathbf{x} - \widehat{\mathbf{x_A}}) \right) +\sum_{k = 1}^{K_{Klip}} <  \epsilon  A(\mathbf{x} - \mathbf{x_A})   - \widehat{\epsilon}   A(\mathbf{x} - \widehat{\mathbf{x_A}}), Z_k(\mathbf{x})>_{\mathcal{S}}  Z_k(\mathbf{x}) 
\end{equation}
\end{itemize}
Plugging these expressions into Eq.~\ref{Eq:RefFreeForwardModelling} yields: 
\begin{equation}
(\tilde{\epsilon},\mathbf{\tilde{x}_A})  = \arg \min_{(\widehat{\mathbf{x_A}},\widehat{\epsilon})} || P(\mathbf{x}) - \widehat{\epsilon} \mathbf{G}(\mathbf{x} - \widehat{\mathbf{x_A}}) ||^2,
\label{Eq:ForwardVerySimple}
\end{equation}
where $\widehat{\epsilon} \mathbf{G}(\mathbf{x} - \widehat{\mathbf{x_A}}) $ corresponds to the negative synthetic source propagated through KLIP. We have thus established that Forward Modeling with RDI simply consists of inverting the ``transfer function'' of the algorithm: namely solving for $(\widehat{\epsilon}, \widehat{\mathbf{x_A}})$ in the least-squares sense. When the algorithm parameters have been well chosen, the noise term $P_{spe}(\mathbf{x})$ is indeed zero mean, gaussian, and does not feature spatial correlations (see\cite{marois08} for in-depth discussions regarding these hypothesis). On the other hand, when these parameters are ill chosen, then some systematics may remain and residual speckle noise may bias the estimation of $(\widehat{\epsilon}, \widehat{\mathbf{x_A}})$. However it is important to note that such biases stem from the speckle noise not being properly subtracted; they are not due to over-subtraction. They can be reduced by choosing a fitting zone, $\mathcal{F}$ over which it has been empirically determined that PSF subtraction residuals are zero mean and do not feature spatial correlations. Alternatively, one can explore other algorithm parameters such as geometries of the $\mathcal{S}$ zone. The example in Figure~\ref{fig:ForwardRDI} shows that this well-behaved regime can be achieved by simply increasing $K_{Klip}$. Here we do not discuss algorithmic parametric searches aimed at minimizing the residual speckle noise, and work under the assumption that  this latter source of uncertainty is well behaved. Appendix D shows how one can solve Eq.~\ref{Eq:ForwardVerySimple} analytically, without any numerical optimization, under the assumption that $\mathcal{F} = \mathcal{S} $. This yields:
 \begin{eqnarray}
&&\mathbf{\tilde{x}_A} =  \arg \max_{\widehat{\mathbf{x_A}}}  <P(\mathbf{x}),A(\mathbf{x} -\widehat{\mathbf{x_A}})>_{\mathcal{S}} \label{Eq::RefFreeForwardModellingAstro}\\
&& \tilde{\epsilon}= \frac{ <P(\mathbf{x}),A(\mathbf{x} -\mathbf{\tilde{x}_A})>_{\mathcal{S}} } {||A(\mathbf{x})||_{\mathcal{S}}^2- \sum_{k = 1}^{K_{Klip}} < A(\mathbf{x} - \mathbf{\tilde{x}_A})  ,Z_k(\mathbf{x})>_{\mathcal{S}}^2}.
\label{Eq::RefFreeForwardModellingPhoto}
 \end{eqnarray}
 %
Eq.~\ref{Eq::RefFreeForwardModellingAstro} implies that  PCA-based algorithms  (on the RDI case) do not bias the astrometry of a point source (under the well-behaved residual speckles assumption).
Eq.~\ref{Eq::RefFreeForwardModellingPhoto} can then be used for photometric estimation or to estimate algorithmic throughput when calculating detection limits. 
\subsection{Calculating correlations with Fourier Transforms}
Note that in practice, this algorithm can be implemented very efficiently at the subpixel precision level using Fourier Transforms. Indeed, the correlation between the two images $I_1(\mathbf{x})$ and $I_2(\mathbf{x})$ over $\mathcal{S}$ can be written as:
 \begin{equation}
<I_1(\mathbf{x}),I_2(\mathbf{x} - \widehat{\mathbf{x_A}})>_{\mathcal{S}} = MFT \left \{  FFT[m_{\mathcal{S}} I_1(\mathbf{x}) ] FFT[m_{\mathcal{S}} I_2(\mathbf{x}) ] \right \} (\widehat{\mathbf{x_A}})
 \label{Eq:FourierShit}
 \end{equation}
 where $m_{\mathcal{S}}$ denotes a mask over the image that is zero everywhere but in  $\mathcal{S}$,  FFT is a fast Fourier Transform and  MFT is the Matrix Fourier Transform, discussed in \cite{2007OExpr..1515935S}, calculated over a subpixel grid of $\widehat{\mathbf{x_A}}$ centered around an initial guess of the position of the detected point source. As a consequence, the entire 2D grid of correlations necessary to solve for the location of the point source  can be calculated without resorting to CPU expensive loops. 

\clearpage

\section{Appendix D: Linear algebra notations underlying astrometry and photometry of a point source with KLIP-FM}
This Appendix details the linear algebra associated with the derivation of Eq.~\ref{Eq::RefFreeForwardModellingPhoto} and Eq.~\ref{Eq::RefFreeForwardModellingAstro}. It also sets up the stage for the mechanics underlying the spectral estimation algorithm discussed in Appendix F. 

\subsection{General Case}
We consider the general case of minimizing the following Forward Modeling cost function:
\begin{equation}
 \arg \min_{(f_0, \mathbf{x_0})} || b(\mathbf{x}) - f_0^T \mathbf{G} (\mathbf{x} - \mathbf{x_0})||^2
 \label{Eq:ForwardLS}
\end{equation}
where $f_0$ is a $N_{\lambda} \times 1$ column vector, $\mathbf{G} (\mathbf{x} - \mathbf{x_0})$ is a  $N_{\lambda} \times N_{pix}$ matrix, and $b(\mathbf{x})$ is a $1 \times N_{pix}$ line vector. This is the most general case for point sources whose both position,  spectrum and astrometry are being estimated. Then, the pair $(\widetilde{f_0}, \widetilde{\mathbf{x_0}})$ is a solution to this least-squares minimization problem if and only if:
\begin{equation}
\widetilde{f_0} = \arg \min_{f_0} || b(\mathbf{x}) - f_0^T \mathbf{G} (\mathbf{x} - \widetilde{\mathbf{x_0}})||^2
\label{Eq:minPhot}
\end{equation}
 and
 \begin{equation}
\widetilde{\mathbf{x_0}} = \arg \min_{x_0} || b(\mathbf{x}) - \widetilde{f_0}^T \mathbf{G} (\mathbf{x} - \mathbf{x_0})||^2
\label{Eq:minAstro}
\end{equation}
Using the shorthanded notation $\mathbf{G}_{\widetilde{\mathbf{x_0}}} = \mathbf{G} (\mathbf{x} - \widetilde{\mathbf{x_0}})$, we can write Eq.~\ref{Eq:minPhot} as:
\begin{equation}
\widetilde{f_0} = \arg \min_{f_0} \left( b b^T + f_0^T \mathbf{G}_{\widetilde{\mathbf{x_0}}} \mathbf{G}_{\widetilde{\mathbf{x_0}}}^T  f_0 - 2  f_0^T \mathbf{G}_{\widetilde{\mathbf{x_0}}} b^T \right).
\end{equation}
Taking the first derivative with respect to $f_0$ yields:
\begin{equation}
\mathbf{G}_{\widetilde{\mathbf{x_0}}} \mathbf{G}_{\widetilde{\mathbf{x_0}}}^T  \widetilde{f_0}  = \mathbf{G}_{\widetilde{\mathbf{x_0}}} b^T,
\label{Eq:FinalPhot}
\end{equation}
 which can be substituted into Eq.~\ref{Eq:minAstro} and yield after simplification:
 \begin{equation}
 \widetilde{\mathbf{x_0}}  = \arg \max_{x_0} \widetilde{f_0}^T\mathbf{G} (\mathbf{x} - \mathbf{x_0}) b^T.
 \label{Eq:FinalAstro}
 \end{equation}
Thus when seeking to minimize a quadratic cost function of a functional form similar to Eq.~\ref{Eq:ForwardLS} the spatial offset $\mathbf{x_0}$ can thus be first calculated by maximizing the cross-correlation in Eq.~\ref{Eq:FinalAstro}, assuming a first guess for the spectrum. Based on this value of $\mathbf{x_0}$ one can update $\widetilde{f_0}$ using Eq.~\ref{Eq:FinalPhot} and iterate until astrometry and photometry have converged. 
\subsection{Case of RDI}
In the case of Eq.~\ref{Eq:RefFreeForwardModelling} we can write:
\begin{eqnarray}
 b(\mathbf{x}) =P(\mathbf{x})  &=& P_{spe}(\mathbf{x}) +  \epsilon (   A(\mathbf{x} - \mathbf{x_A})   -  \sum_{k = 1}^{K_{Klip}} <    A(\mathbf{x} - \mathbf{x_A}) , Z_k(\mathbf{x})>_{\mathcal{S}}  Z_k(\mathbf{x}) )\\
f_0   &=& \widehat{\epsilon}\\
\mathbf{x_0} &=& \widehat{\mathbf{x_A}}\\
\mathbf{G} (\mathbf{x} - \mathbf{x_0}) &=& A(\mathbf{x}- \mathbf{x_0})  - \sum_{k = 1}^{K_{Klip}} <   A(\mathbf{x} - \mathbf{x_0}) , Z_k(\mathbf{x})>_{\mathcal{S}}  Z_k(\mathbf{x})
\end{eqnarray}
We also work under the assumption that $\mathcal{F} = \mathcal{S}$ (i.e., the zone over which the signal is estimated is the same as the one over which the principal components are calculated). In this case:
\begin{eqnarray}
 && < P(\mathbf{x}), \sum_{k = 1}^{K_{Klip}} <  \epsilon  A(\mathbf{x} ) , Z_k(\mathbf{x})>_{\mathcal{S}}  Z_k(\mathbf{x})>_{\mathcal{S}} = 0\\
&& ||\sum_{k = 1}^{K_{Klip}} <  \epsilon  A(\mathbf{x} ) , Z_k(\mathbf{x})>_{\mathcal{S}}  Z_k(\mathbf{x}) ||^2  =  \sum_{k = 1}^{K_{Klip}} <  \epsilon  A(\mathbf{x} ) , Z_k(\mathbf{x})>_{\mathcal{S}}^2, 
\end{eqnarray}
 Eqs.~\ref{Eq::RefFreeForwardModellingAstro} and \ref{Eq::RefFreeForwardModellingPhoto} can be directly derived from estimating the astrometry using  Eq.~\ref{Eq:FinalAstro} and then plugging this estimate into  Eq.~\ref{Eq:FinalPhot}.

\clearpage

\section{Appendix E: analytical propagation of the astrophysical signal through a PCA}
\label{sect:AnalExpansion}

This appendix describes the derivation of the main theoretical result of this paper.

\subsection{Linear Expansion of the covariance matrix}
The expression of a target image and its associated references are:
\begin{eqnarray}
&& T(\mathbf{x}) =  S_{\psi_{\lambda_0,t_0}}(\frac{\mathbf{x}}{\lambda_{p_0}}) +\epsilon a_{\lambda_{0}} A_{\lambda_{0}}(\mathit{R}_{\theta_{0}}[\mathbf{x}]) \\
&& R_{\lambda,t}(\mathbf{x})=S_{\psi_{\lambda,t}}(\frac{\mathbf{x}}{\lambda_0}) +\epsilon a A_{\lambda}(\mathit{R}_{\theta_t}[\mathbf{x} \frac{\lambda}{\lambda_0}])
\end{eqnarray}
where the references have been chosen among all images in the observing sequence according to the parameters $(N \delta_{\theta}, N \delta_{\lambda}^{+}, N \delta_{\lambda}^{-}, N_{Corr},N_{r},N_{\phi})$ . We drop this dependence for simplicity and use the following shorthand notations:
\begin{itemize}
\item $\mathbf{R}(\mathbf{x}) $ is the $N_{\mathcal{R}}$ by $N_{Pix}$ matrix whose $k$ th line entry is $R_{k}(\mathbf{x}) =R_{\lambda_k,t_k}(\mathbf{x})$. Note that each line entry of $\mathbf{R}(\mathbf{x}) $  is zero mean over $\mathcal{S}$.
\item $\mathbf{S}(\mathbf{x}) $ is the $N_{\mathcal{R}}$ by $N_{Pix}$ matrix whose $k$ th line entry is $S_{k}(\mathbf{x}) =S_{\lambda_k,t_k}(\mathbf{x})$. Note that each line entry of $\mathbf{S}(\mathbf{x}) $  is zero mean over $\mathcal{S}$.
\item $\mathbf{a}$ the $N_{\mathcal{R}}$ diagonal matrix whose $k$ th diagonal entry is $a_{\lambda_{k}}$ (the normalized astrophysical flux at the wavelength corresponding to the $k$ th reference).
\item $\mathbf{A_{\delta}}(\mathbf{x}) $ is the $N_{\mathcal{R}}$ by $N_{Pix}$ matrix whose $k$ th line entry is $A_{\delta_k} (\mathbf{x})=  A_{\lambda_k}(\mathit{R}_{\theta_{k}}[\mathbf{x} \frac{\lambda_k}{\lambda_0}])$. Note that each line entry of $\mathbf{A_{\delta}}(\mathbf{x}) $  is zero mean over $\mathcal{S}$. In the case of a point source and neglecting the field dependence of the PSF, this can be further simplified as:
\begin{equation}
A_{\lambda_k}(\mathit{R}_{\theta_{k}}[\mathbf{x} \frac{\lambda_k}{\lambda_0}] )= PSF_{\lambda_k}(  \mathbf{x}  - \mathbf{x_\mathcal{S}} - \mathbf{ \delta^{(\lambda_{0},t,\lambda)} x_\mathcal{S}}) 
\end{equation}

\end{itemize}
In this framework the reference library can be written in a matrix form:
\begin{equation}
\mathbf{R}(\mathbf{x}) = \mathbf{S}(\mathbf{x}) + \epsilon \mathbf{a} \mathbf{A_{\delta}}(\mathbf{x}) . 
\end{equation}
When using PCA-based algorithms the next step is to calculate the Karhune Loeve transform of  this ensemble of references (e.g. Eq.~\ref{Eq:KLdefinition}). Our  goal is to evaluate how the signal in the references --e.g $\epsilon \mathbf{a} \mathbf{A_{\delta}}(\mathbf{x}) $-- propagates through this Principal Component decomposition. To do so, we write the covariance matrix in the presence of an astrophysical signal as the sum of the speckles covariance and the cross-term between the astrophysical signal and the speckle noise:
%
\begin{eqnarray}
\mathbf{C_{RR}}  &=&  \mathbf{R}(\mathbf{x}) \; \mathbf{R}(\mathbf{x})^{T} \nonumber \\
\mathbf{C_{RR}}  & =& \mathbf{S}(\mathbf{x}) \; \mathbf{S}(\mathbf{x})^{T} + \epsilon \; \mathbf{a} \mathbf{A_{\delta}}(\mathbf{x})  \mathbf{S}(\mathbf{x})^{T}   +\epsilon \;  \mathbf{S}(\mathbf{x}) \mathbf{A_{\delta}}(\mathbf{x}) ^{T}   \mathbf{a}^{T}+ \epsilon^2 \mathbf{a} \mathbf{A_{\delta}}(\mathbf{x})  \mathbf{A_{\delta}}(\mathbf{x}) ^{T} \mathbf{a}^T \nonumber \\
\mathbf{C_{RR}}  & =& \mathbf{C_{SS}}+  \epsilon \; \mathbf{C_{A_{\delta} S}}+ \mathcal{O}(\epsilon^2).
\label{Eq:ApproxLinCov}
\end{eqnarray}
Note that here we dropped the $1/\sqrt{N_{Corr}-1}$ factor in front of the covariance matrix. Thus the presence of an astrophysical signal in the PSF library becomes a quadratic (scaling as $\epsilon^2$) perturbation of the reference's covariance matrix:  this non-linearity, associated with the eigenmodes truncation, is the source of  the {\em self-subtraction biases} for astrophysical estimates (over-subtraction occurs regardless even when $\epsilon = 0$ in the references).  However, when: 
\begin{equation}
\epsilon^2 \mathbf{a} \mathbf{A_{\delta}}(\mathbf{x})  \mathbf{A_{\delta}}(\mathbf{x}) ^{T} \mathbf{a}^T  \ll \epsilon \; \mathbf{C_{A_{\delta} S}},
\end{equation}
is true for each entry in these matrice, then the dependence on the astrophysical signal becomes linear and it can be modeled a posteriori  in a tractable fashion. This argument is the crux of the analysis presented in this paper. This inequality  is true when: 
\begin{equation}
 \epsilon \ll \max \left ( \frac{\mathbf{\Lambda} [\mathbf{C_{A_{\delta} S}}]} {\mathbf{\Lambda} [\mathbf{a} \mathbf{A_{\delta}}(\mathbf{x})  \mathbf{A_{\delta}}(\mathbf{x}) ^{T} \mathbf{a}^T] }\right) \label{Eq:MainApprox}
 \end{equation}
 where $\mathbf{\Lambda} [M]$ denotes the operator that calculates the eigenvalue of a matrix $M$. As a consequence, the linear approximation holds either when the astrophysical signal $\epsilon$ is small with respect to the local speckles {\em or} when the algorithm parameters are chosen so that the correlations of astrophysical images across the PSF library --$\mathbf{a} \mathbf{A_{\delta}}(\mathbf{x})  \mathbf{A_{\delta}}(\mathbf{x}) ^{T} \mathbf{a}^T$-- have much smaller eigenvalues than the cross-term between the astrophysical signal and the speckle noise --$\mathbf{C_{A_{\delta} S}}$. This latter case occurs when the algorithm parameters are chosen not to be aggressive ($N \delta_{\theta}, N \delta_{\lambda}^{+}, N \delta_{\lambda}^{-}$  larger than the FWHM of a point source PSF for instance). Note however that while this condition is necessary (e.g when it is true the linear approximation holds), it is not sufficient: there exist cases where this inequality is not strictly true but where the linear model holds. In practice it is preferable to use a metric based on numerical evaluation of the eigen-modes or eigenvalues such as the one presented on Fig.~\ref{fig:FigureEigenValues}. The case of small $\epsilon$  is of most interest in the framework of high-contrast imaging. Indeed when a source is brighter than the speckles, PCA-based algorithm might not be necessary (or can be tuned so that $N \delta_{\theta}, N \delta_{\lambda}^{+}, N \delta_{\lambda}^{-}$  is large enough), and thus the sophistications presented in this manuscript can be circumvented.\\

\subsection{Perturbed Eigenpair of the covariance matrix}

We remind the reader that $V_k = \left[ v_{k}[1] ... v_{k}[N_{\mathcal{R}}] \right]$  are the eigenvectors of $\mathbf{C_{SS}}$ and $\Lambda_k$ its eigenvalues (see discussion associated with Eq.~\ref{Eq:KLdefinition}).  Under the linear approximation described by Eq.~\ref{Eq:ApproxLinCov} we now propagate the astrophysical signal in the PSF library through the calculation of eigenvalues/vectors of the covariance matrix. This identity is a standard linear algebra result, however because it is the cornerstone of KLIP-FM, we recall here the main steps of its derivation. We seek to express the perturbed eigenpair $\Gamma_k, U_{k}$ of $\mathbf{C_{SS}} + \epsilon \mathbf{C_{A_{\delta} S}}$ as:
\begin{eqnarray}
\Gamma_k  &=&  \Lambda_k + \epsilon \delta \Lambda_k  \\
U_k &=& V_k + \epsilon \sum_{p = 1}^{N_{\mathcal{R}}} c_{k,p} V_p 
\end{eqnarray}
This implies that:
\begin{eqnarray}
\left( \mathbf{C_{SS}} + \epsilon \mathbf{C_{A_{\delta} S}} \right) U_k  &=& \Gamma_k  U_{k} \nonumber \\
\left( \mathbf{C_{SS}} + \epsilon \mathbf{C_{A_{\delta} S}} \right) \left( V_k + \epsilon \sum_{p = 1}^{N_{\mathcal{R}}} c_{k,p} V_p  \right)  &=& \left(\Lambda_k + \epsilon \delta \Lambda_k  \right) \left( V_k + \epsilon \sum_{p = 1}^{N_{\mathcal{R}}} c_{k,p} V_p  \right) \nonumber\\
  \epsilon \mathbf{C_{A_{\delta} S}} V_k +  \epsilon \sum_{p = 1}^{N_{\mathcal{R}}} c_{k,p}  \mathbf{C_{SS}} V_p  +\mathcal{O}(\epsilon^2) & = &  \epsilon \delta \Lambda_k V_k + \epsilon \Lambda_k \sum_{p = 1}^{N_{\mathcal{R}}} c_{k,p}   V_p +\mathcal{O}(\epsilon^2)  \nonumber \\
    \epsilon \mathbf{C_{A_{\delta} S}} V_k +  \epsilon \sum_{p = 1}^{N_{\mathcal{R}}} c_{k,p}  \Lambda_p V_p  +\mathcal{O}(\epsilon^2) & = &  \epsilon \delta \Lambda_k V_k + \epsilon \Lambda_k \sum_{p = 1}^{N_{\mathcal{R}}} c_{k,p}   V_p +\mathcal{O}(\epsilon^2) 
  \label{Eq:BasicPert}
\end{eqnarray}
where we have used $  \mathbf{C_{SS}} V_k = \Lambda_k V_k $. We neglect the terms of order $\epsilon^2$  , left multiply Eq.~\ref{Eq:BasicPert} by $V_k^T$, and use the fact that $V_p^T V_q = \delta_{p,q}$ (e.g. the eigenbasis $\{V_{k} \}_{k = 1 ... N_\mathcal{R}}$ is orthonromal). We find:
\begin{equation}
\delta \Lambda_k = V_k ^T \mathbf{C_{A_{\delta} S}} V_k 
\end{equation}
Similarly, left multiplying Eq.~\ref{Eq:BasicPert} by $V_j^T$, $ j \neq k$ yields:
\begin{eqnarray}
V_j^T \mathbf{C_{A_{\delta} S}} V_k  + c_{k,j} \Lambda_j &=&  c_{k,j} \Lambda_k \\
c_{k,j} &=& \frac{V_j^T \mathbf{C_{A_{\delta} S}} V_k }{\Lambda_k  - \Lambda_j }
\end{eqnarray}
Finally, forcing the normalization of $\{U_{k} \}_{k = 1 ... N_\mathcal{R}}$ yields $a_{k,k} = 0$. We finally find that the eigenpair of $\mathbf{C_{RR}}$ can be written as a small perturbation of eigenvalues/vectors of the signal-free reference covariance matrix:
\begin{eqnarray}
\Gamma_k &=& \Lambda_k + \epsilon \; V_k^T \mathbf{C_{A_{\delta} S}} V_k \\
U_k &=& V_k +  \epsilon \sum_{j = 1, j \neq k}^{N_{\mathcal{R}}} \frac{V_j^T \mathbf{C_{A_{\delta} S}} V_k}{ \Lambda_k  - \Lambda_j} V_j,
\label{Eq:PerturbationEigen}
\end{eqnarray}
Note that in this framework the perturbation to both $\Lambda_k $ and $V_{k}$ is a linear function of the astrophysical signal. The validity of Eq.~\ref{Eq:PerturbationEigen} depends both on the validity of the inequality in Eq.~\ref{Eq:MainApprox} and on the magnitude of the terms in $\epsilon^2$.

\subsection{Validity of the Eigenpair expansion}

\begin{figure}[t]
\begin{center}
\includegraphics[width=1\textwidth]{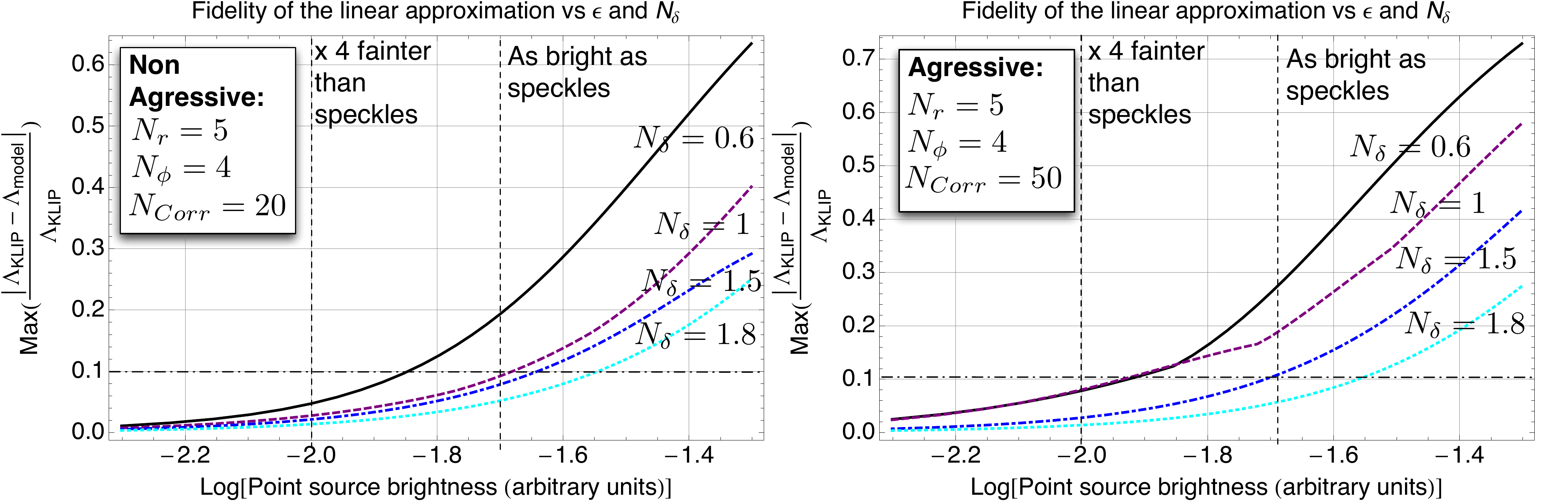} 
\caption{{\bf Using the eigenvalues of the reference correlation matrix as a proxy for the fidelity of the the linear approximation:} our metric is the largest of all possible $N_{Corr}$ relative differences between the true eigenvalues of $\mathbf{C_{RR}}$ and the ones calculated using Eq.~\ref{Eq:PerturbationEigen}. The linear approximation holds very well for small values of $\epsilon$. Making the algorithm parameters less aggressive, by increasing $N_{\delta}$ or reducing $N_{Corr}$, results in a wider range of $\epsilon$ for which the approximation is well behaved. For point source $\time 4$ fainter than the speckles, the linear approximation in  Eq.~\ref{Eq:PerturbationEigen} is always below the $10 \%$ level across all ranges of parameters tested. Similar levels of fidelity can be achieved with a point source as bright as the speckles, provided that $N_{\delta}$ is large enough (less aggressive algorithm).}
\label{fig:FigureEigenValues}
\end{center}
\end{figure}

Figure~\ref{fig:FigureEigenValues} shows how this approximation fares when changing the brightness $\epsilon$ of a synthetic point source injected in IFS coronagraph data and varying algorithm parameters. To do so we used the same public Gemini Planet Imager J-band data on the source Beta Pictoris used in the body of the manuscript. Our proxy to quantify the fidelity of the approximation is the largest of all possible $N_{Corr}$ relative differences between the true eigenvalues of $\mathbf{C_{RR}}$ and the ones calculated using Eq.~\ref{Eq:PerturbationEigen}. We find that indeed the linear approximation holds very well for small values of $\epsilon$ and that making the algorithm parameters less aggressive by increasing $N_{\delta}$ or reducing $N_{Corr}$ results in a wider range of $\epsilon$ for which the approximation is well behaved. It is also important to note that this metric is somewhat conservative. Indeed, in most cases we found that for $K_{Klip}$ smaller than $20 \%$ of the dimensionality of the reference library, the approximation holds very well even with bright point sources. Indeed when using $ K_{Klip} \gtrsim 0.2 \times N_{Corr}$) we find that in most cases, even with a point source brighter than the local speckles, our eigenvalues-based metric remains below $10 \%$.

\subsection{From perturbed Eigenpair to perturbed of the Principal Components}
Next, we propagate the linear expansion in Eq.~\ref{Eq:PerturbationEigen} into the Principal Components. This is achieved by plugging Eq.~\ref{Eq:PerturbationEigen} into Eq.~\ref{Eq:KLdefinition}: 
\begin{eqnarray}
 Y_k(\mathbf{x}) & = & \frac{1}{\sqrt{\Lambda_k}}\sum_{m=1}^{N_{\mathcal{R}}} u_k[m]  R_m(\mathbf{x}) \\
 Y_k(\mathbf{x}) & = & \frac{1}{\sqrt{\Lambda_k +  \epsilon \; V_k^T \mathbf{C_{A_{\delta} S}} V_k}} \times \nonumber\\
 &&  \sum_{m=1}^{N_{\mathcal{R}}} \left( v_k[m] +  \epsilon \sum_{j = 1, j \neq k}^{N_{\mathcal{R}}} \frac{V_j^T \mathbf{C_{A_{\delta} I}} V_k}{ \Lambda_k  - \Lambda_j} v_j[m] \right)  \times \left( S_{m}(\mathbf{x}) +\epsilon a_{\lambda_m} A_{\delta m} (\mathbf{x})
\right). 
  \label{Eq:PerturbePCAAll}
\end{eqnarray}

This expression can be simplified by replacing the definition of the unperturbed Principal Components  --e.g., Eq.~\ref{Eq:KLdefinition}-- into Eq.~\ref{Eq:PerturbePCAAll}, in order to express the $ Y_k(\mathbf{x})$ as a function of the $Z_k(\mathbf{x})$. This finally yields:

\begin{eqnarray}
 Y_k(\mathbf{x}) & = & Z_k(\mathbf{x}) + \epsilon \Delta Z_k(\mathbf{x})     \label{Eq:PerturbePCASimple} \\
\Delta Z_k(\mathbf{x})& = &  - \frac{1}{2 \Lambda_k}V_k^T \mathbf{C_{A_{\delta} S}} V_k \;  Z_k(\mathbf{x})    + \sum_{j = 1, j \neq k}^{N_{\mathcal{R}}}\sqrt{ \frac{\Lambda_j}{\Lambda_k}} \frac{V_j^T \mathbf{C_{A_{\delta} S}} V_k}{ \Lambda_k  - \Lambda_j} Z_j(\mathbf{x})  + \frac{1}{ \sqrt{\Lambda_k}}\; V_k^T \mathbf{a} \mathbf{A_{\delta}}(\mathbf{x}),   \nonumber
\end{eqnarray}

where again we have neglected the terms of $\mathcal{O}(\epsilon^2) $. Eq.~\ref{Eq:PerturbePCASimple} captures how the presence of an astrophysical signal in the reference library translates into a small perturbation of the Principal Components. When neglecting the $\mathcal{O}(\epsilon^2) $ terms, this perturbation is linear with respect the signal's photometry $\epsilon$ and/or its spectrum $f$. Here we identify the three terms discussed in \S 2.4: 
\begin{itemize}
\item {\em Over-subtraction}, scaling at $\sim 1$
 
  $Z_k(\mathbf{x}) $

 \item {\em Direct self-subtraction}, scaling as $\sim \epsilon /\sqrt{\Lambda_k}$
 
  $ \epsilon \frac{1}{ \sqrt{\Lambda_k}}\; V_k^T \mathbf{a} \mathbf{A_{\delta}}(\mathbf{x}) $
    
 \item {\em Indirect self-subtraction}, scaling as $\sim \epsilon / \Lambda_k$
\begin{equation}
 \epsilon \left(  - \frac{1}{2 \Lambda_k}V_k^T \mathbf{C_{A_{\delta} S}} V_k \;  Z_k(\mathbf{x})    +  \sum_{j = 1, j \neq k}^{N_{\mathcal{R}}}\sqrt{ \frac{\Lambda_j}{\Lambda_k}} \frac{V_j^T \mathbf{C_{A_{\delta} S}} V_k}{ \Lambda_k  - \Lambda_j} Z_j(\mathbf{x})  \right). \nonumber
\end{equation}
\end{itemize}

\subsection{Linearity of the perturbed  Principal Components when using IFS data}
One of the main features of Eq.~\ref{Eq:PerturbePCASimple} is that the presence of the astrophysical signal in the reference library is captured in a linear fashion. This has important consequences when using IFS data, because the high dimensionality of the astrophysical unknowns makes it very difficult to carry out the Forward Modeling minimization in  Eq.~\ref{Eq:BasicForwardModelling} {\em ``as is''} for SSDI observations.  Here we rewrite the expression of the  $\Delta Z_k(\mathbf{x})$  in a way that highlights this linear dependence on $ a= [a_1 .... a_{N_{\lambda}}]$. To do so, we use the following linear algebra identity, which is true for any eigenvector $V_i$ of the signal-less covariance matrix:
\begin{equation}
V_i^T \mathbf{a} \mathbf{A_{\delta}}(\mathbf{x})  = \bar{a}^T  \mathbf{Sel_{\lambda}}\mathbf{V_i} \mathbf{A_{\delta}}(\mathbf{x}). 
\label{Eq:StupidIdentity}
\end{equation}
This stems from: 
\begin{itemize}
\item defining  $\mathbf{L}$ as the $ N_{\lambda} \times  (N_{\lambda} N_{Exp}) $ rectangular matrix that relates each slice of each exposure to its wavelength --$\mathbf{L}= \left[ \begin{array}{ccc} \mathbb{I}_{N_{\lambda}}... \mathbb{I}_{N_{\lambda}}\end{array} \right]$. We then write $ \mathbf{Sel_{\lambda}} =  \mathbf{L} \; \mathbf{Sel}^T$.
\item recognizing that $\bar{a}^T \mathbf{Sel_{\lambda}}$ is an $N_{\mathcal{R}}$ dimensional vector whose $k$ th entry is $a_{\lambda_{k}}$ (the normalized astrophysical flux at the wavelength corresponding to the $k$ th reference).
\item recognizing that  for any $N_{\mathcal{R}}$ dimensional vector $b$: $b \mathbf{V_i} =  V_i^T \mathbf{b}$, where $\mathbf{V_i}$ and $\mathbf{b}$ denote the matrices whose diagonal elements are populated with the $N_{\mathcal{R}}$ entries of  the vectors $V_i$  and  $b$. Applying this identity to $\bar{a}^T  \mathbf{Sel_{\lambda}}\mathbf{V_i}$ yields Eq.~\ref{Eq:StupidIdentity}.
\end{itemize}
We then use Eq.~\ref{Eq:StupidIdentity} to simplify Eq.~\ref{Eq:PerturbePCASimple}. Some linear algebra  manipulations finally yield the main theoretical result of this manuscript:
\begin{eqnarray}
\epsilon \Delta Z_k(\mathbf{x})  & = & \epsilon a^T \mathbf{\Delta Z_k}^{\lambda}  (\mathbf{x}) = f^T \mathbf{\Delta Z_k}^{\lambda}  (\mathbf{x})   \label{Eq:LinearKLPerturbations} \\
\mathbf{\Delta Z_k}^{\lambda}  (\mathbf{x}) &=&  \frac{\mathbf{Sel_{\lambda}}}{\sqrt{\Lambda_k}} \left[ -\frac{1}{\sqrt{\Lambda_k}} \mathbf{V_k} \mathbf{A_{\delta}}(\mathbf{x})  \mathbf{S}(\mathbf{x})^{T} V_k \; Z_k(\mathbf{x}) + \mathbf{V_k} \mathbf{A_{\delta}}(\mathbf{x}) \sum_{j = 1, j \neq k}^{N_{\mathcal{R}}}  \frac{ \sqrt{\Lambda_j} }{ \Lambda_k  - \Lambda_j} ( \mathbf{V_k} \mathbf{A_{\delta}}(\mathbf{x})  \mathbf{S}(\mathbf{x})^{T} V_j + \mathbf{V_j} \mathbf{A_{\delta}}(\mathbf{x})  \mathbf{S}(\mathbf{x})^{T} V_k) Z_j(\mathbf{x})  \right] \nonumber
\end{eqnarray}
 While  Eq.~\ref{Eq:LinearKLPerturbations} can seem daunting at first, we emphasize that for a given unperturbed basis set $Z_{k}(\mathbf{x})$ and a given model of the image of the astrophysical source,  then $\mathbf{\Delta Z_k}^{(\lambda)}  (\mathbf{x})$ only needs to be pre-computed once and stored. Then the propagation of hypothetical sources of any arbitrary spectra $f$ through the least-squares speckle subtraction algorithm can be evaluated using a {\bf simple matrix multiplication}.  This has considerable advantages for statistical inference in the case of non-detections and for the estimation of astrophysical observables when faint substellar companions are detected.

\subsection{Generalization to perturbed LOCI coefficients}

While this manuscript used the formalism laid out in \cite{2012ApJ...755L..28S}, we can also use the equivalency between Eq.~18 and Eq.~29 in \cite{2015ApJ...800..100S} under the assumption of full rank correlation matrices, to apply our formalism to the case of LOCI. In the absence of astrophysical signal in the PSF library, the processed image using LOCI can be written as:
\begin{equation}
 \mathcal{P}_{LOCI \mathcal{R}} [T(\mathbf{x})]^T = T(\mathbf{x})^T - \mathbf{S}(\mathbf{x})^T \mathbf{c_{LOCI}}
\end{equation}
where the LOCI coefficients are given by Eq.~18 and Eq.~29  in \cite{2015ApJ...800..100S}. Note that here we have transposed all quantities in \cite{2015ApJ...800..100S} in order to follow the array conventions in \cite{2012ApJ...755L..28S}. 
\begin{equation}
\mathbf{c_{LOCI}} = \mathbf{V}^T  \mathbf{\Lambda^{-1}} \mathbf{V}\mathbf{S} T(\mathbf{x})^T
\end{equation} 
We use the the following notations for the perturbed problem:
\begin{itemize}
\item $\mathbf{V}, \mathbf{\delta V}$ are, respectively, the matrices whose column entries are $V_k, \delta V_k$,
\item $\mathbf{S},\mathbf{A_{\delta}}$ are as defined in the body of the manuscript,
\item $\mathbf{\Lambda^{-1}},\mathbf{\delta \Lambda^{-1}} $  are, respectively, the diagonal matrices whose diagonal entries are $1/\Lambda_k,1/ \delta \Lambda_k$. When using an eigenvalue truncation for LOCI, and keeping the first $K_{Klip}$ modes, then the last $K_{Klip}$ diagonal terms of both matrices are set to zero.
\end{itemize} 
In the presence of an astrophysical signal then the reduced image becomes:
\begin{equation}
 \mathcal{P}_{LOCI \mathcal{R}} [T(\mathbf{x})]^T = T(\mathbf{x})^T - \mathbf{S}(\mathbf{x})^T \mathbf{c_{LOCI}}  - \epsilon \mathbf{A_{\delta}}(\mathbf{x})^T \mathbf{c_{LOCI}} -  \epsilon \mathbf{S}(\mathbf{x})^T \mathbf{\delta c_{LOCI}}
\end{equation}
and we can use the formalism of the present manuscript to write $ \mathbf{\delta c_{LOCI}}$ 
\begin{equation}
\mathbf{\delta c_{LOCI}}= \left(   \mathbf{V}^T  \mathbf{\Lambda^{-1}} \mathbf{\delta V}\mathbf{S} +    \mathbf{\delta V}^T  \mathbf{\Lambda^{-1}} \mathbf{V}\mathbf{S} +   \mathbf{\delta V}^T  \mathbf{\Lambda^{-1}} \mathbf{V}\mathbf{S} \mathbf{S}  \mathbf{\delta V}^T  \mathbf{\Lambda^{-1}} \mathbf{V}\mathbf{A_{\delta}}^T - \mathbf{S}  \mathbf{V}^T  \mathbf{\delta \Lambda^{-1}} \mathbf{\delta V}\mathbf{S}^T \right) T(\mathbf{x})^T
\end{equation}
where $ \mathbf{\delta V}$ and $\mathbf{\delta \Lambda^{-1}} $ can be calculated using Eq.~\ref{Eq:PerturbationEigen}. The formalism developed herein is also applicable to LOCI implementation of least-squares speckle fitting algorithms.

 \clearpage
 
 \section{Appendix F: KLIP-FM in when an astrophysical signal is present in the reference library: case of IFS observations}

Here we describe the mathematical formalism that takes advantage of Eq.~\ref{Eq:LinearKLPerturbations} to generalize the RDI Forward Modeling concepts discussed in Appendix D to the case of IFS spectroscopy of faint point sources with ADI+SSDI. We assume that the location of the point source is known and seek to estimate its spectrum. We leave the astrometric estimation to further investigation.

\subsubsection{Forward modeling cost function in the presence of astrophysical signal in the reference library}


We start with Eq.~\ref{Eq:BasicForwardModelling}. Because of the presence of an astrophysical signal in the PSF library, the data analysis operator $\mathcal{LSQ}_{\mathcal{R}(\mathcal{A},\widehat{ \mathcal{A}})}$ cannot be simplified as straightforwardly as in the case of RDI.  Moreover, in the most general case the dependence on  $\mathcal{A}$ and $\widehat{ \mathcal{A}}$ is nonlinear. Fortunately in the case of  small perturbations (both for the actual signal and for its synthetic negative counterpart) the linear approximation in Eq.~\ref{Eq:LinearKLPerturbations} holds and the data analysis operator can be simplified as follows:
%
%
\begin{eqnarray}
\mathcal{LSQ}_{\mathcal{R}(\mathcal{A}) } [I(x)] &=& I(x)   - \sum_{k = 1}^{K_{Klip}} <I(x) ,Z_k(\mathbf{x}) + f^T \mathbf{\Delta Z_k}^{\lambda}  (\mathbf{x})  >_{\mathcal{S}}  (Z_k(\mathbf{x})+f^T \mathbf{\Delta Z_k}^{\lambda}  (\mathbf{x}) ) \\
\mathcal{LSQ}_{\mathcal{R}(\mathcal{A}) } [I(x)] &=& I(x)   - \sum_{k = 1}^{K_{Klip}} <I(x) ,Y_k(\mathbf{x})  >_{\mathcal{S}}  Y_k (\mathbf{x}) \\
\mathcal{LSQ}_{\mathcal{R}(\mathcal{A},\widehat{ \mathcal{A}}) } [I(x)] &=& I(x)   - \sum_{k = 1}^{K_{Klip}} <I(x) ,Y_k(\mathbf{x}) - \widehat{f}^T \widehat{\mathbf{\Delta Y_k}^{\lambda}  (\mathbf{x})}  >_{\mathcal{S}}  (Y_k(\mathbf{x})-\widehat{f}^T \widehat{\mathbf{\Delta Y_k}^{\lambda}  (\mathbf{x}) )} \label{Eq:KLIPwithNegative}
\end{eqnarray}

{\color{black} Of course in reality $f$ and $\mathbf{\Delta Z_k}^{\lambda}  (\mathbf{x})$ are unknown.  We can however apply the exact same treatment derived for the actual astrophysical signal in Eq.~\ref{Eq:LinearKLPerturbations} to the negative synthetic source. Plugging Eq.~\ref{Eq:KLIPwithNegative} into Eq.~\ref{Eq:BasicForwardModelling} yields the single wavelength Forward Modeling cost function at $\lambda_0$:
\begin{equation}
||  T_{\lambda_0}(\mathbf{x}) - \widehat{f_{\lambda_0}} A_{\lambda_0}(\mathbf{x}) -  \sum_{k = 1}^{K_{Klip}}  \left( < [T_{\lambda_0}(\mathbf{x}) -\widehat{ f_{\lambda_0}} A_{\lambda_0}(\mathbf{x})], [ Y_{k}^{\lambda_0}(\mathbf{x}) - \widehat{f} \widehat{\mathbf{\Delta Y_k^{\lambda_0}} (\mathbf{x})}]>   [ Y_{k}^{\lambda_0}(\mathbf{x}) - \widehat{f} \widehat{\mathbf{\Delta Y_k^{\lambda_0}} (\mathbf{x})}] \right)  ||_{\mathcal{F}}^2. 
\end{equation} 
Adding this cost function for all wavelengths finally yields the spectral extraction cost function over the full range of wavelengths covered by the IFS:
\begin{equation}
\sum_{p = 1}^{N_{\lambda}}||  T_{\lambda_p}(\mathbf{x}) - \widehat{f_{\lambda_p}} A_{\lambda_p}(\mathbf{x}) -  \sum_{k = 1}^{K_{Klip}} \left( <(T_{\lambda_p}(\mathbf{x}) -\widehat{ f_{\lambda_p}} A_{\lambda_p}(\mathbf{x})), ( Y_{k}^{\lambda_p}(\mathbf{x}) - \widehat{f} \widehat{\mathbf{\Delta Y_k^{\lambda_p}} (\mathbf{x})})>   ( Y_{k}^{\lambda_p}(\mathbf{x}) - \widehat{f} \widehat{\mathbf{\Delta Y_k^{\lambda_p}} (\mathbf{x})}) \right)  ||_{\mathcal{F}}^2. 
\label{Eq:RefFreeForwardModellingSDI}
\end{equation}
 Note that in the most rigorous case, each wavelength should to be weighted by its noise and possible correlation between wavelengths ought to be included. This is critical when deriving confidence intervals. However, the applications presented in this paper are only focused on biases associated with the most likely estimated spectrum. We leave further sophistications related to the calculations of confidence intervals to a future communication (Wang et al. 2016, in preparation).  In \S 3. we show that even using this simple approach yields unbiased estimated spectra (albeit without confidence intervals). Here we emphasize that the Principal Components considered are the sum of two terms:
\begin{itemize}
\item the principal components of the data itself, $Y_{k}^{\lambda_p}(\mathbf{x}) $, which contain (or do not contain) perturbations due to the presence of a hypothetical point source. Unfortunately the contribution of these perturbations cannot be evaluated a priori. 
\item a term corresponding to the the perturbation of $Y_{k}^{\lambda_p}(\mathbf{x}) $ due to the injection of the synthetic injected negative  point source --$\widehat{f_{\lambda}} \widehat{\mathbf{\Delta Y_k} (\mathbf{x})}$. This synthetic source is used for Forward Modeling purposes. Even if this term corresponds to the  injection of a non-physical source for the purpose of astrophysical inference, its impact can still be quantified using the result in Eq.~\ref{Eq:LinearKLPerturbations}. 
\end{itemize}
}
\subsubsection{Linearized  problem}
We now simplify Eq.~\ref{Eq:RefFreeForwardModellingSDI} in order to reduce the problem to the functional form described in Appendix D (Eq.~\ref{Eq:ForwardLS}, which is more amenable to astrophysical inference). We use a fitting region $\mathcal{F}$ whose size is independent of the size of the search area  $\mathcal{S}$. Assuming that the point source has been detected but its position is only known at the $\sim 1-2$ pixels level, this fitting region can be an aperture the size of the FWHM of the PSF centered around this rough position. This is what we use in practice in the examples in this paper. We introduce the following notations:
\begin{eqnarray}
&&P_{\lambda_p} (\mathbf{x})=  T_{\lambda_p}  (\mathbf{x}) - \sum_{k = 1}^{K_{Klip}} <T_{\lambda_p}  (\mathbf{x}),Y_{k}^{\lambda_p}  (\mathbf{x})>_{\mathcal{S}} Y_{k}^{\lambda_p} (\mathbf{x}) \nonumber \\
&& F_{\lambda_p} (\mathbf{x})  =\widehat{f_{\lambda_p}} \left( A_{\lambda_p}  (\mathbf{x}) - \sum_{k = 1}^{K_{Klip}}  <A_{\lambda_p} (\mathbf{x}) ,Y_{k}^{\lambda_p}  (\mathbf{x})>_{\mathcal{S}}  Y_{k}^{\lambda_p}  (\mathbf{x})  \right) \nonumber \\
&& -\widehat{f} \sum_{k = 1}^{K_{Klip}} \left( <T_{\lambda_p}  (\mathbf{x}) ,Y_{k}^{\lambda_p}  (\mathbf{x}) >  \widehat{\mathbf{\Delta Y_k^{\lambda_p}}} (\mathbf{x})  + <T_{\lambda_p}  (\mathbf{x}),  \widehat{\mathbf{\Delta Y_k^{\lambda_p}}}  (\mathbf{x})>_{\mathcal{S}} Y_{k}^{\lambda_p} (\mathbf{x}) \right). \nonumber 
\end{eqnarray}
This latter term contains all the contributions of the synthetic negative point source that are linear in $ \widehat{f}$ (e.g. neglecting the terms in $\mathcal{O}(\widehat{\epsilon}^2)$ in the Forward Modeling cost function). This expression for $F_{\lambda_p} (\mathbf{x})$ captures both over-subtraction and self-subtraction. Here again, $F_{\lambda_p} (\mathbf{x})$ can be written as a simple matrix multiplication:
\begin{equation}
F_{\lambda_p} (\mathbf{x}) = \widehat{f} \mathbf{F_{\lambda_p} }(\mathbf{x})
\end{equation}
with $\mathbf{F_{\lambda_p} }$ being a $N_{\lambda} \times N_{pix}$ matrix whose $q$ th line entry is defined by:
\begin{eqnarray}
&& \mathbf{F_{\lambda_p} }[q](\mathbf{x}) =\delta_{p,q} \left( A_{\lambda_p} (\mathbf{x})  - \sum_{k = 1}^{K_{Klip}}  <A_{\lambda_p}(\mathbf{x}) ,Y_{k}^{\lambda_p}(\mathbf{x})>  Y_{k}^{\lambda_p}(\mathbf{x}) \right) \nonumber  \\
& & ... - \sum_{k = 1}^{K_{Klip}} \left( <T_{\lambda_p}(\mathbf{x}),Y_{k}^{\lambda_p}(\mathbf{x})>  \widehat{\mathbf{\Delta Y_k^{\lambda_p}}}[q](\mathbf{x}) +  <T_{\lambda_p}(\mathbf{x}),  \widehat{\mathbf{\Delta Y_k^{\lambda_p}}}[q](\mathbf{x})> Y_{k}^{\lambda_p}(\mathbf{x}) \right) \label{Eq:DefLambdaMatrix}
\end{eqnarray}
where $\widehat{\mathbf{\Delta Y_k^{\lambda_p}}}[q](\mathbf{x})$ is the $q$ th line entry in $\widehat{\mathbf{\Delta Y_k^{\lambda_p}}}(\mathbf{x})$ . This yields the following simplified form for the Forward Modeling cost function:
\begin{equation}
\widetilde{f}= \arg \min_{\widehat{f_{\lambda}}}  \sum_{p = 1}^{N_{\lambda}} || P_{\lambda_p}(\mathbf{x}) - \widehat{f} \mathbf{F_{\lambda_p} }(\mathbf{x})||_{\mathcal{F}}^2
\label{Eq:ForwardModellingSpectra}
\end{equation}

 \subsection{Spectral extraction algoritm}
Eq.~\ref{Eq:ForwardModellingSpectra} can seem daunting at first but it can be easily implemented following the steps below:
\begin{itemize}
\item Once a faint point source has been identified, carry out a KLIP reduction using geometric parameters that keep the point source at the center of the subtraction $\mathcal{S}$ zone. If combining ADI and SSDI, derotate each exposure and sum over time to obtain:
\begin{equation}
P_{\lambda_p}(\mathbf{x})  = \sum_t P_{\lambda_p, t}(\mathit{R}_{-\theta_t}[\mathbf{x}]) 
 \end{equation}
\item Assuming that the location of the point source and the off-axis PSF of the instrument are known, build a model of the motion of the point source across wavelengths and exposures $A_{\lambda}(\mathit{R}_{\theta_t}[\mathbf{x} \frac{\lambda}{\lambda_0}])$.
\item For each $Z_k(\mathbf{x}),V_k,\lambda_k, \mathbf{Sel_{\lambda}}$ associated with the reduction of each exposure at each wavelength, use this model in conjunction with Eq.~\ref{Eq:LinearKLPerturbations} and Eq.~\ref{Eq:DefLambdaMatrix} to calculate $\widehat{\mathbf{\Delta Y_k^{\lambda_p}}}(\mathbf{x})$ and $ \mathbf{F_{\lambda_p} }(\mathbf{x})$. When using ADI, derotation and summation over time are necessary:
\begin{equation}
\mathbf{F}_{\lambda_p}(\mathbf{x})  = \sum_t \mathbf{F}_{\lambda_p,t}(\mathit{R}_{-\theta_t}[\mathbf{x}]) 
 \end{equation}
\end{itemize}
These three steps only consist of basic linear algebra operations  (associated with some image rotations over of a small subregion $\mathcal{F}$ of the image, usually ranging from a PSF FWHM to the entire $\mathcal{S}$ zone). Once they have been carried out, the spectrum of the point source can be retrieved using any quadratic optimization algorithm to minimize Eq.~\ref{Eq:ForwardModellingSpectra}. Here we only consider the most simple route and recognize that Eq.~\ref{Eq:ForwardModellingSpectra} is similar to Eq.~\ref{Eq:minPhot} (up to a summation) and thus the estimated spectrum can be obtained by solving the following inverse problem:
\begin{equation}
\left(   \sum_{p = 1}^{N_{\lambda}} \mathbf{F_{\lambda_p} }(\mathbf{x}) \mathbf{F_{\lambda_p} }(\mathbf{x}) ^T \right)  \widehat{f} =  \sum_{p = 1}^{N_{\lambda}} \mathbf{F_{\lambda_p} }(\mathbf{x}) P_{\lambda_p}(\mathbf{x})^T
\label{Eq:SimpleMat}
\end{equation}

 \subsection{Detection limits in IFS data}
Here we show that the algorithm discussed above can also be used to obtain detection limits that vary as a function of the hypothetical nature of the point sources that have not been detected in IFS data. Quantifying completeness as a function of underlying spectral type is most often carried by injecting a series of point sources featuring the various hypothetical spectra that the experiment is expecting to be sensitive to (see the example of \cite{2014PNAS..11112661M} for an illustration using the with and without methane hypothesis). Because this involves analyzing multiple data sets with synthetic sources, very often only a small finite number of hypothesis are tested due to practical (CPU time) limitations. Using Eq.~\ref{Eq:LinearKLPerturbations} completely circumvents this problem because it enables the injection of a ``generic'' synthetic planet in the data: the reduced cubes/images can then be obtained from this ``generic data set'' via a simple matrix multiplication. Then any observer can be used to generate the ROCs and calculate completeness. In practice, this method follows the steps described below for a standard ADI+SSDI sequence:
\begin{itemize}
\item For each cube and each wavelength, calculate the reduced image $P_{\lambda,t} (\mathbf{x})$. This image can be calculated by splitting the field of view in multiple subtraction regions,  $\mathcal{S}$. 
\item For each cube and each wavelength, add an astrophysical scene $A_{\lambda}(\mathit{R}_{\theta_{t}}[\mathbf{x}])$ to the processed image. This scene can be composed of a series of point sources separated in radius and azimuth so that there is only one synthetic source for each $\mathcal{S}$ zone. 
\item For each point source in the astrophysical scene (and thus each corresponding  $\mathcal{S}$ zone), each cube, and each wavelength, calculate the associated perturbed Principal Components $\mathbf{\Delta Z_k}^{\lambda,t} (\mathbf{x})$ 
\item Pick a value of $K_{Klip}$ and calculate $\mathbf{F}_{\lambda,t} (\mathbf{x})$ associated with each synthetic point source.  
\item For each underlying hypothetical spectrum, create a synthetic observed image integrated over the entire observing sequence, at each wavelength based on these quantities:
\begin{equation}
 P_{\lambda}^{Syn} (\mathbf{x}) = \sum_t \left ( P_{\lambda} (\mathit{R}_{-\theta_{t}}[\mathbf{x}]) +  A_{\lambda} (\mathbf{x}) \right) + f \sum_t \mathbf{F}_{\lambda,t} (\mathit{R}_{-\theta_{t}}[\mathbf{x}]) 
\label{Eq:KLIPFMNoDetec}
\end{equation}
(Note that here we have dropped the summation over the number of synthetic sources included in the dataset).  
\item Once the various terms in Eq.~\ref{Eq:KLIPFMNoDetec}, $ P_{\lambda}^{Syn} (\mathbf{x}) $ have been evaluated, then generating a reduced image for a given hypothetical spectrum $\widehat{f}$ can be done via simple matrix multiplication. The observer of choice can the be applied to a wide variety of spectra and completeness calculated for each hypothesis. 
\end{itemize}
This method tests a wide variety of non-detection hypothesis without having to resort to a CPU-costly matrix inversion (associated with the Principal Components calculation) for each astrophysical scenario. It does involves a computational overhead of $N_{\lambda}$ extra image rotations when compared to the analysis of an entire data set simply using KLIP. However, this overheard can be mitigated by only carrying out these rotations locally in regions surrounding the injected planets. Moreover, most implementations of PCA-based methods calculate reduced images for a range of $K_{Klip}$ all the way to $N_{Corr}$. Because the linear model in Eq.~\ref{Eq:KLIPFMNoDetec} is valid for all $K_{Klip}$ in the case of  synthetics planets at the detection limit, it is sufficient to apply only the procedure described above for one value of  $K_{Klip}$. \\


 \end{document}